\documentclass[rmp,aps,twocolumn,showpacs,amsmath,amssymb,floatfix]{revtex4}

\usepackage{graphicx}
\graphicspath{{fig/}}
\usepackage{dcolumn}
\usepackage{bm}
\usepackage{natbib} 
\usepackage{multirow} 
\usepackage{enumerate}
\usepackage{amssymb,pifont} 
\usepackage{epstopdf}

\newcommand\T{\rule{0pt}{2.6ex}} 

\begin{document}

\title{Liquid Xenon Detectors for Particle Physics and Astrophysics} 

\author{E. Aprile}
\affiliation{Department of Physics, Columbia University, New York, NY 10027, USA}

\author{T. Doke}
\affiliation{Advanced Research Institute for Science and Engineering, Waseda University, Tokyo 169-8555, Japan}

\date{\today} 

\begin{abstract}
This article reviews the progress made over the last 20 years in the development and applications of liquid xenon detectors in particle physics, astrophysics and medical imaging experiments. We begin  with a summary of the fundamental properties of liquid xenon as radiation detection medium, in light of the most current theoretical and experimental information. After a brief introduction of the different type of liquid xenon detectors, we continue with a review of past, current and future experiments using liquid xenon to search for rare processes and to image radiation in space and in medicine. 
We will introduce each application  with a brief survey of the underlying scientific motivation and experimental requirements, before reviewing the basic characteristics and expected performance of each experiment. Within this decade it appears likely that large volume liquid xenon detectors operated in different modes will contribute to answering some of the most fundamental questions in particle physics, astrophysics and cosmology, fulfilling the most demanding detection  challenges. From experiments like MEG, currently the largest liquid xenon scintillation detector in operation, dedicated to the rare ``$\mu \rightarrow$ e$\gamma$'' decay,  to the future XMASS which also exploits only liquid xenon scintillation to address an ambitious program of rare event searches,  to the class of time projection chambers like XENON and EXO which exploit both scintillation and ionization of liquid xenon for dark matter and neutrinoless double beta decay, respectively, we anticipate unrivaled performance and important contributions to physics in the next few years. 
\end{abstract}

\maketitle
\tableofcontents

\section{Introduction}
This article is concerned with liquid xenon as radiation detector medium, and the experiments which use it in search of answers to some of the most intriguing  questions in physics today.  It is a tribute to the vision of those investigators who recognized early on the opportunity that this material would offer for particle detectors, and who have continued their effort in the field for more than 30 years, inspiring new generations.  Their vision and persistent work has significantly contributed  to the successful realization of today's experiments using large scale liquid xenon detectors.  

Historically, the advantages of liquid xenon for radiation detection were first recognized by Alvarez in 1968 \cite{Alvarez:1968}. Simultaneously with the development of the first liquid xenon ionization detector by the Berkeley group, independent groups in Russia and Japan have also carried out experiments to   study the fundamental properties of liquid xenon for radiation detection. From the seminal work of Dolgoshein \cite{Dolgoshein:1967}, Doke, Kubota \cite{Doke:1981} and co-workers, to the first development of a large volume liquid xenon time projection chamber as a Compton telescope \cite {EAprile:89:SPIE}, we have seen an evolution in liquid detectors and their application in different fields. Exploiting the original idea by Dolgoshein of electron emission from the liquid to the gas, two-phase time projection chambers with single electron detection capability have been developed to search for dark matter particles.  From space-based detectors for astrophysical gamma-rays, to detectors located at accelerators or deep underground in search of rare decays expected from physics beyond the standard model, liquid xenon remains the preferred medium for many reasons. 
Among liquid rare gases, liquid xenon has the highest stopping power for penetrating radiation, thanks to its high atomic number and density. It also has the highest ionization and scintillation yield, the latter comparable to that of NaI(Tl) and with a faster time response. Compared to all other detector media, liquid rare gases have the unique feature of responding to radiation with both ionization electrons and scintillation photons. Detectors which use both signals with high detection efficiency have a significant advantage in the measurement of the properties of the radiation. In recent years, much progress has been made in the development of photodetectors with high quantum efficiency at the 178 nm wavelength of the liquid xenon scintillation. Simultaneously, progress in the development of cryo-coolers with sufficient power to liquefy and maintain the liquid temperature has enabled reliable detector's operation. The relatively warm temperature of liquid xenon, compared to other liquid rare gases, makes the removal of electronegative contaminants more difficult to achieve, since at colder temperatures, many impurities freeze out. On the other hand, purification systems with continuous cleaning through commercial purifiers, have shown the required level of purity in large volume liquid xenon detectors. The current challenge for several experiments is in the stringent radio-purity requirement for all materials in contact with the liquid, from containment vessels and cryo-coolers to  photon and charge read-out sensors. Through materials selection and screening, the background level and the sensitivity of liquid xenon experiments for rare event searches continues to improve. Twenty years ago the standard size of a liquid xenon detector was limited to a few hundred grams of mass. Ten year ago the  LXeGRIT and the RAPID time projection chambers had a mass of 30 kg and 60 kg, respectively. Today the XENON100 and the EXO time projection chambers have a mass of about 200 kg. The MEG experiment is currently operating the largest liquid xenon scintillating calorimeter with a mass of 2.7 tons. Several other detectors from 300 kg to 3 tons are under construction or are planned for the next few years.  Most of the success of liquid xenon detectors can be attributed to the richness of the information contained in its charge and light signals. Looking back at the development over the last twenty years, it seems appropriate to summarize our current understanding and the status of the field. 

The article is organized as follows. In \ref{sec:fundamental} we summarize the properties of liquid xenon as detector medium, based on well established experimental knowledge, some dating back more than fifty years, but accounting for the most recent works. In \ref{sec:det_type}, we review the basic operating principle of different type of detectors while the last chapter covers the applications of liquid xenon detectors in  physics, astrophysics and medical imaging experiments. Some selection was necessary, largely to comply with the length limit for this review, but we have tried to include the majority of current and planned experiments. We have included our own research work on many of the topics and experiments discussed in this review, but have attempted also to cover all important contributions from other groups and authors and apologize for any omission.  

\section{Properties of Liquid Xenon as Detector Medium}
\label{sec:fundamental}

We review here the most important physical and operational properties of liquid xenon (LXe) as a detector medium for different types of particles. A unique and important feature of LXe, shared only by liquid argon (LAr) among liquid rare gases, and specific to this class of materials, is the production of both charge carriers and scintillation photons in response to radiation. The charge and light signals are highly complementary and anti-correlated. When detected simultaneously and with high efficiency, the two signals enable a precise measurement of the particle's properties from its energy and interactions history in space and time, to its type. 

\subsection{Physical Properties of Liquid Xenon}
Table \ref{Table1} summarizes the physical properties of LXe critical to its function as a detector
medium \citep{Hallet:1961, Crawford:1977,Gruhn:1977,Schmidt:2001}.
The high atomic number (54) and high density ($\sim$ 3 g/cm$^3$) of LXe make it very efficient to stop penetrating radiation. Compared to a crystal scintillator such as NaI (Tl) or to a semiconductor such as Ge, LXe offers the high stopping power benefit in a single large and homogeneous volume, which is not easily possible with the other media. The presence of many isotopes in natural Xe, with different spin and at a non negligible isotopic abundance, is of interest for different physics applications, as we will discuss in \ref{sec:applications}. 

\begin{figure} 
\begin{center} 
\includegraphics*[width=.4\textwidth]{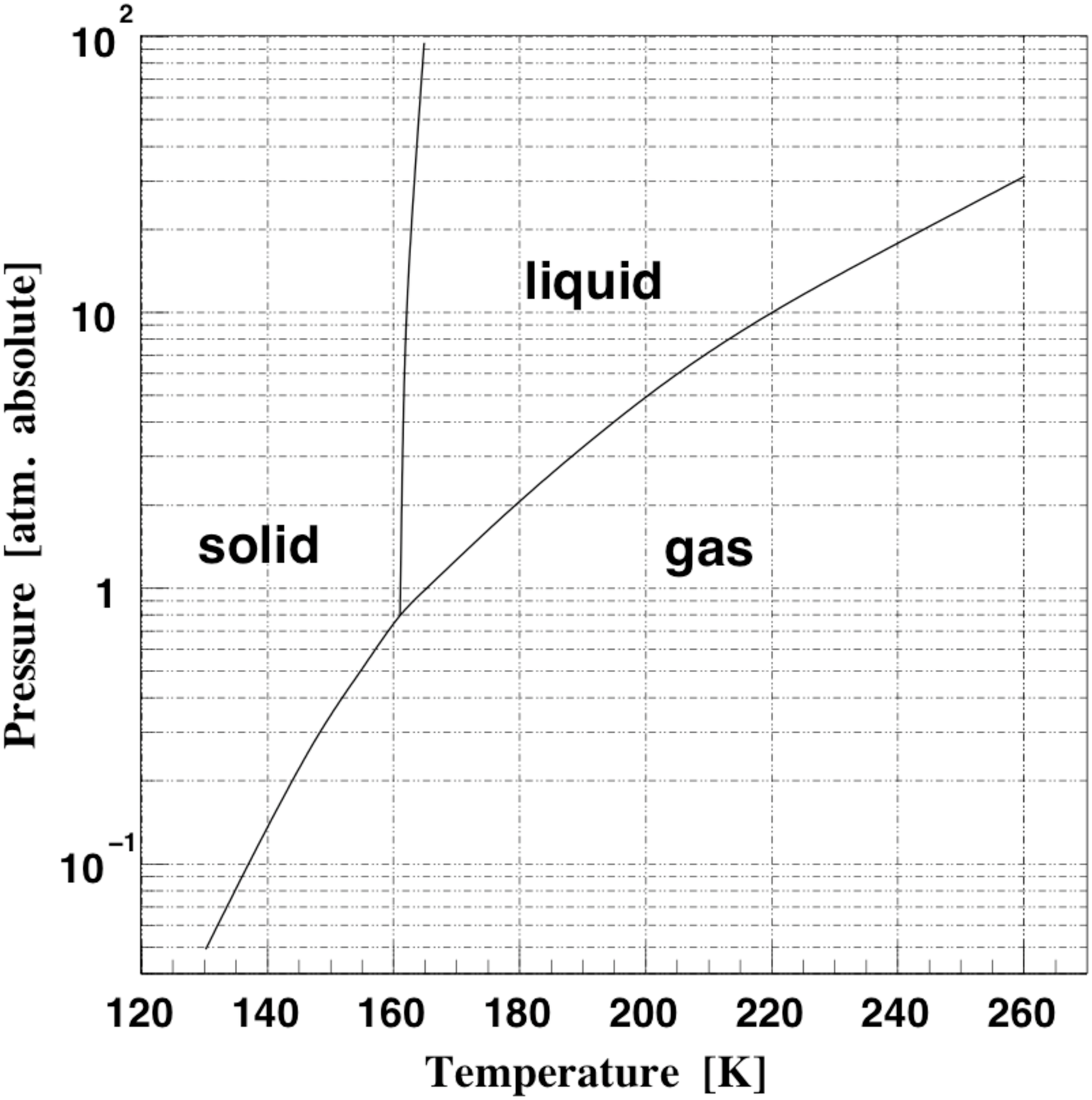} 
\caption{Phase diagram of xenon.} 
\label{fig:phases} 
\end{center}
\end{figure}

Figure \ref{fig:phases} shows a phase diagram of crystalline, liquid, and gaseous xenon \citep{Hallet:1961}. At
atmospheric pressure, the liquid phase of Xe extends over a narrow temperature range, from about 162 K to 165
K. The relatively high boiling point temperature of LXe, compared to other liquid rare gases, requires a modest cryogenics system for gas liquefaction.  

\subsection{Electronic Structure}

\begin{figure} 
\begin{center}
\includegraphics*[width=.4\textwidth]{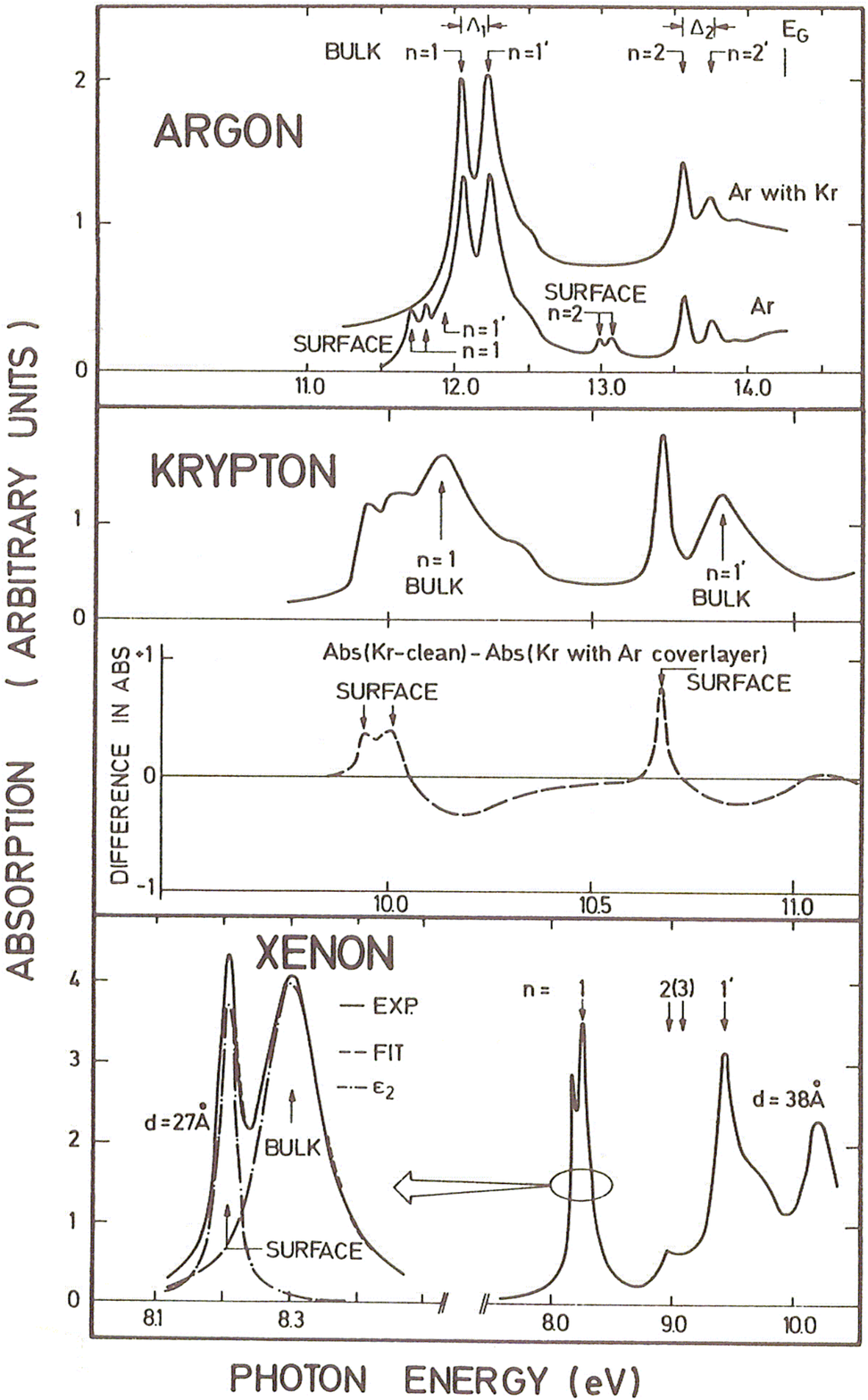} 
\caption{High-resolution absorption spectra for solid Ar, Kr, and Xe in the range of the valence excitons. Volume and surface excitons are observed for all three samples. For Ar and Kr the results of surface coverage experiments are also shown. For Xe the experimentally determined spectrum in the range of the n = 1 surface and volume exciton is displayed on an expanded scale together with a lineshape analysis. Reprinted with permission from \cite{Schwenter:1985}.} 
\label{fig:Abs_Spectra} 
\end{center}
\end{figure}

The electronic band structure picture is useful for understanding the ionization and the scintillation processes in LXe. Solid rare gases are excellent insulators. They are crystals, with a face-centered cubic
structure. With the
exception of solid helium, they have electronic band structures, despite the very weak atomic interactions due
to van der Waals forces. Absorption spectroscopy provides the most direct method to
measure the band gap energy, $E_g$, which is the energy difference between the bottom of the
conduction band and the top the valence band, as in a semiconductor or an insulator.  Figure \ref{fig:Abs_Spectra} shows the high resolution absorption spectra
for solid Ar, Kr, and Xe from which the band gaps have been determined (see \cite{Schwenter:1985} and references therein). Exciton peaks were clearly observed in these spectra, providing direct evidence of the band structure of the solid rare gases. The observation of the exciton level \cite{Beaglehole:1965,Steinberger:1973} and the direct measurement of
the band gap energy of LXe, LAr and LKr \cite{Asaf:1974,Reininger:1984,Bernstorff:1983,Reininger:1982},
determined that the liquids also have a band structure (see Table \ref{W-value}). Given the large band gap, liquid rare gases are also excellent insulators. 

\begin{center}
\begin{table}[htbp]
\caption{Physical Properties of Liquid Xenon}
\label{Table1}
\begin{tabular}{|c|c|}
\hline
{\bf  Property} & {\bf Value} \T \\ 
\hline
Atomic number Z \T & $54$\\
\multirow{4}{*}{Isotopes} &
$^{124}\mathrm{Xe}(0.09\%)$, $^{126}\mathrm{Xe}(0.09\%)$,\\
& $^{128}\mathrm{Xe}(1.92\%)$, $^{129}\mathrm{Xe}(26.44\%)$\\
& $^{130}\mathrm{Xe}(4.08\%)$, $^{131}\mathrm{Xe}(21.18\%)$\\
& $^{132}\mathrm{Xe}(26.89\%)$, $^{134}\mathrm{Xe}(10.44\%)$\\
& $^{136}\mathrm{Xe}(8.87\%)$ \\
Mean atomic weight A& $131.30$ \\
Density & 3 g$\cdot$cm$^{-3}$ \\
\multirow{2}{*}{Boiling point} & $T_b = 165.05$ K, $P_b = 1$ atm \\
& $\rho_b = 3.057$ g$\cdot$cm$^{-3}$ \\
\multirow{2}{*}{Critical point} & $T_c = 289.72$ K, $P_c = 58.4$ bar \\
& $\rho_c = 1.11$ g$\cdot$cm$^{-3}$ \\
\multirow{2}{*}{Triple point}& $T_t = 161.3$ K, $P_t = 0.805$ bar \\
& $\rho_t = 2.96$ g$\cdot$cm$^{-3}$ \\
\multirow{2}{*}{Volume ratio ($\rho_{liquid}/\rho_{gas}$)}&\\
& 519\\
\hline
{\bf Thermal properties} &  \\
Heat capacity & $10.65$ cal$\cdot$g$\cdot$mol$^{-1}\cdot$K$^{-1}$ \\
& for $163-166$ K \\
Thermal conductivity & $16.8 \times 10^{-3}$ cal$\cdot$s$^{-1}\cdot$cm$^{-1} \cdot$K$^{-1}$ \\
Latent heat of & \\
 a) evaporation & \multirow{2}{*}{$3048$ cal$\cdot$g$\cdot$mol$^{-1}$} \\
at triple point &\\
 b) fusion & \multirow{2}{*}{$548.5$ cal$\cdot$g$\cdot$mol$^{-1}$} \\
at triple point &\\
\hline
{\bf Electronic properties} &  \\
Dieletric constant & $\epsilon _r=1.95$ \\
\hline
\end{tabular}
\end{table}
\end{center}

\subsection{Ionization Process}

In rare gases, the energy deposited by radiation is expended in the production of a number of electron-ion pairs, $N_i$, excited atoms, $N_{ex}$, and free electrons with a kinetic energy lower than the energy of the first excited level, known as sub-excitation electrons. 

We can express the transfer of the deposited energy $E_0$ into ionization, excitation, and sub-excitation
electrons by an energy balance equation, as originally proposed by Platzman for rare
gases \cite{Platzman:1961}: 
\begin{equation}
	E_0 = N_i E_i + N_{ex} E_x + N_i \epsilon
\end{equation}
where $N_i$ is the number of electron-ion pairs produced at an average expenditure of energy $E_i$, $N_{ex}$
is the number of excited atoms at an average expenditure of energy $E_x$, and $\epsilon$ is the average
kinetic energy of sub-excitation electrons. The $W$-value is defined as the average energy required to produce
one electron-ion pair, and is  given as:
\begin{equation}
	W = E_0/N_i = E_i + E_x \left({N_{ex}/N_i}\right) + \epsilon \label{eq:traritraraa}
\end{equation}
In solid or liquid rare gases, the established existence of an electronic band structure, allows us to rewrite the Platzman equation with
the band gap energy, $E_g$, replacing the ionization potential of the gas:
\begin{equation}
	W/E_g = E_i/E_g + \left({E_x/E_g}\right) \left({N_{ex}/N_i}\right) + \epsilon/E_g
\end{equation}
To calculate $W/E_g$ for LXe, the ratios $E_x/E_g$ and $N_{ex}/N_i$ were estimated by
using the oscillator strength spectrum of solid Xe obtained from photo-absorption data, in the optical
approximation \cite{Takahashi:1975}. For E$_i$, the data of Rossler \cite{Rossler:1971} are used, assuming
the width of the valence band to be negligibly small. For an estimate of $\epsilon$, the Shockley model \cite{Doke:1976, Shockley:1961} was used. The calculated ratio $W/E_g$  are about 1.65 for LXe, LAr and LKr, in good agreement with the measured values of about 1.6 for all three liquids, reported in Table \ref{W-value}. This  supports the electronic
band structure assumption for the liquid rare gases heavier than Ne. 

\subsubsection{Ionization Yield}

The ionization yield is defined as the number of electron-ion pairs produced per unit absorbed energy. In radiation chemistry, the
$G$-value is usually used as such unit, defined to be the average number of electron-ion pairs produced per 100
eV of absorbed energy. In physics, however, we prefer to use the $W$-value, which is inversely proportional to
$G$. Since the $W$-value depends weakly on the type and the energy of the radiation, except for very low energies, we
consider it to be almost constant. Therefore the ionization signal produced in a LXe detector can be used to
measure the deposited energy. To correctly measure the  number of electron-ion pairs produced by
radiation in LXe, one needs (a) to minimize the loss of charge carriers by attachment to impurities, i.e. the liquid
has to be ultra pure; (b) to minimize the recombination of electron-ion pairs and thus collect all the original charge carriers produced, i.e. by applying a very high electric field and (c) to estimate the deposited energy correctly. Measurements of the ionization yield in LXe have been carried out with small gridded ionization chambers that met these requirements, irradiated with electrons and gamma-rays from internal radioactive sources. From these measurements, the $W$-value is inferred by extrapolation to infinite field. Table \ref{W-value} summarizes the measured $W$-values in LAr, LKr and LXe \citep{Aprile:1993,Doke:1969,Miyajima:1974,Takahashi:1975}; they are smaller than the corresponding $W$-values in gaseous Ar, Kr, and Xe (also shown, along with the ionization potential of the gas). LXe has the smallest $W$-value, hence the the largest ionization yield, of all liquid rare gases.

\begin{table}[t]
\caption{Ionization potentials or gap energies and W-values in liquid argon, krypton and xenon.  a(Doke, 1969); b(Miyajima, 1974); cÄ(Aprile, 1993); d(Takahashi, 1975). }
\label{W-value}
\center
\begin{tabular}{|l|l|l|l|}
\hline
Material & Ar & Kr & Xe \\
\hline
\it{\bf{Gas}} & & & \\
\ Ionization potential {\it I} (eV) & 15.75 & 14.00 & 12.13 \\
\ W-values (eV) & 26.4$^{\rm a}$ & 24.2$^{\rm a}$ & 22.0$^{\rm a}$ \\

\it{\bf{Liquid}} & & & \\
\ Gap energy (eV) & 14.3 & 11.7 & 9.28 \\
\ W-value (eV) & 23.6$\pm$0.3$^{\rm b}$ & 18.4$\pm$0.3$^{\rm c}$ & 15.6$\pm$0.3$^{\rm d}$ \\
\hline
\end{tabular}
\end{table}

As discussed above, the
energy lost by radiation in LXe is expended in ionization, excitation, and sub-excitation electrons. The
average energy lost in the ionization process is slightly larger than the ionization potential or the gap
energy because it includes multiple ionization processes. The average energy lost in the excitation process is comparatively small ($\sim$5\% \citep{Aprile:1993, Takahashi:1975}). The energy transferred to sub-excitation
electrons is larger than 30\% of the ionization potential or gap energy. As a result, the ratio of the $W$-value
to the ionization potential or the gap energy is 1.4 for rare gases \cite{Platzman:1961}, 1.6$\sim$1.7 for liquid rare gases
\citep{Aprile:1993, Takahashi:1975} and $\sim$2.8 for semiconductors with a wide conduction and valence band
\citep{Klein:1968}.
In 1992, Seguinot et al. reported a small $W$-value of 9.76$\pm$0.76 eV for LXe, using an
electron beam with a kinetic energy of 100 keV \citep{Seguinot:1992}. This value is just slightly higher than the
band gap energy of solid xenon (9.33 eV \citep{AIPH:1982}), and the ratio W/E$_g$ = 1.046 is too small and inconsistent with the above considerations. See \citep{Miyajima:1995} for a critical analysis of these data. 

\begin{figure}
\includegraphics[width=1\columnwidth,clip,trim=10 0 10 0]{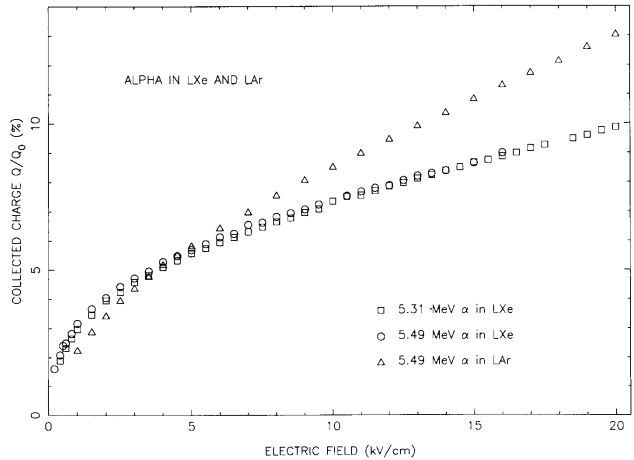}
\caption{\label{fig:alphas} Collected charge (Q/Q$_0$ \%) as a function of electric field for $^{210}$Po in liquid xenon (squares) and $^{241}$Am in liquid xenon (circles) and liquid argon (triangles)\citep{Aprile:1991_307}.}
\end{figure}

\begin{figure} 
\begin{center}
\includegraphics*[width=1\columnwidth,clip,trim=20 0 20 0]{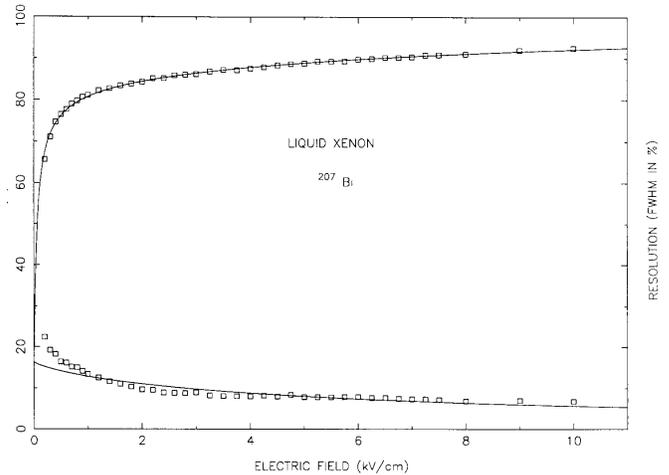} 
\caption{Energy resolution and collected charge for 570 keV gamma-rays in LXe as a function of electric field \citep{Aprile:1991}.} 
\label{fig:electrons} 
\end{center}
\end{figure}

Measurements of the ionization yield of heavily ionizing alpha particles in LXe have also been carried out by several authors (see \citep{Aprile:1991_307} and references therein) as they provide important information on the electron-ion recombination process. 
Alpha particles have a cylindrical track in which most of the energy is lost in a dense core with a high recombination rate, surrounded by a sparse ``penumbra" of delta rays.  Complete charge collection for alpha particles is thus difficult, and indeed less than 10\% of the total charge is typically collected at an electric field as high as 20 kV/cm.  Figure~\ref{fig:alphas} shows the characteristic non-saturation feature of the ionization yield of alpha particles in LXe and 
LAr, for comparison \citep{Aprile:1991_307}. The different ionization densities of alpha tracks in the two liquids explain the different slope of the saturation curves. On the other hand, almost complete charge collection at modest electric fields, is possible for minimum ionizing particles.  Figure~\ref{fig:electrons} shows a typical saturation curve, or charge collected as a function of field, measured for 570 keV gamma-rays in a gridded ionization chamber \citep{Aprile:1991}. The corresponding energy resolution is also shown. 

\begin{figure}
\includegraphics[width=1\columnwidth,clip,trim=50 0 50 0]{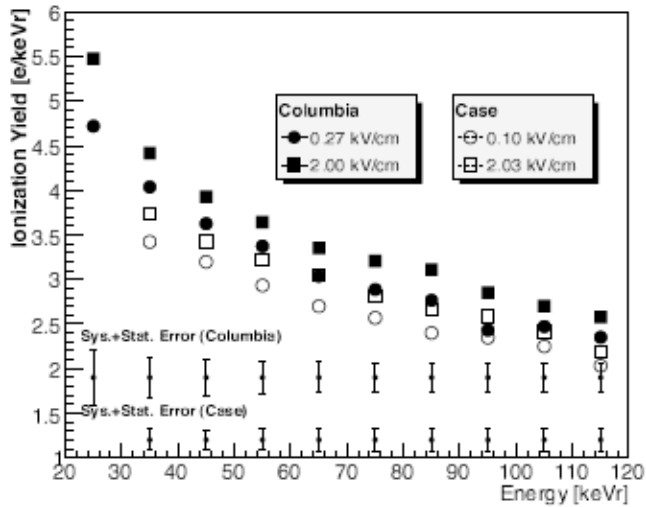}
\caption{\label{fig:NR} Ionization yield from nuclear recoils measured with small scale two-phase xenon detectors (labeled Columbia and Case), at different electric field \cite{Aprile:2006}.}
\end{figure}

Existing models \cite{Jaffe:1913, Kramers:1952, Onsager:1938, Thomas:1988, Mozumder:1995} explain most of the experimental data in various special cases.
Recently, the interest in LXe as target and detector medium for dark matter searches has triggered  measurements of the ionization yield of low energy nuclear recoils in LXe, as there was no prior experimental data. The ionization yield, defined as the number of observed electrons per unit recoil energy ($\rm e^-/keV_r$), was measured for the first time by \cite{Aprile:2006} as a function of electric field and recoil energy. It is shown in Figure~\ref{fig:NR}.

\begin{figure}
\includegraphics[width=.5\textwidth]{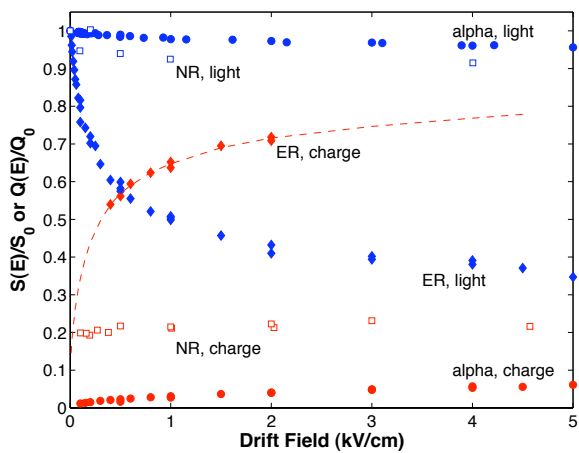}
\caption{\label{fig:Yields} Field dependence of scintillation and ionization yield
in LXe for 122 keV electron recoils (ER), 56.5 keVr nuclear
recoils (NR) and 5.5 MeV alphas, relative to the yield with no drift field \cite{Aprile:2006}.}
\end{figure}

\begin{figure}
\includegraphics[width=1\columnwidth,clip,trim=20 0 20 0]{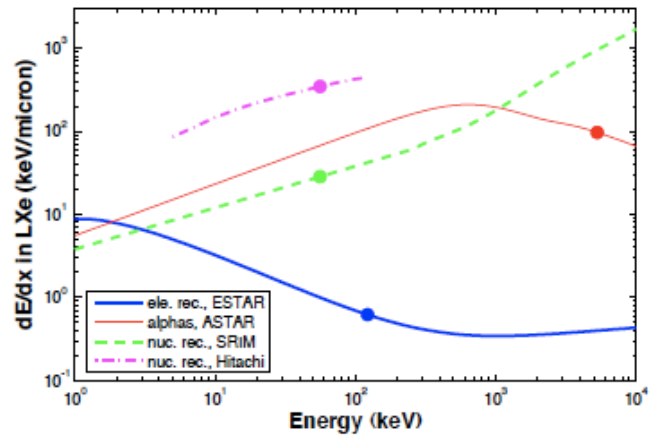}
\caption{\label{xetest:dedx} Predicted electronic stopping power, dE/dx, for different particles in LXe, based on various references. The circles refer to the particle energies discussed in \cite{Aprile:2006}.}
\end{figure}

It was expected that the stronger recombination rate along the dense track of low energy Xe ions would result in a much smaller number of electron-ion pairs, compared to that produced by electron-type recoils of the same energy. It was not expected, however, that the number of carriers would increase with decreasing energy. Also, it was not expected that the ionization yield would be largely unaffected by the applied electric field. Figure~\ref{fig:Yields}, also from \cite{Aprile:2006}, shows the field dependence of both the ionization and the scintillation yields of 56.5 keV$_r$ nuclear recoils, as well as for electron recoils (122 keV gamma-rays from $^{57}$Co), and alpha recoils (5.5 MeV from $^{241}$Am). The observed field dependence may be explained by the different rate of recombination, which depends on the electric field but also on the ionization density along the particle's track, with stronger recombination at low fields and in denser tracks. Simulations of low energy nuclear recoils in LXe show that most energy is lost to a large number of secondary branches, each having substantially lower energy than the initial recoil. The recombination in the very sparse ends of the many secondary branches is strongly reduced at all fields. This situation is quite different from that of an alpha particle. A rough measure of the ionization density is the electronic stopping power, shown in Fig. \ref{xetest:dedx} \cite{Aprile:2006}, for alphas, electrons, and Xe nuclei, respectively. Also shown is a recent calculation by Hitachi \cite{Hitachi:2005} of the total energy lost to electronic excitation per path-length for Xe nuclei, which differs from the electronic stopping power in that it includes energy lost via electronic stopping of secondary recoils. At very low energies, the ionization yield appears to be high both for very slow recoil Xe atoms and for electrons. This seems to be confirmed by the observation of very low energy electron recoils with two-phase Xe detectors such as XENON10 \cite{Angle:2008}. 

\subsubsection{Fano-limit of Energy Resolution} 
In 1947, Fano \cite{Fano:1947} demonstrated that the standard deviation, $\delta$, in the fluctuation of
electron-ion pairs produced by an ionizing particle when all its energy is absorbed in a stopping material, is
not given by Poisson statistics, but by the following formula:
\begin{equation}
	\delta^2 = \langle{\left({N-N_i}\right)^2}\rangle = F \times N_i
\end{equation}
where $F$ is a constant less than 1, known as the Fano factor, and depends on the stopping material. When
$F=1$, the distribution is Poisson-like. The calculation of the Fano factor for LXe and other liquid rare
gases was carried out by Doke \cite{Doke:1981}, in the optical approximation. With the
known Fano factor and $W$-value, the ultimate energy resolution of a LXe detector is given by:
\begin{equation}
	\Delta E\!\left({\mathrm{keV}}\right) = 2.35\sqrt{F\,W\!\left({\mathrm{eV}}\right) E\!\left({\mathrm{MeV}}\right)}
\label{int_res}
\end{equation}
where $\Delta E$ is the energy resolution, expressed as full width at half maximum (FWHM: keV), and $E$ is
the energy of the ionizing radiation, in MeV. This energy resolution is often called the Fano-limit of
the energy resolution. Table \ref{FanoFactor} shows 
$F$ and $FW$ 
for electrons or gamma-rays in LAr, LKr and LXe \citep{Alkhazov:1972,Policarpo:1974, de Lima:1982,Doke:1976}.

\begin{table}[t]
\caption{Calculated Fano factor $F$ and $FW$
in gaseous state and liquid states. a(Alkhazov,1972); b(Policaropo,1974); c(de Lima,1982); d(Doke,1976)}
\label{FanoFactor}
\center
\begin{tabular}{|l|l|l|l|}
\hline
Material & Ar & Kr & Xe \\
\hline
$\bf{Gas}$ & & & \\
\ $F$ & 0.16$^{\rm a}$ & 0.17$^{\rm b}$ & 0.15$^{\rm c}$ \\
\ $FW$(eV) & 4.22 & 4.11 & 3.30 \\
$\bf{Liquid}$ & & & \\
\ $F$ & 0.116$^{\rm d}$ & 0.070$^{\rm d}$ & 0.059$^{\rm d}$ \\
\ $FW$(eV) & 2.74 & 1.29 & 0.92 \\
\hline
\end{tabular}
\end{table}

The Fano-limit of the resolution of LXe, which is comparable to that measured with a Ge- or Si-detector
\citep{Doke:1969,Owen:2002}, has however not yet been achieved. In fact, the best energy resolution measured with a LXe ionization chamber is even worse than the Poisson limit, with the value of 30 keV for $^{207}$Bi 554 keV gamma-rays, measured at the highest field of 17 kV/cm \citep{Takahashi:1975,Doke:1981}.  
Recently,  a similar value was measured \citep{Aprile:2006} at a field of 1kV/cm, using the summed
signals of ionization and scintillation measured simultaneously. 

\subsubsection{Experimental Energy Resolution}

\begin{center}
\begin{figure} 
\leavevmode 
\includegraphics*[width=1\columnwidth,clip,trim=0 0 0 0]{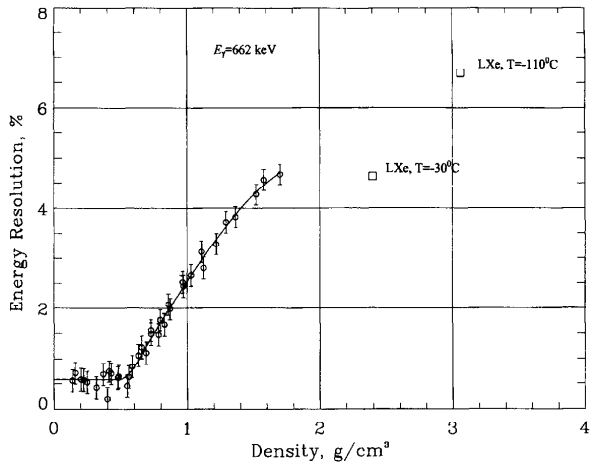} 
\caption{Density dependence of the energy resolution (\%FWHM) measured for 662 keV
	$\gamma$-rays \cite{Bolotnikov:1997}.} 
\label{fig:densitydep} 
\end{figure}
\end{center}

Large fluctuations in the number of such $\delta$-rays are considered to be the main contribution to the spread in energy resolution, as proposed by Egorov et al. in 1982 \citep{Egorov:1982}. Thomas et al. \citep{Thomas:1988} attempted to explain the experimental energy resolution measured from the ionization signal with a recombination model between electrons and ions produced along $\delta$-ray tracks. 
Aprile et al. \citep{Aprile:1991} also attempted to explain their experimentally obtained energy resolution on
the basis of the recombination model (see Figure~\ref{fig:electrons}), while  Obodovsky et al. presented the concept of dual $W$-value in LXe, one for
high energy electrons and one for X-rays (or electrons) with lower energy \citep{Obodovsky:2003}. 
The conclusion that recombination is the primary reason for the poor energy
resolution measured with pure LXe contradicts, however, the measured improvement of the energy resolution in LXe doped with photo-ionizing molecules,
as discussed in \citep{Shibamura:1995}. 
The energy resolution of compressed Xe gas measured by Bolotnikov et al. \cite{Bolotnikov:1997}
also does not support the recombination model, since the resolution improves at lower gas density without an
increase in collected charge. Figure \ref{fig:densitydep}
shows the density dependence of the energy resolution (\% FWHM) measured for 662 keV $\gamma$-rays. The resolution improves when reducing the density from 1.4 g/cm$^3$ to 0.5 g/cm$^3$, where it is close to the Fano limit. 

To date, the reasons for the discrepancy between the experimental and theoretical energy resolution of LXe and
other liquid rare gases remain unclear, pointing to the need for more data to reach a complete understanding
of the ionization process in liquid rare gases.

\subsubsection{Drift Velocity of Electrons and Ions}

\begin{center}
\begin{figure} 
\includegraphics*[width=1\columnwidth,clip,trim=0 0 40 0]{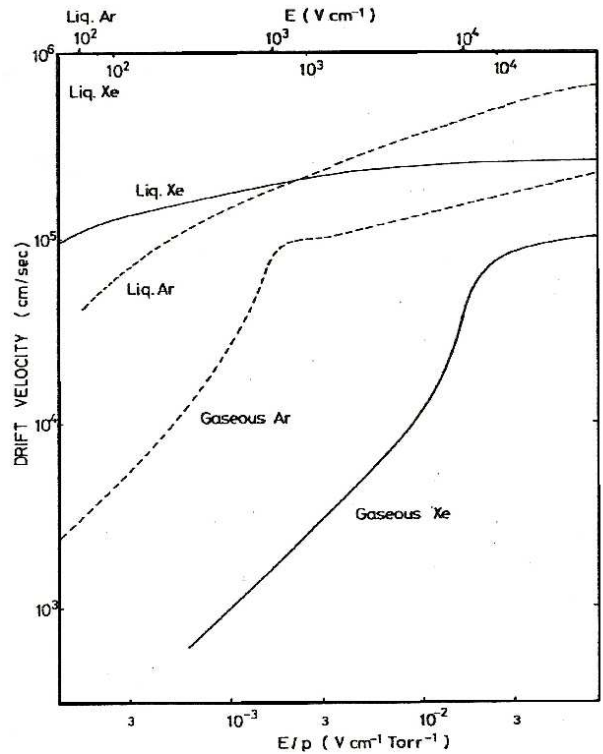} 
\caption{Electron drift velocity in gaseous and liquid xenon and argon, as a function of
	reduced electric field \citep{Pack:1962,Miller:1968,Yoshino:1976}.} 
\label{fig:driftv} 
\end{figure}
\end{center}

LXe has a distinct band structure which consists of a conduction band and a valence band.  Electrons
are excited to the conduction band from the valence band by high energy radiation and become free electrons.
As a result, holes are formed in the top of the valence band. The motion of these carriers under an external electric field has been studied as a function of field strength, concentration of impurities in the liquid,  and liquid temperature by several authors \citep{Pack:1962,Miller:1968,Yoshino:1976, Gushchin:1982}.
Figure
\ref{fig:driftv} \citep{Pack:1962,Miller:1968,Yoshino:1976} shows the dependence of the electrons drift velocity on the
applied electric field in liquid and gaseous argon and xenon.

At low fields, the electron drift velocity, $v_d$, is almost proportional
to the field strength, E, with the electron mobility, $\mu$, as the proportionality constant ( $v_d = \mu E$). The electron
mobility in LXe is about 2000 cm$^2$ V$^{-1}$s$^{-1}$, which is near the mobility of electrons in silicon. At high fields, the  electron drift velocity saturates (becomes independent of electric field strength). Figure \ref{fig:driftv} also shows that the drift velocity of electrons in
LAr and LXe is much higher than the velocity in the corresponding gas phase.  Specifically, the difference is more
marked in LXe.

\begin{center}
\begin{figure} 
\includegraphics*[width=.4\textwidth]{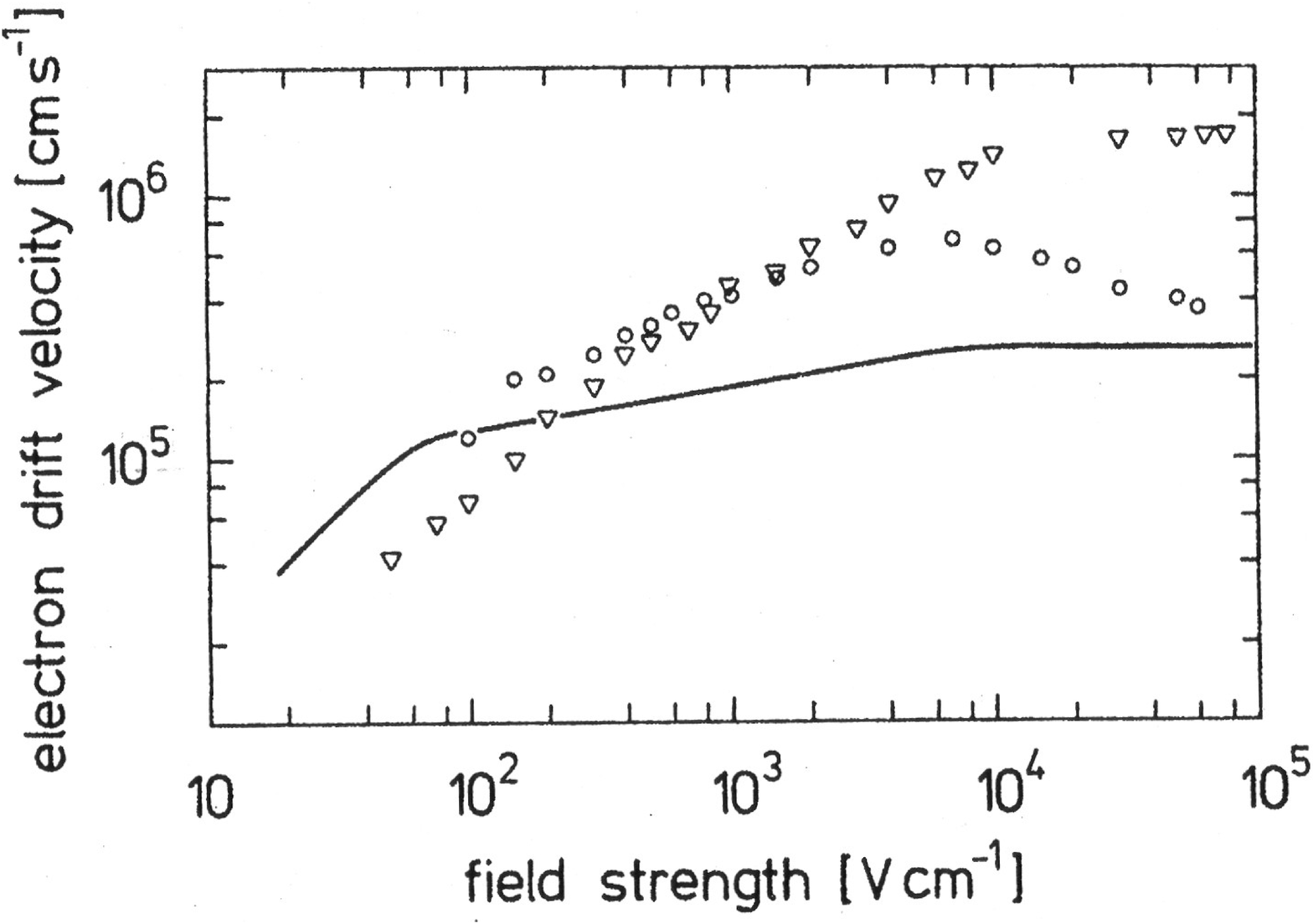} 
\caption{Influence of butane on the electron drift velocity in liquid xenon. Butane ($\circ$: 0.09\%),
	($\triangle$: 1.4\%), and solid line: pure xenon \citep{Yoshino:1976}.} 
\label{fig:driftvbut} 
\end{figure}
\end{center}

The drift velocity of electrons in LXe increases by doping organic materials into the pure liquid, as
it is also found in LAr \citep{Shibamura:1975}. Collisions with the organic molecules cause this effect by reducing
the average excitation energy of the electrons. Figure \ref{fig:driftvbut} shows one example involving LXe
doped with butane \citep{Yoshino:1976}. However, pure LXe is preferred as a detector medium because its purification is less difficult than the purification of organic molecules. 

The electron drift velocity in LXe
depends slightly on the temperature of the liquid and is almost inversely proportional to the liquid temperature with a rate of about 0.5\% per $^\circ$C
\citep{Masuda:1981a}. Aprile et al. \citep{Aprile:1991_300} measured the electron velocity at the
relatively high temperature of 195 K, from which a zero-field mobility of $4230\pm400$ cm$^2$V$^{-1}$s$^{-1}$
was obtained. 

\begin{center}
\begin{figure} 
\includegraphics*[width=1\columnwidth,clip,trim=0 0 20 0]{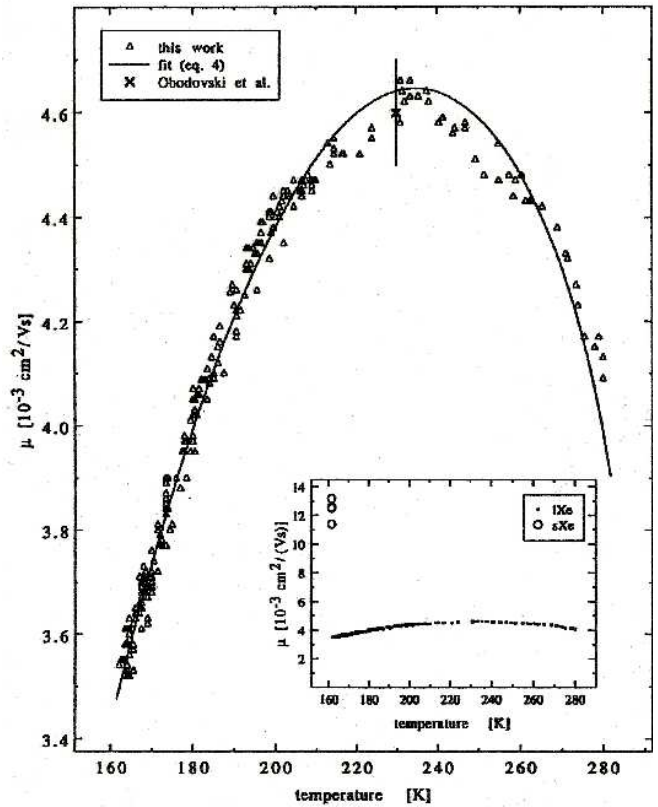} 
\caption{Mobility of positive holes in liquid xenon as a function of temperature \citep{Hilt:1994}.} 
\label{fig:mobility} 
\end{figure}
\end{center}

In LXe, positive carriers are holes, with a mobility that is several orders of magnitude less than the afore-mentioned electron mobility. Its temperature dependence is shown in Figure \ref{fig:mobility} \citep{Hilt:1994}.

\subsubsection{Electron Diffusion}

\begin{center}
\begin{figure} 
\includegraphics*[width=.9\columnwidth,clip,trim=0 0 0 0]{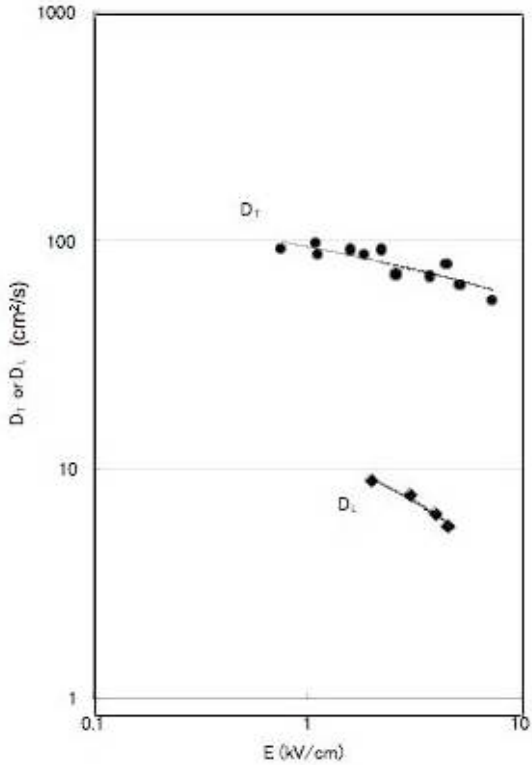} 
\caption{Transverse and longitudinal diffusion coefficients for electrons in LXe as a function of
	electric field \citep{Doke:1981,Shibamura:2009}.} 
\label{fig:diffusion} 
\end{figure}
\end{center}

The spread of the electron cloud due to drift in an electric field is determined by the diffusion rate. This
spread determines the intrinsic limit to the position resolution of a LXe detector operated in the Time
Projection Mode, discussed in \ref{ss:TPC:mode}. The electron diffusion coefficient depends on the direction of the
electric field. Specifically, the diffusion coefficient, $D_L$, in the electric field direction, is different from that in the direction transverse to the drift field, $D_T$. The former is
much smaller than the latter. The experimental results on these diffusion coefficients for LXe are shown in
Figure \ref{fig:diffusion} \citep{Doke:1981,Shibamura:2009}. 

The longitudinal diffusion coefficient  is about 1/10 of the transverse diffusion coefficient. This is
theoretically expected \citep{Robson:1972}. The transverse spread of an electron cloud drifting over a distance $d$ is given by 

\begin{equation}
 \sigma_{D_T} = \sqrt{D_T t_d} 
\end{equation}
 where $t_d$=$d/v_d$ is the electron drift time at the drift velocity ${v_d}$. This spread contributes a negligible amount to the position resolution of a LXe detector, even over a long distance.

\subsubsection{Electron Multiplication and Proportional Scintillation}

In LXe, under a sufficiently high electric field (higher than $10^6$ V/cm) a phenomenon known as
electron-multiplication (or avalanche) occurs. 

\begin{center}
\begin{figure} 
\includegraphics*[width=.4\textwidth]{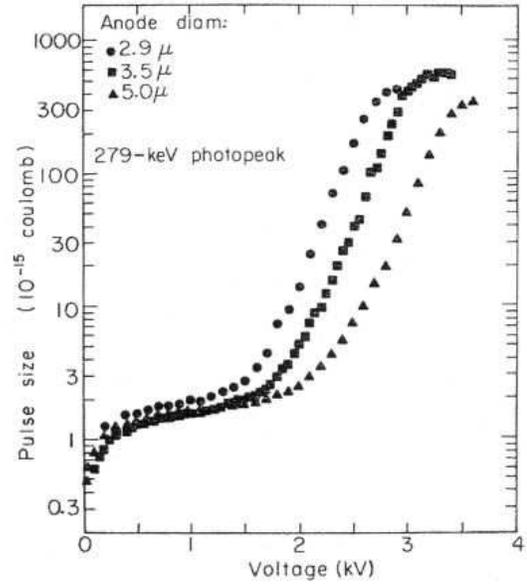} 
\caption{Pulse height of 279 keV photopeak as a function of voltage for 2.9-, 3.5-,and 5.0-$\mu$m diameter anode wires \citep{Derenzo:1974}.} 
\label{fig:avalanche} 
\end{figure}
\end{center}

The first observation of this process
was made by the Berkeley's group \citep{Muller:1971}; this group later quantitatively measured the
electron avalanche \citep{Derenzo:1974}. Their results were confirmed by the Grenoble and Waseda groups
\citep{Prunier:1973, Miyajima:1976}. Figure \ref{fig:avalanche} shows the variation in the gain of
multiplication as a function of applied voltage to an anode wire of 2.9, 3.5 or 5.0 $\mu$m in diameter in a
cylindrical counter \citep{Derenzo:1974}. The maximum gain measured in these cylindrical counters is a factor of
several hundred and the energy resolution deteriorates with an increase in gain. 

\begin{center}
\begin{figure} 
\includegraphics*[width=.9\columnwidth]{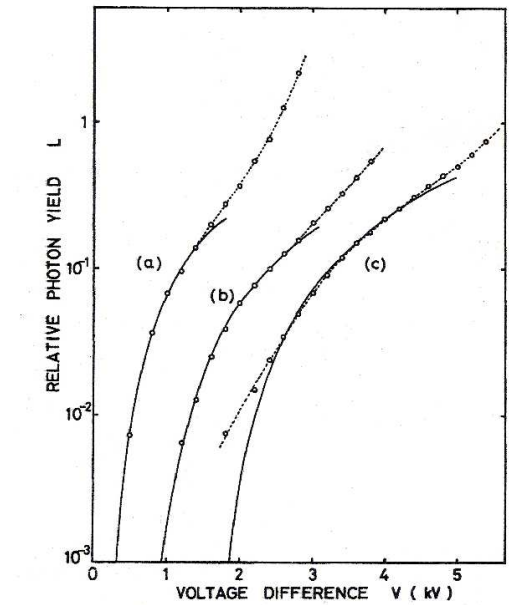} 
\caption{Relative photon yield of proportional scintillation and charge gain as a function of applied voltage to anode wires of different diameter. Solid lines are results of the fit in \citep{Derenzo:1974} to the data points in the region of the ionization chamber mode: (a) for 4 $\mu$m diameter wire, (b) for 10 $\mu$m, (c) for 20 $\mu$m\citep{Masuda:1979}.} 
\label{fig:propscint} 
\end{figure}
\end{center}

Another process occurring in LXe, with an electric field lower than the threshold for electron avalanche, is photon-multiplication, also known as ``proportional scintillation". The process was first observed by
the Saclay group \citep{Lansiart:1976} and systematically investigated by the Waseda group. In Figure \ref{fig:propscint} \citep{Masuda:1979}, the solid lines show the
variation of the relative photon yield as a function of applied voltage to anode wires of different
diameters. The dotted lines show the variation in charge gain as a function of
applied voltage to the anode wires.

The measured energy spectrum does not deteriorate with applied voltage as long as the counter is operated below electron multiplication. The maximum photon
gain for 20 $\mu$m diameter wire at 5 kV is estimated to be about 5 photons per electron \citep{Doke:1982}.

\subsubsection{Electron Attachment to Impurities}
\label{sec:impurity}
For stable observation of ionization signals with high ionization yield, the concentration of electro-negative impurities in LXe has to be below the 1 part per billion (ppb) level of  O$_2$ equivalent substances. The reaction of an electron with an impurity $S$ leads to the formation of a negative ion,
\begin{equation}
e + S \rightarrow S^{-}
\end{equation}
 and the decrease of the electron concentration [$e$] is given as  
\begin{equation} 
d[e]/dt = -k_S [e][S] 
\end{equation}
where $S$ is the concentration of impurity given in mol/$l$ and $k_S$ is the attachment rate constant in $l$/(mol s). The temporal variation of the electron concentration [$e$] is then given as 
\begin{equation}
[e(t)] = e(0) \; \mathrm{exp} (-k_S [S] t)
\end{equation}
 and the time 
 \begin{equation}
 \tau = (k_S [S])^{-1}
 \end{equation}
is called the electron lifetime. Usually, in the detection of impurities in a sample of LXe, both $k_S$ and [$S$] are unknown. One measures an exponential decay in time of the electron concentration, through the measurement of the direct current induced by the drift of the electrons or its integrated value of collected charge. Attachment leads to a decrease of the current with time, or to a reduction of the collected charge. Instead of quoting $\tau$, one often quotes the attenuation length of the electrons,
\begin{equation}
\lambda_{att} = {\mu}E{\tau}
\end{equation}
where $\mu$ is the electron mobility and \textit{E} is the field. 

\begin{center}
\begin{figure} 
\includegraphics*[width=.4\textwidth]{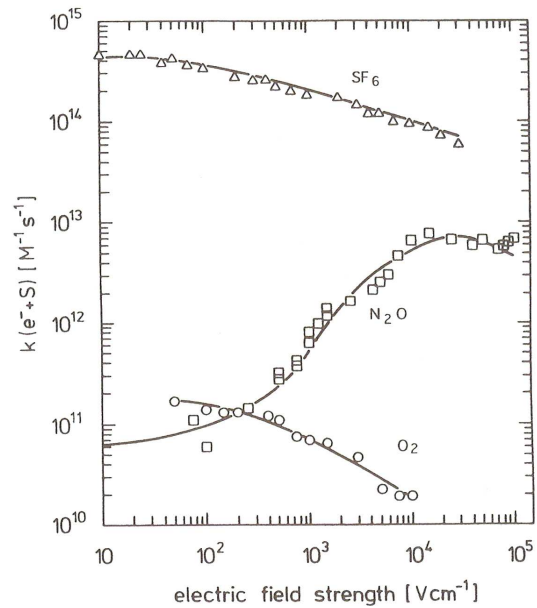} 
\caption{Rate constant for the attachment of electrons in liquid xenon(T=167$^\circ$K) to several solutes: ($
	\triangle $) SF$_6$, ($ \square$) N$_2O$, ($\circ$) O$_2$\citep{Bakale:1976}.} 
\label{fig:attachment} 
\end{figure}
\end{center}

Two types of electro-negative impurities are normally
found in LXe: those with an attachment cross-section that decreases with increasing electric field
and those with a cross-section that increases with increasing field. Figure \ref{fig:attachment} shows the
attachment cross-section as a function of applied electric field for some electro-negative gases,
SF$_6$, N$_2$O, and O$_2$ \citep{Bakale:1976}. Here, SF$_6$ and O$_2$ correspond to the first type of
impurity and N$_2$O to the second.

A concentration of 1 ppb (oxygen equivalent) corresponds to an attenuation length of 1 meter. 
Several methods have been developed for the removal of impurities, and the choice of the most appropriate method depends not only on the volume of liquid to be purified, but also on the application. They include adsorption and chemical reaction methods, filtration, separation, electrical discharges and irradiation with gamma-rays. Outgassing from the liquid containment vessel and other detector's materials contribute to the total concentration of electron-attaching impurities in the liquid, through a diffusion process. The outgassing rates and the diffusion time scale depend on the type of material and on the surface finish. 
For LXe a purity at the ppb level has been achieved by using hot metal getters and/or cold molecular sieves. For details of purification
methods and different purification systems used to date with LXe detectors we refer to \citep{Prunier:1973,Masuda:1981b,Aprile:1991_300,Barabash:1993,Ichige:1993, Carugno:1993, Angle:2008}.

\subsubsection{Photo-ionization}

Some organic molecules have a high photo-sensitivity. If the dopant molecules have an ionization potential lower than the energy of the scintillation photons emitted from LXe, these photons will ionize the molecules, producing additional charge carrier pairs. This idea of increasing the ionization yield of liquid rare gases with photo-sensitive dopants, was first proposed by \cite{Policarpo:1982} and first observed by Anderson in LAr \cite{Anderson:1986}. For LXe, results on the increased ionization yield with appropriate photo-sensitive molecules, were first reported by Suzuki et al. \citep{Suzuki:1986}.

\begin{center}
\begin{figure} 
\includegraphics*[width=1\columnwidth,clip,trim=20 0 0 0]{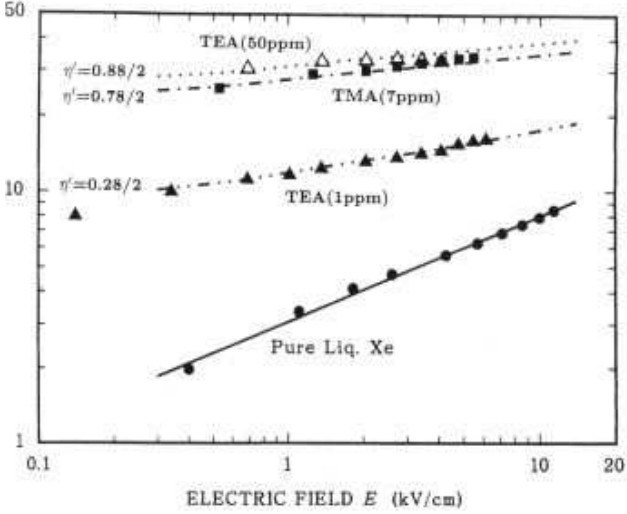} 
\caption{Collected charge Q/N$_i$, where N$_i$ is the total ionization (number of electrons) for P$_o$
	$\alpha$-particles as a function of applied electric field. The curves represent pure liquid xenon
	($\bullet$) and liquid xenon doped with TMA ($\blacksquare$ : 7 ppm) and TEA ($\blacktriangle$ : 1 ppm ;
	$\triangle$: 50 ppm). The curves give the calculated results from eq. (3) in \citep{Bakale:1976}. Figure from  \citep{Hitachi:1997}.} 
\label{fig:collcharge} 
\end{figure}
\end{center}

 Figure \ref{fig:collcharge} shows the 
collected charge for alpha-particles as a function of electric field in liquid xenon doped with TEA
(triethylamine) or TMA (tri-methylamine) \citep{Hitachi:1997}. 

As already shown in Figure~\ref{fig:alphas}, the collected charge in pure LXe for alpha particles is only 1\% to 10\% of the total charge estimated from the
$W$-value, over the applied electric field of 1 to 10 kV/cm, reflecting the strong electron-ion recombination rate.  However, doping liquid xenon with 1$\sim$50 ppm of TEA, recovers part of the charge lost in the recombination process, and the collected charge increases to almost 40\% of the total, at the maximum applied field.  On the basis of this data, the quantum efficiency of photo-ionization of TEA in LXe is estimated to be 80\%. 

\begin{center}
\begin{figure}
\includegraphics*[width=1\columnwidth,clip,trim=10 280 0 300]{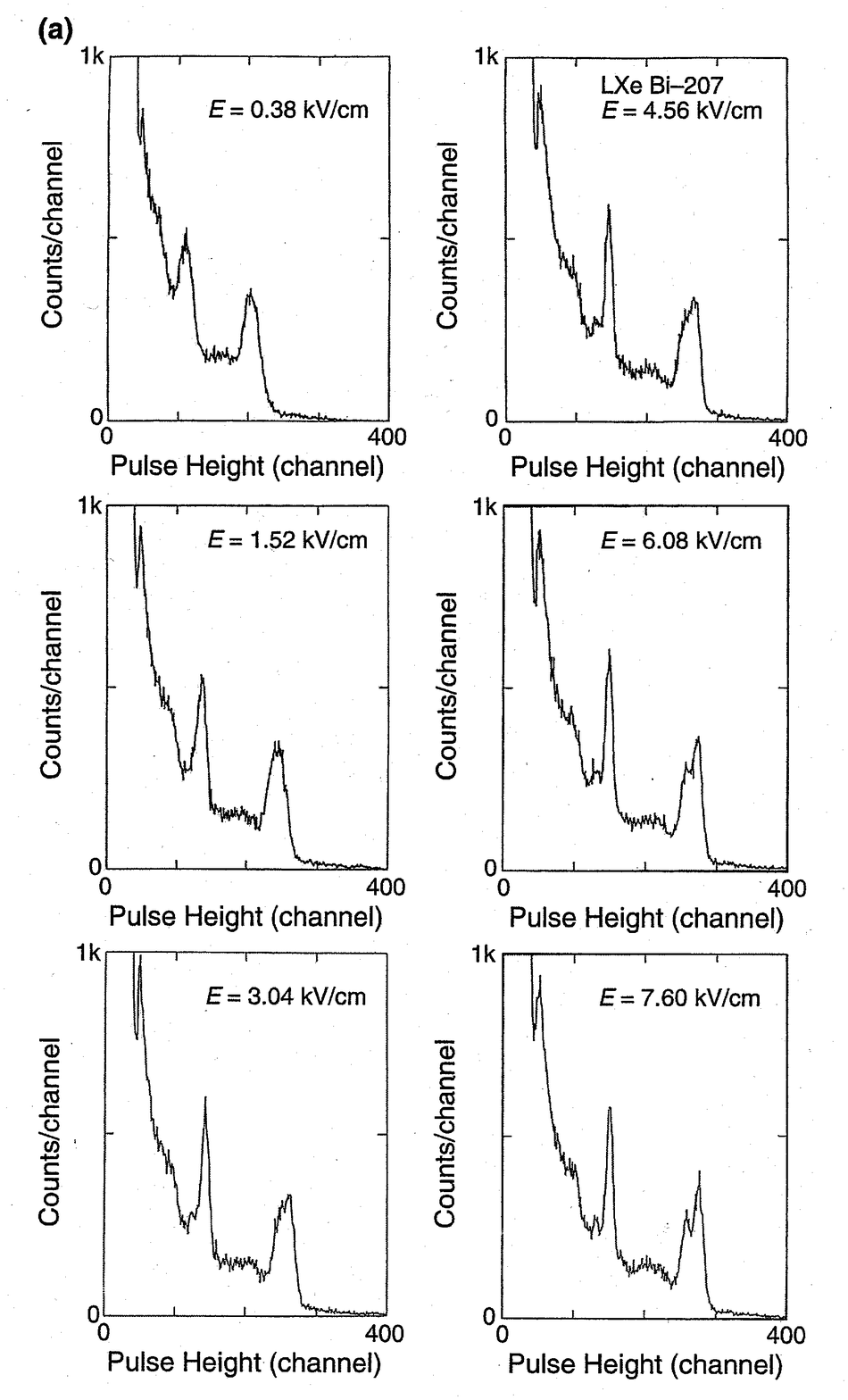}
\includegraphics*[width=1\columnwidth,clip,trim=70 270 20 320,angle=1]{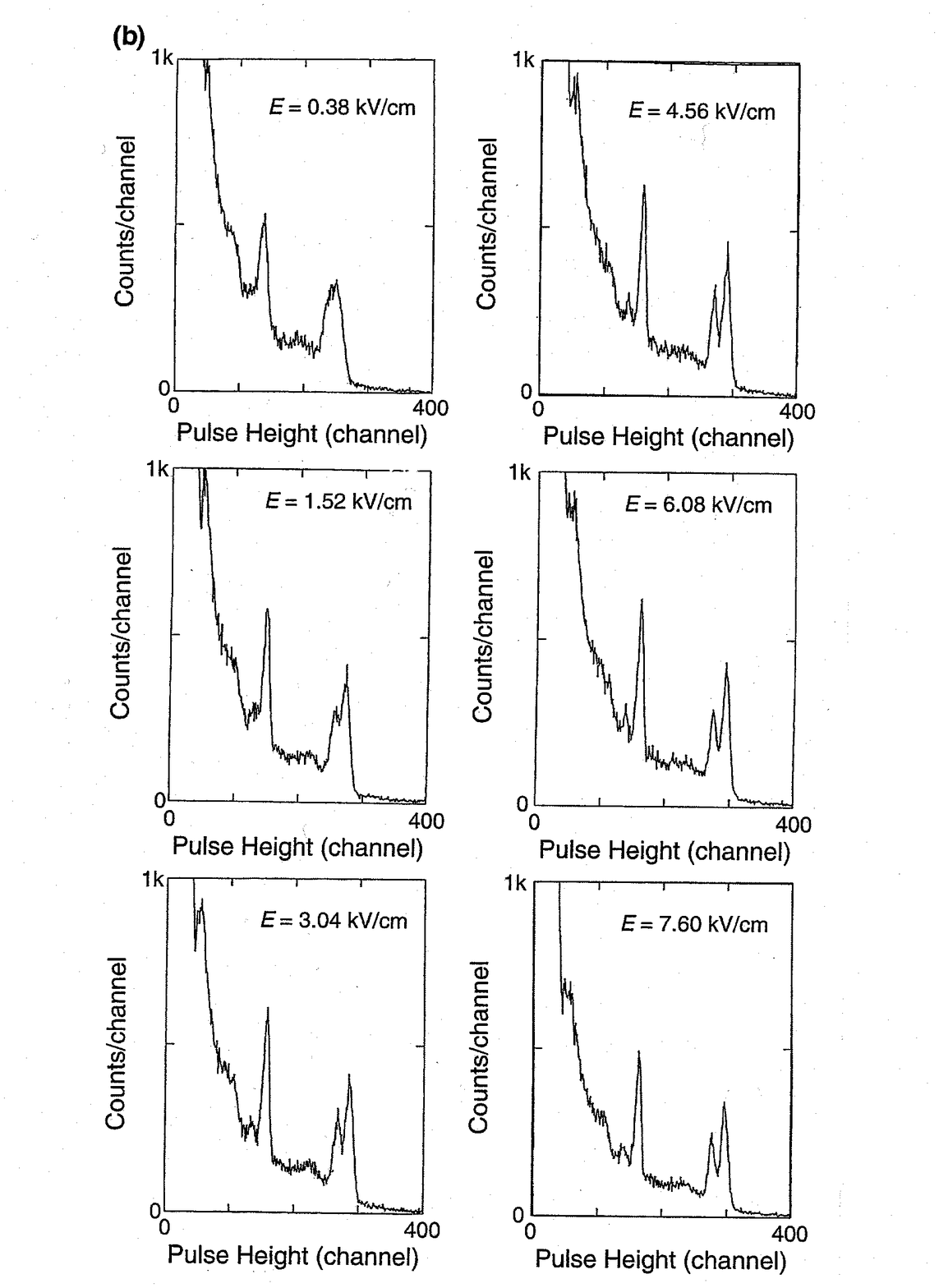}
\caption{Comparison between the energy spectra of $^{207}$Bi  internal conversion electrons and gamma-rays measured  from ionization: (top) in pure liquid xenon and (bottom) in TEA-doped liquid xenon.}
\label{fig:Fig28}
\end{figure}
\end{center}

The photo-ionization effect was used to improve the energy resolution of a LXe ionization chamber irradiated with gamma-rays \citep{Ichinose:1992}, as illustrated in Figure \ref{fig:Fig28}. The separation of the internal conversion electron peak (976 keV) and the gamma-ray peak (1047 keV) can be seen to be much better in the TEA doped LXe when compared to pure LXe. An electric field higher than 10kV/cm is necessary to realize the separation of the two peaks in pure LXe.

In TEA-doped LXe,
photon-mediated electron multiplication also occurs, at the electric field where proportional scintillation
takes place \citep{Sano:1989}. The maximum gain of photo-mediated multiplication is about 200, which is almost
equal to that of electron multiplication but with poorer energy resolution.

\subsection{Scintillation Process}
\begin{center}
\begin{figure}
\includegraphics*[width=0.42\textwidth]{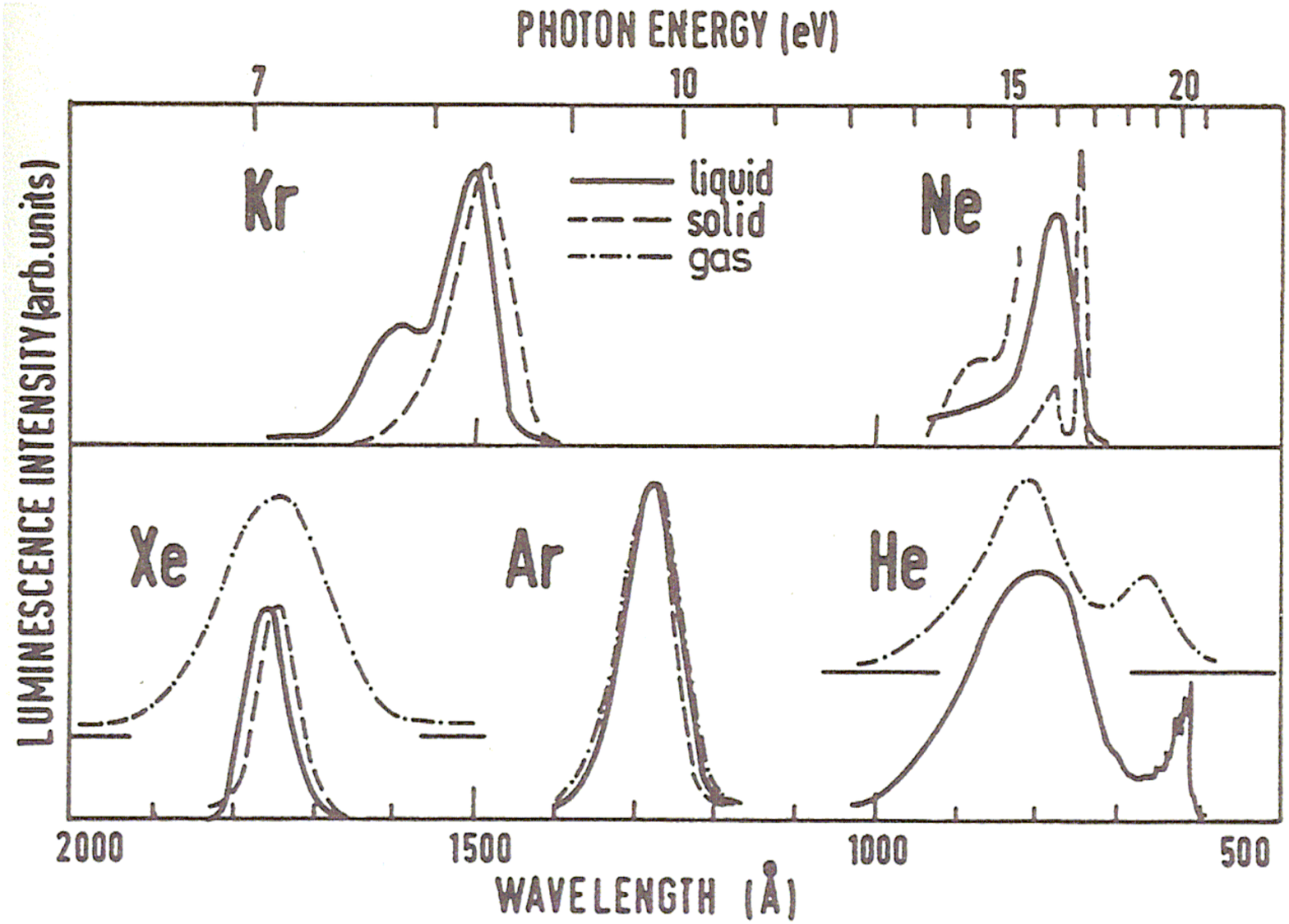}
\caption{Emission bands in liquid rare gases, together with
solid- and gas-phase spectra \citep{Schwenter:1985, Jortner:1965}.}
\label{fig:light}
\end{figure}
\end{center}

The emission of luminescence, also called scintillation, from liquid rare gases is attributed to the decay of excited dimers (excimers, in short) to the ground state.
The luminescence emission bands for Ar, Kr, Ne and He in liquid, solid and gas- phase are shown in Figure \ref{fig:light} \citep{Schwenter:1985,Jortner:1965}. We note that the emission bands in the three phases of Xe, Ar and Kr are almost identical. In contrast, the emission band for liquid Ne differs dramatically from that of solid Ne.
For LXe, the wavelength of the scintillation photons is centered at 177.6 nm.

\subsubsection {Origin of Scintillation}
Both direct excitation of atoms and electron-ion recombination lead to the formation of excited dimers, $\rm Xe_{2}^{\ast}$. Thus, the origin of the vacuum ultraviolet (VUV) scintillation light in LXe is attributed to two separate processes involving excited atoms
 (Xe$^*$) and ions (Xe$^+$), both produced by ionizing
 radiation \cite{Kubota:1978}:\\
\begin{eqnarray}
 \rm{Xe^{\ast}+ Xe+ Xe}  &\rightarrow& \rm{Xe_{2}^{\ast}}+ Xe \nonumber,\\
 \rm{ Xe_{2}^{\ast}} &\rightarrow& \rm{2Xe} + \mathit{h\nu}
 \label{eq:excite}
\end{eqnarray}
\begin{eqnarray}
 \rm{ Xe^{+}+Xe}     &\rightarrow& \rm{Xe_{2}^{+}} \nonumber,\\
 \rm{ Xe_{2}^{+}+e^-}&\rightarrow& \rm{Xe^{\ast \ast} + Xe}
\nonumber\\
 \rm{ Xe^{\ast \ast}}&\rightarrow& \rm{ Xe^{\ast} + heat}
 \nonumber\\
 \rm{ Xe^{\ast}+Xe+ Xe}  &\rightarrow& \rm{ Xe_{2}^{\ast}+ Xe}, \nonumber\\
 \rm{Xe_{2}^{\ast}}  &\rightarrow& \rm{2Xe} + \mathit{h\nu}
 \label{eq:recom}
\end{eqnarray}

\subsubsection{Scintillation Pulse Shape}
\begin{center}
\begin{figure} 
\includegraphics*[width=1\columnwidth,clip,trim=0 0 20 0]{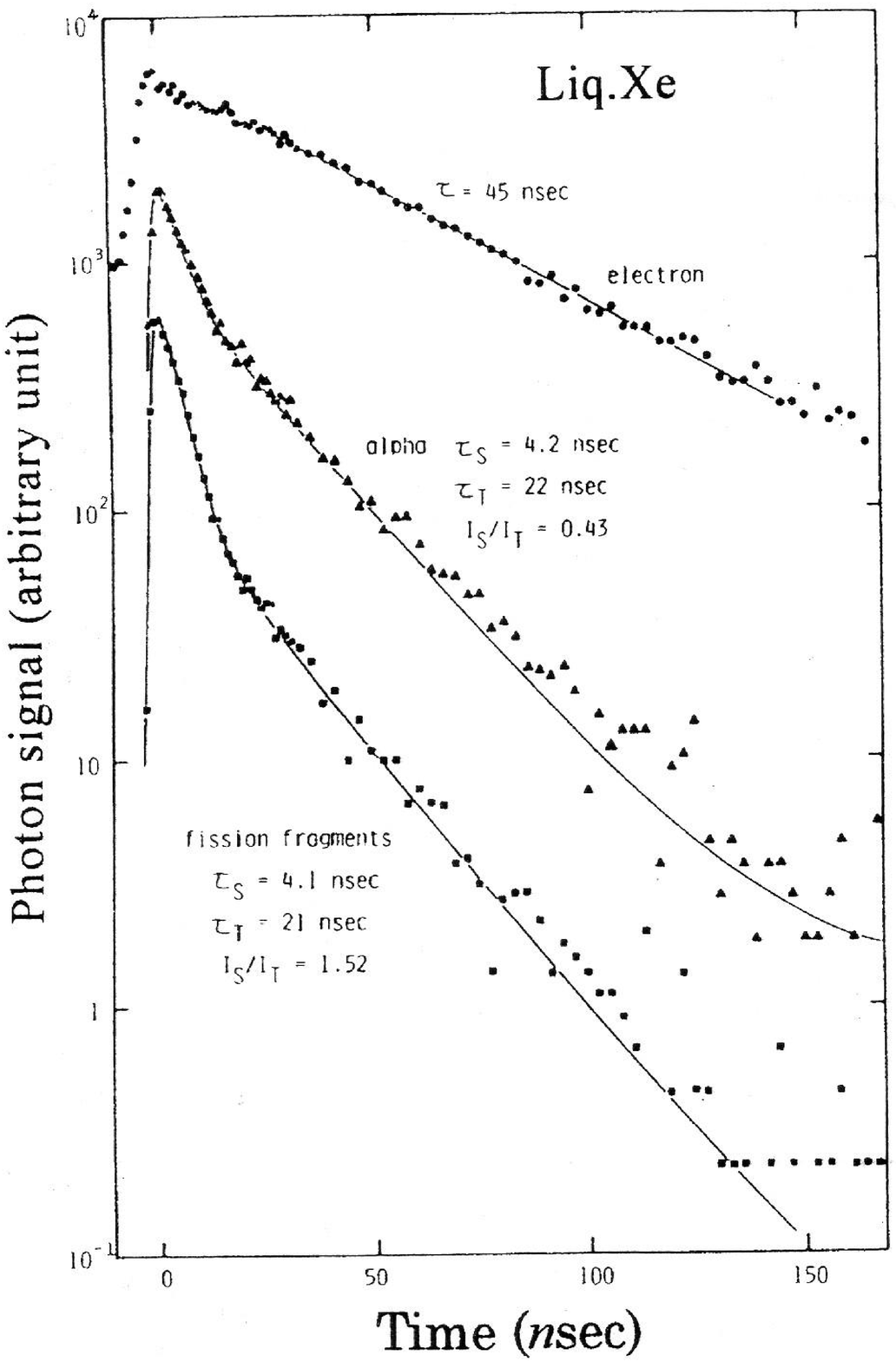} 
\caption{Decay curves of scintillation from liquid xenon excited by electrons, $\alpha$-particles and
	fission fragments, without an applied electric field \citep{Hitachi:1983,Kubota:1978}.} 
\label{fig:decaynofield} 
\end{figure}
\end{center}

The scintillation light from pure LXe  has two decay components due to
de-excitation of singlet and triplet states of the excited dimer Xe$_{2}^{\ast}$. Figure \ref{fig:decaynofield} \citep{Hitachi:1983}  shows the decay curves of
the scintillation light for electrons, alpha-particles and fission fragments in LXe, without an electric field. The
decay shapes for $\alpha$-particles and fission fragments have two components. The shorter decay shape is
produced by the de-excitation of singlet states and the longer one from the de-excitation of triplet states.
Specifically, the short and long decay times are 4.2 and 22 ns for alpha-particles. For fission fragments, the values are 4.1 and 21 ns, respectively. These decay times make LXe the fastest of all liquid rare gas scintillators. While the singlet and triplet lifetimes depend only weakly on the density of excited species, the intensity ratio of
singlet to triplet states is larger at higher deposited energy density.

\begin{center}
\begin{figure} 
\includegraphics*[width=1\columnwidth,clip,trim=0 0 0 0]{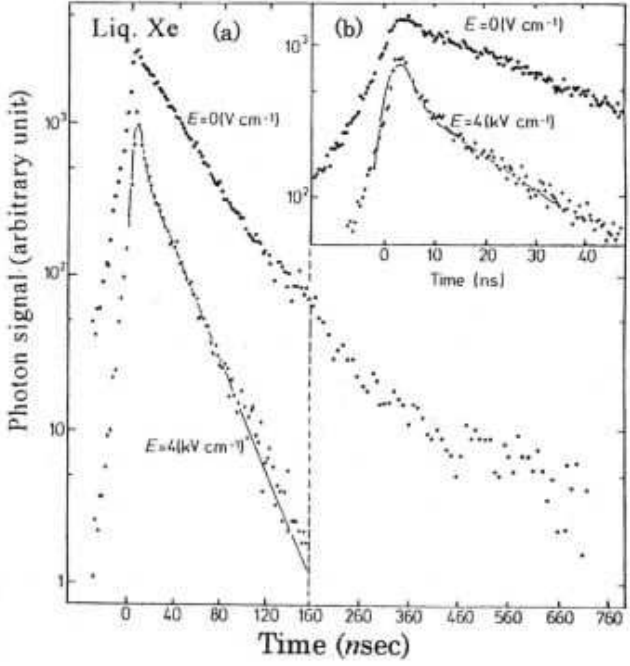} 
\caption{Decay curves of scintillation from liquid xenon with and without an electric
	field of 4kV/cm, over a long time scale (a) and a short time scale (b). Note the change of time scale at 160
	ns in (a) \citep{Kubota:1978}.} 
\label{fig:decaywithfield} 
\end{figure}
\end{center}

For relativistic electrons, the scintillation has only one decay component, with a decay time of 45 ns \cite{Kubota:1979,Hitachi:1983}. Since this component disappears with an applied electric field, it is likely due to the slow recombination between electrons and ions produced by relativistic electrons. Figure \ref{fig:decaywithfield} \citep{Kubota:1978} shows the decay curves of LXe scintillation light, with an electric field of 4 kV/cm, with two distinct decay components. From this
figure, it is estimated that the short decay time for relativistic electrons is 2.2$\pm$0.3 ns and the long
decay time is 27$\pm$1 ns.  
The difference in the scintillation pulse decay shape for different types of particle in liquid rare gases can be used to effectively discriminate these particles. Pulse shape discrimination (PSD) is however difficult for LXe given the small time separation of the two decay components. On the other hand, PSD is very effective for LAr, given the larger time separation of the two components with values of 5.0 ns and 1590 ns, respectively \citep{Hitachi:1983, Lippincott:2008}. 

\subsubsection{Scintillation Yield}

If $E$ is the energy deposited by the ionizing radiation, the maximum scintillation yield is given as $E/W_{ph}$, where $W_{ph}$-value is the average energy required for the production of
a single photon. Assuming the absence of photon
reduction (or quenching) processes, $W_{ph}$ can be obtained from \ref{eq:traritraraa} as \citep{Doke:2002}
\begin{center}
\begin{equation}
{W_{ph}} = {W}/(1 + {N_{ex}/N_i)}
\label{Eqn1}
\end{equation}
\end{center}
where $W$ is the average energy required for an electron-ion pair production, discussed in {Ionization Process}. $N_{ex}$ and $N_i$ are the numbers of excitons and electron-ion pairs, respectively, produced by the ionizing radiation.

\begin{table}[t]
\caption{Most probable values of $N_{\rm ex}$/$N_{\rm i}$, $W_{\rm ph}$(max), $\beta$/$\alpha$ ratio, $W_{\rm ph}(\alpha)$, and $W_{\rm ph}(\beta)$ in liquid argon and xenon, derived from data in figures \ref{fig:letscintar} and \ref{fig:letscintxe}. $\alpha$ and $\beta$ mean $\alpha$-particle and $\beta$-ray and $\beta$/$\alpha$ is the ratio of scintillation yields for $\beta$-ray and $\alpha$-particle. }
\label{NexNi}
\center
\begin{tabular}{|l|l|l|}
\hline
Item & Liq. Ar & Liq. Xe \\
\hline
\ $N_{\rm ex}$/$N_{\rm i}$ & 0.21 & 0.13 \\
\ $W_{\rm ph}$(max)(eV) & 19.5 $\pm$ 1.0 & 13.8 $\pm$ 0.9 \\
\ $\beta$/$\alpha$ & 1.11 $\pm$ 0.05 & 0.81$^{+0.07}_{-0.13}$ \\
\ $W_{\rm ph}(\alpha)$ (eV) & 27.1 & 17.9 \\
\ $W_{\rm ph}(\beta)$ (eV) & 24.4 & 21.6 \\
\hline
\end{tabular}
\end{table}

\begin{center}
\begin{figure} 
\includegraphics*[width=1\columnwidth,clip,trim=10 0 10 0]{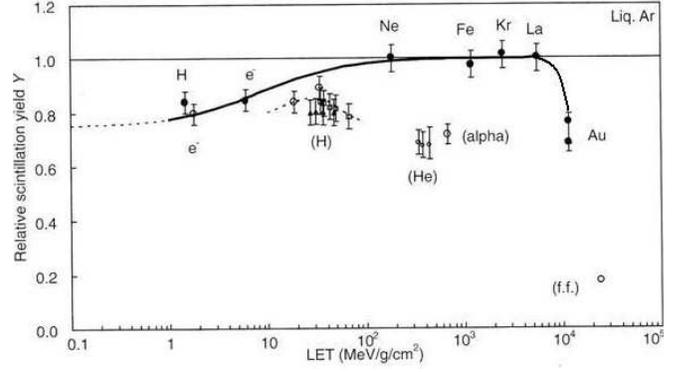} 
\caption{LET dependence of the scintillation yield in liquid argon for various ionizing particles \citep{Doke:1988,Doke:2002}.} 
\label{fig:letscintar} 
\end{figure}
\end{center}

\begin{center}
\begin{figure} 
\includegraphics*[width=1\columnwidth,clip,trim=10 0 10 0]{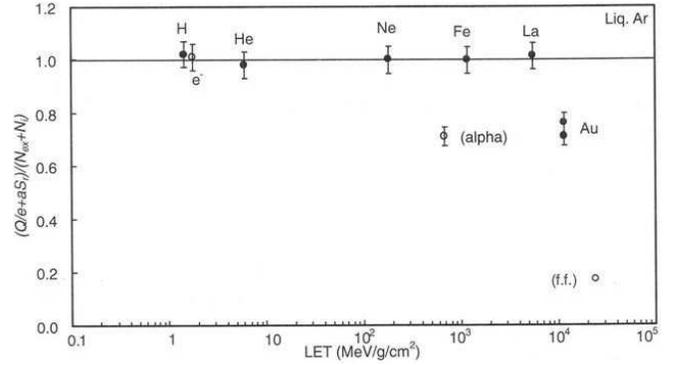} 
\caption{LET dependence of the ratio (Q/e + aS$_r$)/(N$_{ex}$ + N$_i$) in liquid argon. The label (f.f.) stands for fission fragments or non-relativistic particles \citep{Doke:1988,Doke:2002}.} 
\label{fig:letratio} 
\end{figure}
\end{center}

The scintillation yield depends on the linear energy transfer
(LET), that is, the density of electron-ion pairs produced along the track of a particle, because the
recombination probability between electrons and ions increases with the density of electron-ion pairs. 
Figure
\ref{fig:letscintar} \citep{Doke:1988,Doke:2002} shows such an LET dependence of the scintillation yield
in LAr. As seen from the figure, the scintillation yield stays at a maximum value over an extended
region of the LET. In these experiments involving relativistic heavy ions, the scintillation signals were
simultaneously observed with the ionization signals. The sum of ionization and scintillation signals,
properly normalized \citep{Crawford:1987}, and divided by $N_{ex}$ + $N_i$, gives a completely
flat LET dependence, except for alpha-particles, fission fragments, and relativistic Au ions as shown in Figure \ref{fig:letratio} \citep{Doke:1988,Doke:2002}. It should be noted that the flat level in Figure \ref{fig:letscintar} corresponds to the maximum limit of the sum signals. In the lower LET region, the scintillation yield gradually decreases with LET, and in the higher LET region, the scintillation yield decreases as the LET increases. The former behavior is caused by the so-called ``escape electrons", namely, a large number of electrons that do not recombine with parent ions for an extended period of time (on the order of a few ms) in the absence of electric field. The latter behavior is due to the so-called quenching effect. The observed behavior of non-relativistic protons and helium ions is explained by both effects (escape electrons and quenching).  

\begin{center}
\begin{figure} 
\includegraphics*[width=.45\textwidth]{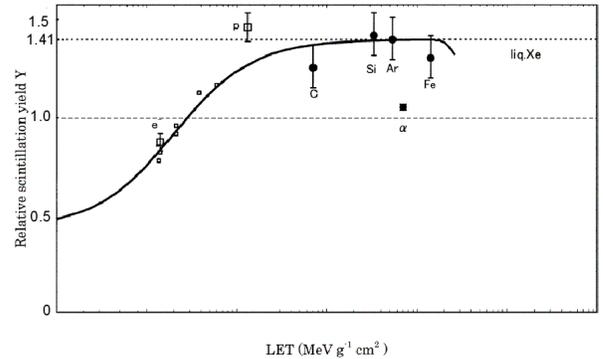} 
\caption{LET dependence of the scintillation yield in liquid xenon for various ionizing particles \citep{Doke:2002,Tanaka:2001}.} 
\label{fig:letscintxe} 
\end{figure}
\end{center}

The LET dependence of scintillation yield in LXe  is shown in Figure \ref{fig:letscintxe}
\citep{Doke:2002,Tanaka:2001}. LAr and LXe display a similar LET dependence. In the case of LXe, however, the
ionization and scintillation signals were not observed simultaneously. Therefore, it is not straightforward  to conclude
on this basis that the flat behavior of the LET in LXe corresponds to the maximum scintillation
yield. However, as discussed in \citep{Doke:2002}, this assumption is consistent with results obtained for LAr and LXe.

On the basis of these experimental results on the maximum
scintillation yield measured over a wide LET region in LXe,  the $W_{ph}$-value was estimated to
be 13.8$\pm$0.9 eV \citep{Doke:2002}. The $W_{ph}$($\beta$)
value for 1 MeV electrons and the $W_{ph}$($\alpha$) for $\alpha$-particles are both lower than the derived maximum $W_{ph}$ because of escape electrons or scintillation quenching. Doke \citep{Doke:2002} estimated  a value of 17.9 eV for $W_{ph}$($\alpha$ ) and a value of 21.6 eV for $W_{ph}$($\beta$). 
$N_{ex}$/$N_i$, $W_{ph}$, $\beta / \alpha$  and
$W_{ph}(\alpha,\beta )$ in LXe are shown in Table \ref{NexNi} \citep{Doke:2002}. The same quantities are given also for LAr.  Compared to all liquid rare gases, LXe has the highest scintillation light yield, similar to that of the best crystal scintillators.  

\subsubsection{Relative Scintillation Efficiency of Nuclear Recoils}
\begin{figure}
\includegraphics[width=1\columnwidth]{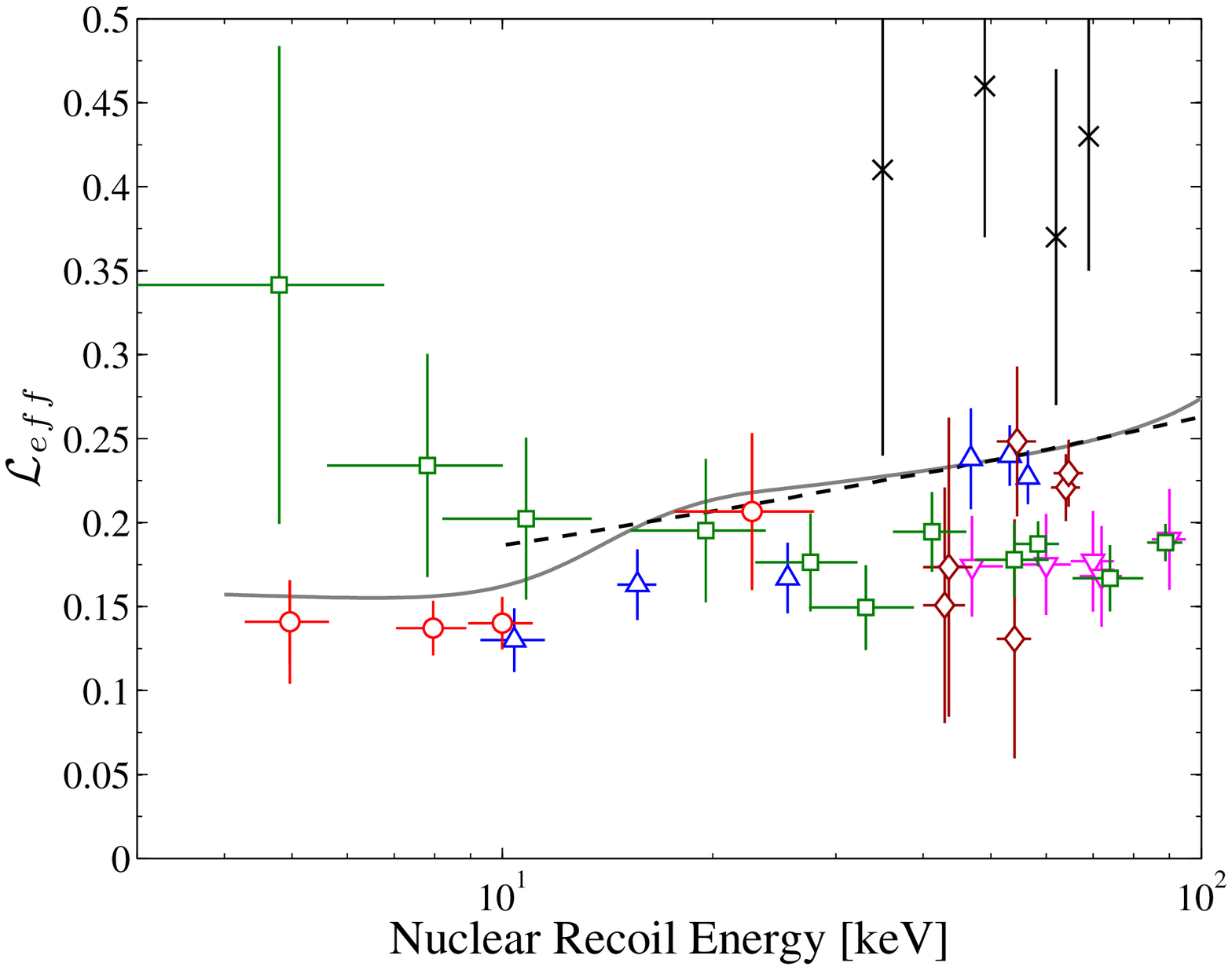}
\caption{Relative scintillation efficiency of nuclear recoils in LXe ~\cite{Aprile:2008c} and references therein.}
\label{NR}
\end{figure}

Since the excitation density of nuclear recoils in LXe is higher
than that of electron recoils of the same energy, the
scintillation light yield is expected to be different for these two types of particle.  Knowledge of the ratio between the two scintillation yields, called relative scintillation efficiency ($L_{eff}$), is important for the determination of the sensitivity of LXe-based detectors to dark matter weakly interacting massive particles (WIMPs), which we discuss in \ref{sec:DM}. The Xe nuclear recoils which result from WIMPs (or neutrons) scattering off Xe nuclei have energies in the range of a few keV up to several tens of keV. Several measurements of  $L_{eff}$ have been carried out \cite{Chepel:2006,Aprile:2005,Akimov:2002,Bernabei:2001,Arneodo:2000}, with the most recent one extending down to 5\,keV$_r$ nuclear recoil energy \cite{Aprile:2008c}. The relative scintillation efficiency for recoils of this energy is 14\%, constant around this value up to 10\,keV$_r$. For higher energy recoils, the value is on average about 19\%. Figure~\ref{NR} summarizes all the measurements to date. A fit through the data and the predicted curve by Hitachi \cite{Hitachi:2005} are also shown as solid and dotted lines, respectively. 

Compared to the scintillation yield of electron or alpha particle excitation, the scintillation yield of nuclear
recoil excitation is significantly reduced due to nuclear quenching \cite{Lindhard:1963}. Hitachi estimates the additional loss in scintillation yield that results from the higher excitation density of nuclear recoils. Rapid recombination in LXe under high LET excitation \cite{Hitachi:1983,Hitachi:1992} provides a mechanism for reducing the scintillation yield of nuclear recoils in addition to that of nuclear quenching treated by Lindhard \cite{Lindhard:1963}. In order to estimate the total scintillation yield, Hitachi \cite{Hitachi:2005, Hitachi:1992} considers biexcitonic collisions, or collisions between two ``free'' excitons that emit an electron with a kinetic energy close to the difference between twice the excitation energy $E_{ex}$ and the band-gap energy $E_{g}$ (i.e. 2$E_{ex}$
-$E_{g}$):
\begin{eqnarray} 
Ê\rm{Xe^{*} + Xe^{*}} &\rightarrow& \rm{Xe+ Xe^{+} + e^{-}}
ÊÊ\label{eq:bi} 
\end{eqnarray} 
The electron then loses its kinetic energy very rapidly before
recombination. This process reduces the number of excitons available
for VUV photons since it requires two excitons to eventually produce
one photon. Hitachi therefore considered this to be the main mechanism responsible
for the reduction of the total scintillation yield in LXe under
irradiation by nuclear recoils.  
The Hitachi model \cite{Hitachi:2005, Hitachi:1992}, however, does not hold at energies below 10\,keV$_r$. 

\subsubsection{Absorption Length and Rayleigh Scattering}
\begin{center}
\begin{figure} 
\includegraphics*[width=.45\textwidth]{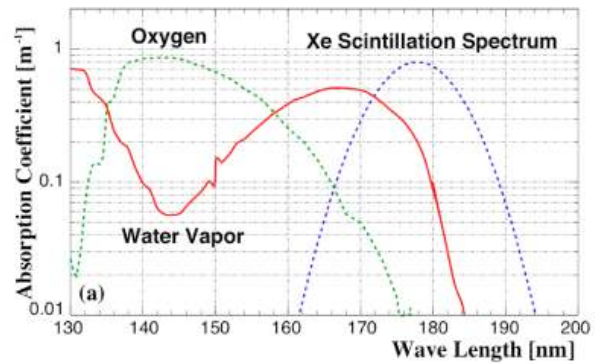} 
\caption{Absorption coefficient for VUV photons in 1 ppm water vapor and oxygen and superimposed Xe emission spectrum \cite{Ozone:2005}.}
\label{fig:Abs_coeff} 
\end{figure}
\end{center}
Impurities dissolved in LXe may absorb the VUV photons, reducing the observed scintillation light yield. Light attenuation can be described by 

\begin{equation}
I(x) = I(0) \mathrm{exp} (-x/\lambda_{att})
\end{equation}
where $\lambda_{att}$ is the photon attenuation length, which consists of two separate components, the absorption length, $\lambda_{abs}$, describing true absorption and loss of photons by impurities,
and the scattering length, $\lambda_{sca}$, that represents elastic scattering of photons without any
loss. The latter is dominated by Rayleigh scattering. The two are related by
 \begin{equation}
 1/\lambda_{att} = 1/\lambda_{abs} + 1/\lambda_{sca}
 \end{equation}
The attenuation length can be measured after removing the contribution from Rayleigh scattering. The Rayleigh scattering length is theoretically estimated to be about 30 cm \citep{Seidel:2002, Baldini:2005a}, which roughly
agrees with the $\lambda_{sca}$ experimentally obtained \citep{Braem:1992,Chepel:1994a,Ishida:1997,Solovov:2004}. 

The wavelength of scintillation light from liquid argon or krypton doped with
xenon is different from that from pure liquid argon or krypton. Accordingly, the measurement of the attenuation length in liquid argon or krypton doped with xenon should
show the wavelength dependency of the attenuation length due to Rayleigh scattering. The attenuation length
due to Rayleigh scattering calculated for liquid Ar + Xe (3\%) and liquid Kr + Xe (3\%) roughly agreed with
the results obtained experimentally \citep{Ishida:1997}.

The most serious impurity for the VUV light of LXe is water vapor, which is largely contributed by the outgassing of the liquid containment vessel and other detector materials placed inside the liquid. Figure \ref{fig:Abs_coeff} shows the absorption coefficient for VUV photons in 1 ppm water vapor and oxygen. The absorption spectra of water and oxygen largely overlap with the xenon scintillation spectrum (\cite{Signorelli:2004}, \cite{Ozone:2005} and references therein). Within the R\&D program for the development of the large LXe scintillation calorimeter for the MEG experiment \citep{PSI:1999}, an absorption length longer than 100 cm at 90\% confidence level was achieved \cite{Baldini:2005a}, corresponding to less than 100 ppb concentration of water vapor. This work has been very useful for other experimenters in the field, bringing out the awareness of the careful detector's preparation to minimize absorption by water vapor. 
 
\subsubsection{Refraction Index for Scintillation Light}

To simulate the number of photons collected in a LXe detector, the knowledge of
the refraction index in the region of the Xe VUV light emission is also relevant. Measurements of this quantity range from 1.54 to 1.69 (Barkov, 1996; Solovov, 2004; Subtil, 1987). 

\subsection{Correlation between Ionization and Scintillation}

As previously mentioned, in heavy liquid rare gases, the two signals of ionization and scintillation can be observed
simultaneously. The amplitude of the two signals are complementary and strongly anti-correlated. In the LET region in which the scintillation yield is maximum (see Figures \ref{fig:letscintar}-\ref{fig:letscintxe}), both signals are
perfectly anti-correlated.

\subsubsection{Correlation in Liquid Argon for Relativistic Heavy Ions}

\begin{center}
\begin{figure} 
\includegraphics*[width=.45\textwidth]{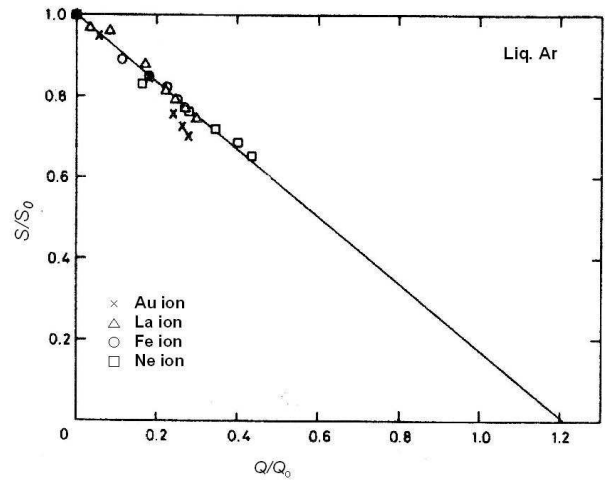} 
\caption{Anti-correlation between scintillation intensity and collected charge in liquid argon for relativistic Ne
	ions($\square$), Fe ions ($\circ$), La ions ($\triangle$ ), and Au ions ($\times$). The solid line is the
	theoretically estimated relation \citep{Masuda:1989}.} 
\label{fig:chargelightar} 
\end{figure}
\end{center}

The simultaneous measurement of ionization and scintillation signals due to relativistic heavy ions has not been carried out for LXe. Yet, measurements of the correlation between ionization and scintillation for relativistic heavy ions in
LAr \citep{Masuda:1989}, provide a good example (see Figure \ref{fig:chargelightar}). 
The scintillation intensity, $S$, normalized to the
scintillation intensity at zero-electric field, $S_0$, is shown as a function of the collected charge, $Q$, normalized to the charge expected for infinite electric field, $Q_\infty$. The figure shows that the
correlation between $S/S_0$ and $Q/Q_\infty$ is perfectly complementary for relativistic Ne, Fe and La ions,
whose scintillation yields are 100\%. On the other hand, for relativistic Au ions, the data deviate from the
straight line corresponding to a perfectly complementary relation, because the scintillation yield of relativistic
Au ions is not 100\% due to quenching. 

\begin{center}
\begin{figure} 
\includegraphics*[width=.4\textwidth]{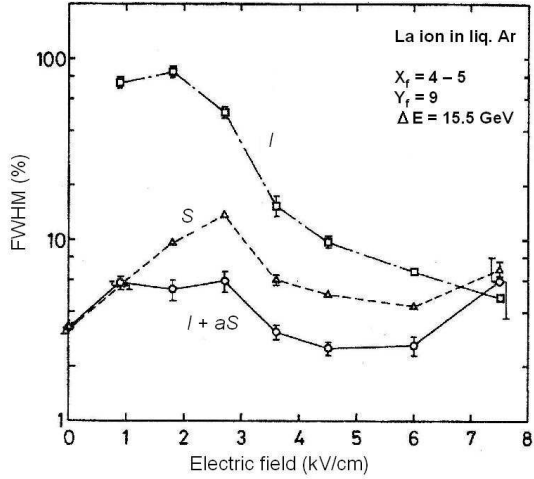} 
\caption{Energy resolution (FWHM) of La ions in LAr as a function of the electric field, measured separately from scintillation, ionization and from their sum \citep{Crawford:1987}.} 
\label{fig:resfield} 
\end{figure}
\end{center}

Figure \ref{fig:resfield} shows the energy resolution of the ionization signal, the scintillation signal, as well as from the two signals combined, measured for relativistic La ions in the above experiment. As expected, the energy resolution of the summed signals is better than that of each individual signal, despite the fact that only a small fraction of the scintillation light was collected in this experiment.

\begin{center}
\begin{figure} 
\includegraphics*[width=.45\textwidth]{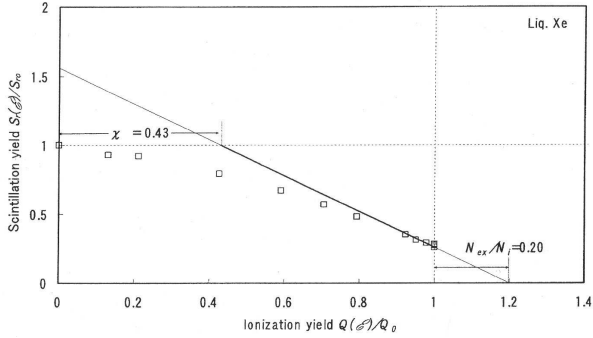} 
\caption{Relation between $S_r/S_{r_0}$ and $Q/Q_0$ in liquid xenon \citep{Doke:2002}. The straight line indicates the perfect anti-correlation of the charge and light signal. $\chi$ is the ratio of number of escaping electrons at zero electric field over the numbers of ion pairs produced by the ionizing radiation.} 
\label{fig:chargelightxe} 
\end{figure}
\end{center}

\subsubsection{Correlation in Liquid Xenon for Relativistic Electrons}

For relativistic electrons whose scintillation yield is lower than 100 \%, the correlation of both signals
does not reflect a perfectly complementary relation. 
Figure \ref{fig:chargelightxe} \citep{Doke:2002} shows the correlation between charge and light signals measured for 1 MeV conversion electrons emitted from $^{207}$Bi in LXe. The solid straight line indicates a perfect anti-correlation of the two signals. Only the data obtained at high electric fields are on the straight line. The more recent data obtained by the Columbia University group for $^{137}$Cs gamma-rays, shows a perfect anti-correlation even at modest fields, as discussed in \ref{sec:sum_signal} and shown in figure \ref{fig:Fig26}. 
These measurements show that the anti-correlation of ionization and scintillation in LXe reduces the fluctuation in the summed signal to a lower level than that in each individual signal, and results in a better energy resolution. 

All fundamental parameters of LXe as detector medium described in this chapter are summarized in Table \ref{LXeProperty}. 
\begin{table}[t]
\caption{Properties of liquid xenon as detector medium.}
\label{LXeProperty}
\center
{\begin{tabular}{|l|l|}
\hline
Item & Values \\
\hline
\bf{Ionization properties} & \\
\ W-value & 15.6 $\pm$ 0.3 eV \\

\bf{Drift velocity and hole mobility} & \\
\ Electron mobility & 2000 $\pm$ 200 cm$^{2}$/V s \\
\ Saturated drift velocity of electrons & 2.6$\times$10$^5$ $\pm$ 10\% cm/s \\
\ \ \ at an electric field of 3-10 kV/cm & \\
\ Hole mobility & 3.6$\times$10$^{-3}$ cm$^2$/V s\\

\bf{Diffusion coefficient of electrons} & \\
\ Transverse diffusion coefficient($D_{\rm T}$) & \\
\ \ \ at an electric field of 1 kV/cm & 80 cm$^2$s$^{-1}$ \\
\ \ \ at electric field of 10 kV/cm & 50 cm$^2$s$^{-1}$ \\
\ Longitudinal diffusion coefficient($D_{\rm L}$) & 0.1 $D_{\rm T}$ \\

\bf{Multiplications} & \\
\bf{Electron multiplication} & \\ 
\ Threshold electric field & 1-2$\times$10$^6$ V/cm \\
\ Maximum gain & 400 \\
\bf{Photon multiplication} & \\
\ Threshold electric field & 4-7$\times$10$^6$ V/cm \\

\bf{Scintillation properties} & \\
\ Wavelength & 177.6 nm \\
\ $W_{\rm ph}$ value & See Table IV \\

\bf{Decay time constants} & \\
\ \bf{from singlet state} & \\
\ \ \ for electrons & 2.2 $\pm$ 0.3 nsec \\
\ \ \ for alpha-particles & 4.3 $\pm$ 0.6 nsec \\
\ \bf{from triplet state} & \\
\ \ \ for electrons & 27 $\pm$ 1 nsec \\
\ \ \ for alpha-particles & 22 $\pm$ 1.5 nsec \\
\ \bf{due to recombination} & \\
\ \ \ for electrons & 45 nsec \\

Radiation length & 2.87 cm \\
Moliere radius & 4.1 cm \\
Refraction index & 1.54 - 1.69 \\
\bf{Attenuation length due to} & \\
\ Rayleigh scattering (theoretical) & 30 cm \\
\ Rayleigh scattering (experimental) & 29 - 50 cm \\
\hline
\end{tabular}}
\end{table}

\section{Liquid Xenon Detector Types}
\label{sec:det_type}
Among liquid rare gases suitable for radiation detection, liquid xenon (LXe) has the highest atomic number and density at a modest boiling temperature, the best ionization and scintillation yields, as well as the highest electron mobility and lowest diffusion. All these are important features for a practical detector. The ionization and scintillation can be detected simultaneously  to provide a precise measurement of the energy, position and time of an event occurring in the sensitive liquid volume. Most of the early LXe detectors have exploited only the ionization process, due to the difficulty of efficiently detecting the VUV scintillation.  
In the mid 1990, the two authors of this review initiated an R\&D program with Hamamatsu Photonics Co, aimed at the development of new VUV sensitive photomultipliers (PMTs), which would operate immersed in LXe, withstanding  up to 5 bar over-pressure. This development,  together with the more recent development of large area avalanche photodiodes (LAAPDs) \citep{Solovov:2002c,Ni:2005}, also able to work in LXe with high sensitivity in the VUV, has had a dramatic impact on the evolution of LXe detectors, leading to an explosion of new detectors exploiting the benefits of the double signature of charge and light in LXe. In this chapter we review the basic operating principle and characteristics of the most common classes of LXe detectors. The applications which use these detectors in specific experiments are covered in the last chapter of this review. 

\subsection{Ionization mode}
\label{sec:ion_chamber}

We consider first a LXe detector sensitive only to the ionization process. The passage of ionizing radiation depositing an energy \textit{E$_0$} in the liquid results in a number of electron-ion pairs, \textit{N$_i$}, given as

\begin{equation}
{N_i} = {E_0}/{W}
\label{Eqn3}
\end{equation}

where \textit{W} is the \textit{W}-value in LXe, defined in the previous chapter. 

The measurement of the total charge associated with an ionizing event \textit{N$_i$} is referred to as the ionization mode, and is usually used only for electrons, $\gamma$-rays or minimum ionizing particles. For heavily ionizing particles like alphas, the electron-ion recombination rate is too high to allow a complete charge collection, as shown in Figure~\ref{fig:alphas}.

\begin{center}
\begin{figure}
\includegraphics*[width=0.45\textwidth]{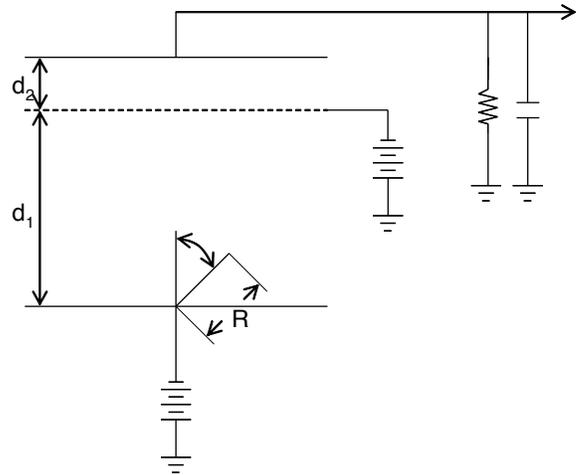}
\caption{Schematic drawing of a gridded ionization chamber. d$_1$ is the drift gap between the cathode (on the bottom) and the Frish grid, d$_2$ is the field region between the Frish grid and the anode (on the top).}
\label{fig:Fig19}
\end{figure}
\end{center}

The most straightforward detector to measure the $N_i$ is a gridded ionization chamber, a device consisting of a cathode and an anode, separated by a third electrode, called a Frish grid \citep{Frish:1945}. The cathode is kept at a more negative potential with respect to the anode, and the grid is at an intermediate potential. Figure \ref{fig:Fig19} shows  the schematic of a gridded parallel plate ionization chamber. The ionizing particle track of length R is oriented at an angle $\theta$ with respect to the electric field created between cathode and grid. The electric field separates the carriers, with the electrons drifting towards the anode, and the ions towards the cathode. The purpose of the grid is two-fold: 1) it allows the electron signal to be induced only after they cross the grid; 2) it shields the anode from the slow motion of the positive ions. Thus the signal on the anode is derived only from the drift of the electrons, with an amplitude which  is proportional to number of electron-ion pairs, but independent of the position where they were created \citep{Knoll:2000}. The calculation of the shielding inefficiency of a Frish grid and the condition for maximum electron transmission has been  carried out by Bunemann et al. \citep{Bunemann:1949}.

To measure the X- and Y-coordinate of the ionizing event, one can use the signal induced on two orthogonal planes of wires, between the Frish grid and the anode. In this case, the Frish grid serves the additional purpose of focusing the electric field lines and maximizing electron transmission. This is the approach used for the time projection chamber (TPC) of the LXeGRIT Compton telescope \citep{Aprile:2000}, discussed in \ref{ss:TPC:mode}. 

\begin{center}
\begin{figure}
\includegraphics*[width=1\columnwidth]{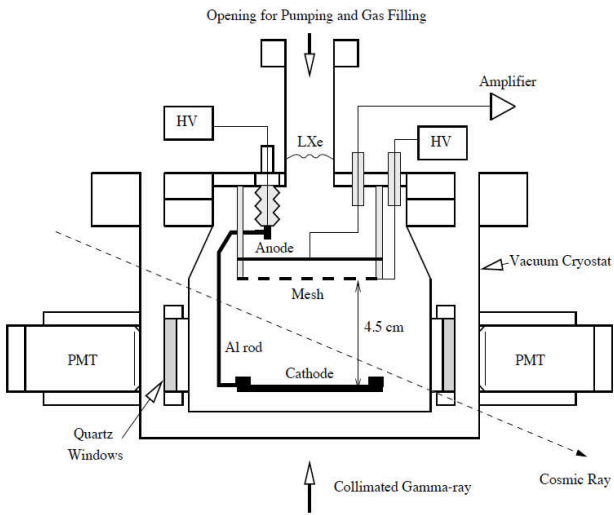}
\caption{Schematic of a 3.5 liter LXe gridded ionization chamber, triggered by the scintillation light \citep{Aprile:2001}. }
\label{fig:gridded_chamber}
\end{figure}
\end{center}

Many of the results obtained by the authors themselves on ionization yield, energy resolution of electrons and gamma-rays in LXe have been obtained with gridded ionization chambers, with drift gaps of several millimeters (see for example \citep{Takahashi:1975, Aprile:1991}). Figure \ref{fig:gridded_chamber} shows a larger gridded ionization chamber, with a  volume of 3.5 liters, used  to measure gamma-rays in the MeV energy range \citep{Aprile:2001}. The drift gap of 4.5 cm is viewed by two VUV-sensitive photomultipliers (PMTs), coupled to the LXe vessel by quartz windows,  to detect the fast scintillation signal as event trigger. 

\begin{center}
\begin{figure}
\includegraphics*[width=0.45\textwidth]{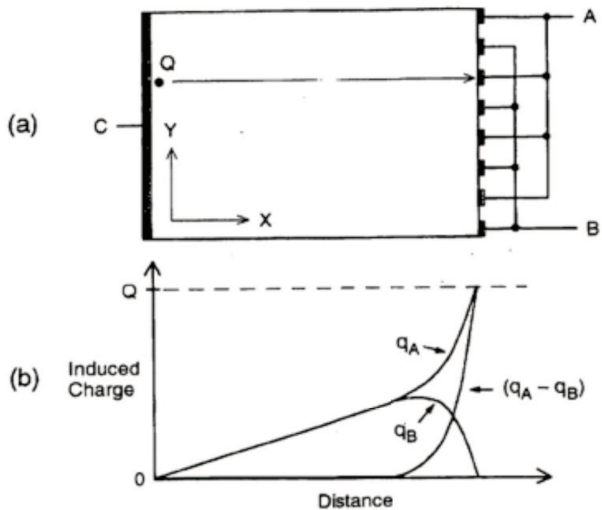}
\caption{(a) Schematic of an ionization chamber with a coplanar anode made of parallel strips \citep{Luke:1995}. (b) Time variation of the induced  charge signals and the resulting signal in case the charge \textit{Q} is collected onto the group of strips marked as A.}
\label{fig:Fig20}
\end{figure}
\end{center}

Another method for eliminating the position dependence of the charge signal in an ionization  chamber was proposed by Luke \citep{Luke:1995} and successfully applied to semiconductor detectors but not yet to liquid rare gases. The anode plate is segmented in parallel strips, grouped into two sets, indicated as A and B in Figure \ref{fig:Fig20}(a); the cathode is \textit{C} and \textit{Q} is a negative charge. The potential of the anode A is adjusted so that all electrons produces by the ionizing radiation can be collected. The shape of the charge signals \textit{q$_A$} and \textit{q$_B$} induced by the motion of \textit{Q} onto each set of strips is shown in Figure \ref{fig:Fig20}(b). The difference \textit{q$_A$}-\textit{q$_B$} is the charge given only by the electrons. Thus, one can measure the total charge  produced by drifting electrons without using a Frish grid. This configuration, known as the virtual Frish grid method \citep{Bolozdynya:2006}, results in much reduced electronic noise, and thus allows for better energy resolution and lower minimum energy threshold. Tests of such devices in LXe are planned for the near future.

\begin{center}
\begin{figure}
\includegraphics*[width=0.4\textwidth]{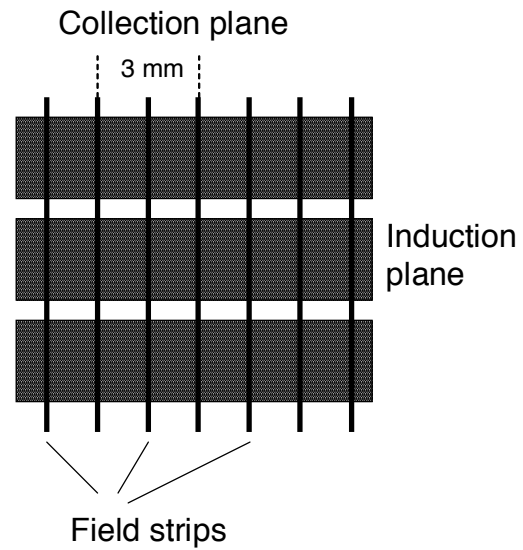}
\caption{An example of two-dimensional position-sensitive electrode, with orthogonal strips  for collection and induction signals \citep{Cennini:1994}.}
\label{fig:Fig21}
\end{figure}
\end{center}

With a two-dimensional strip electrode such as shown in Figure \ref{fig:Fig21}, one can measure the X- and Y-coordinates from the collected and induced charge \citep{Cennini:1994}.
The strips are deposited on opposite sides of a thin glass plate, perpendicular to each other. The collecting strips are very thin and widely spaced, and those deposited on the back are wide, allowing a large induced signal on the back strips for a deposited energy as low as 100 keV. 

\begin{center}
\begin{figure}
\includegraphics*[width=0.4\textwidth]{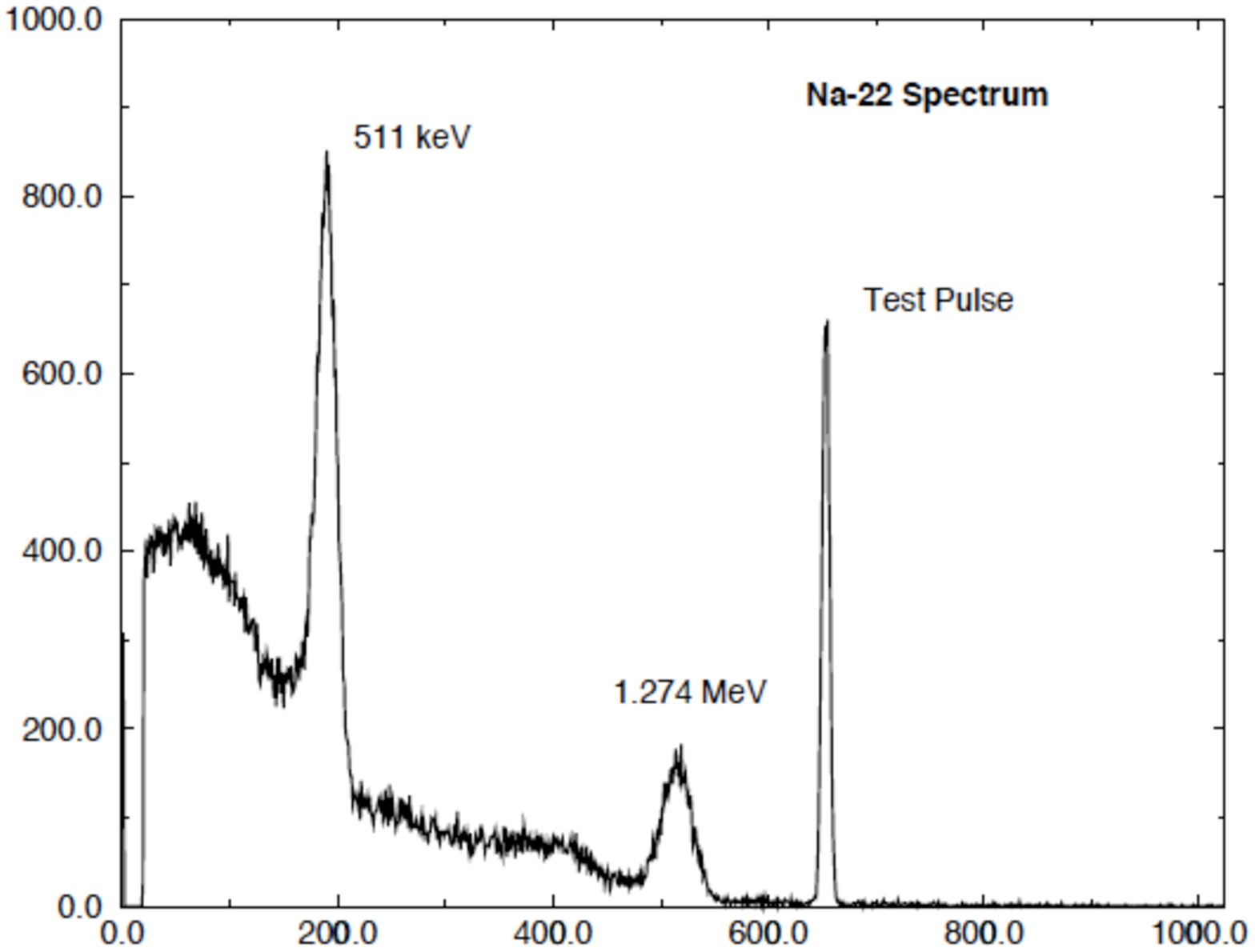}
\caption{The energy spectrum of Na$^{22}$ gamma-rays measured with the gridded ionization chamber shown in Figure \ref{fig:gridded_chamber} at 4 kV/cm drift field \citep{Aprile:2001}. The energy resolution of the 511 keV line is about 8\% (FWHM).}
\label{fig:Na22_spectrum}
\end{figure}
\end{center}
\begin{center}
\begin{figure}
\includegraphics*[width=0.4\textwidth]{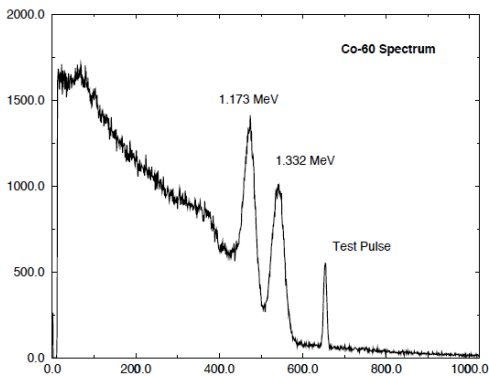}
\caption{The energy spectrum of Co$^{60}$ gamma-rays measured with the gridded ionization chamber shown in Figure \ref{fig:gridded_chamber} at 4 kV/cm drift field \citep{Aprile:2001}. The energy resolution of the 1.3 MeV line is about 5\% (FWHM).}
\label{fig:co60_spectrum}
\end{figure}
\end{center}

Several factors contribute to the spread of the energy measured from the ionization signal detected with a gridded ionization chamber, in addition to the intrinsic fluctuation in the number of electron-ion pairs, expressed by the Fano factor, and discussed in \ref{sec:fundamental}. These include electronic noise, variation in the signal rise time due to different emission angles of the primary particle and shielding inefficiency of the grid. As discussed in \ref{sec:fundamental}, the best energy resolution measured to date with LXe gridded chambers is far from the Fano limit given by Eq.\ref{int_res}.

Figure \ref{fig:Na22_spectrum} and Figure \ref{fig:co60_spectrum} show the energy spectra of $^{22}$Na and $^{60}$Co gamma-rays measured at a field of 4 kV/cm with the ionization chamber of Figure \ref{fig:gridded_chamber} \cite{Aprile:2001}. The noise-subtracted energy resolution achieved with this chamber is consistent with 5.9\% (FWHM) at 1 MeV, scaling as $E^{-1/2}$ over the measured energy range.  

\begin{center}
\begin{figure}
\includegraphics*[width=0.4\textwidth]{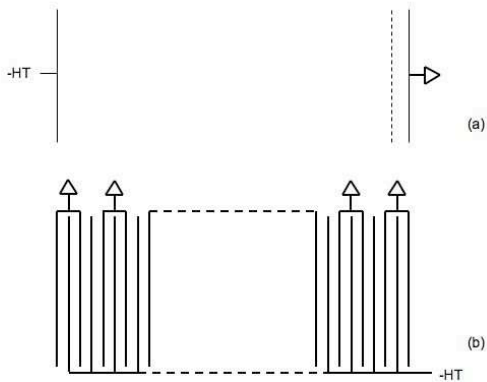}
\caption
{(a) Gridded ionization chamber with a long sensitive region. (b) Multi-parallel plate electrode system for an ionization calorimeter \citep{Doke:1993}.}
\label{fig:Fig22}
\end{figure}
\end{center}

To measure high-energy electrons and gamma rays, which produce  electromagnetic showers, a massive detector is required. A simple LXe gridded ionization chamber is not practical as a calorimeter, since the very long sensitive region required for full energy deposition would involve very high voltage to generate an adequate electric field (see Figure \ref{fig:Fig22}(a)). In addition, the signal from  such a chamber would be extremely slow, given the electron drift velocity in LXe. 
The approach used instead is a multi-parallel plate electrode system such  as illustrated in Figure \ref{fig:Fig22}(b).

If the individual plate electrodes are thin, the energy loss in the electrodes is small and one can measure the energy deposited in the sensitive volume in almost fractional amounts. Such a chamber is referred to as a homogeneous calorimeter \citep{Doke:1993}. Each cell of the calorimeter  consists of two electrodes, with no wire grid. As a result, the charge \textit{Q} induced in each cell can be expressed as
\begin{equation}
\label{Eqn6}
Q/e = \frac{E_0 (1-\Delta E/E_0)}{2W}
\end{equation}
where \textit{E$_0$} is the energy (expressed in eV) deposited in the calorimeter and $\Delta$\textit{E} is the fraction of energy (expressed in eV) deposited in each cell of electrodes. The effective \textit{W}-value under these conditions becomes 2$W$ due to the positive ion effect \citep{Doke:1992a}. In place of such a thin electrode system, a wire electrode system may be used \citep{Doke:1992b}. In addition, sometimes an electrode system with a complicated shape is used, but in such a case, 2$W$ should be used as the \textit{W}-value to estimate the charge produced by penetrating particles.

If the individual plate electrodes were thick or massive, the calorimeter is referred to as a sampling calorimeter. In this case, the calorimeter is more compact and can be used to measure the energy of very high-energy radiation. However, the energy resolution of a sampling calorimeter is significantly inferior to that of a homogeneous calorimeter. 
To date, large volume ionization or sampling calorimeters for high energy physics experiments have used liquid krypton and liquid argon, respectively, but not liquid xenon, largely because of its cost.  

\subsection{Scintillation mode}
Let us now consider a LXe detector which is sensitive only to the scintillation process. Typically, a photomultiplier (PMT) is used to detect the scintillation photons produced by an event depositing an energy $E_0$ in the liquid, and the number of photoelectrons emitted from the photo-cathode of the PMT follows a Poisson distribution. Its spread can be expressed as (\textit{N$_{phe}$})$^{1/2}$, where \textit{N$_{phe}$} is the number of photoelectrons, generally expressed as    
\begin{equation}
\label{Eqn7}
N_{phe}=\eta L_c \times E_0/W_{ph}
\end{equation}
Here $\eta$ is the quantum efficiency of PMT, $L_c$ is the light collection efficiency and \textit{W$_{ph}$} is the average energy required for the production of one photon.

\begin{center}
\begin{figure}
\includegraphics*[angle=270,width=1\columnwidth]{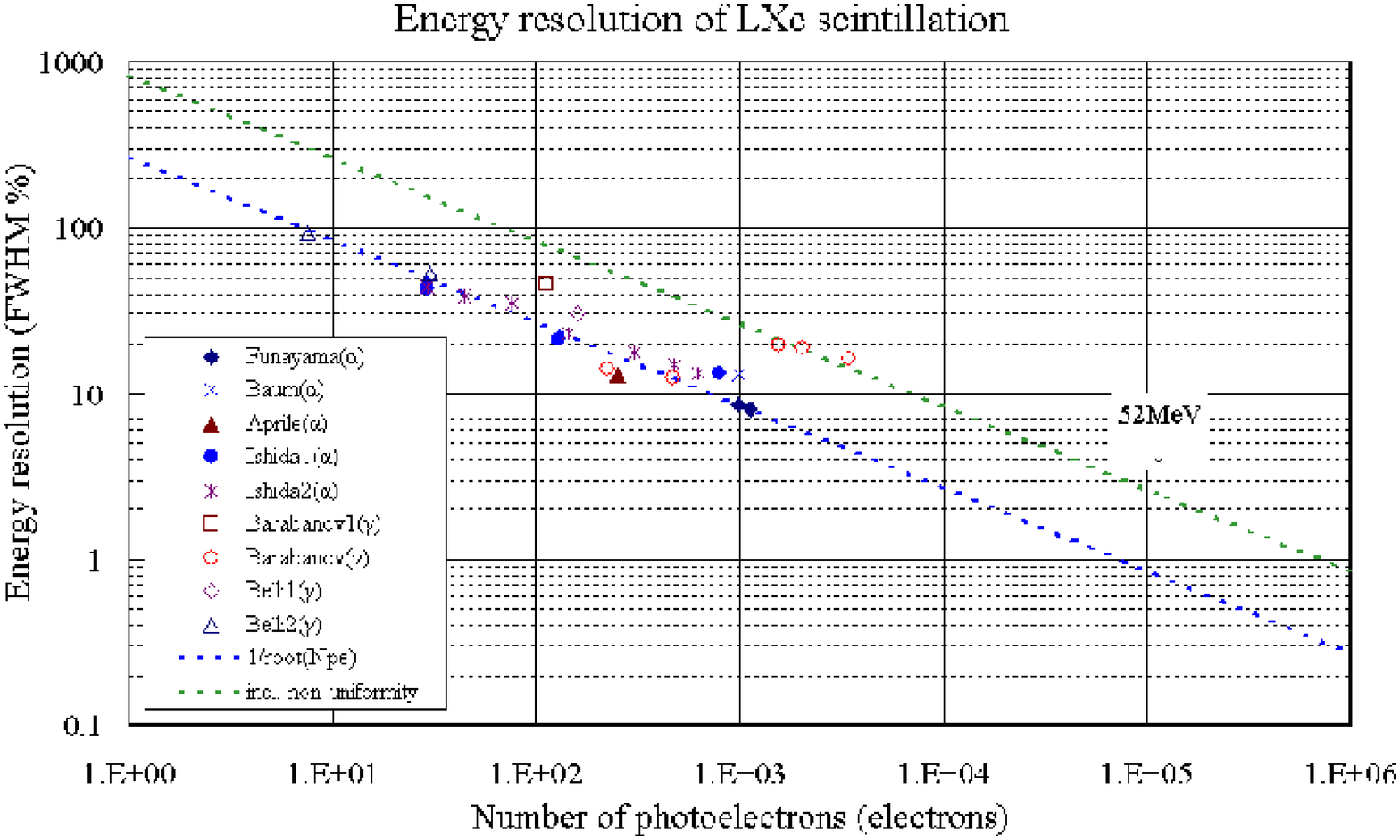}
\caption{Relation between energy resolutions and the estimated number of photoelectrons obtained for $\alpha$-particles and $\gamma$-rays (\citep{Doke:1999} and references therein). The lower line (blue) expresses a relation of energy resolution which is inversely proportional to the square root of the number of photoelectrons. The upper line (green) shows a worse energy resolution expected from the non-uniformity of light collection.}
\label{fig:ene_res_pe}
\end{figure}
\end{center}

An energy resolution better than that of an ionization detector should be obtained if $\eta$ and $L_c$  are 100\%, since for a given energy $E_0$ deposited in LXe, the total  number of scintillation photons (\textit{N$_i$} + \textit{N$_{ex}$}) is greater than the total number of ionization electrons (\textit{N$_i$}), as discussed in \ref{sec:fundamental}. 
However, it is difficult to achieve a complete light collection and to date the best quantum efficiency of PMTs sensitive to the VUV regime is about 35\%. The energy resolution, expressed as

\begin{equation}
\label{Eqn8}
\Delta E/E_0(\mathrm{FWHM}) \approx 2.35(N_{phe})^{1/2}                 
\end{equation}

was confirmed by experiments with LXe scintillation detectors irradiated with alpha particles. Figure \ref{fig:ene_res_pe} illustrates the relation between the measured energy resolution and the estimated number of photoelectrons for alpha particles  and for gamma-rays \citep{Doke:1999}. The data points in the log-log plot for alpha particles are almost on a line of 1/(\textit{N$_{phe}$})$^{1/2}$. In contrast, the data points for the gamma-rays are scattered, and the best data were obtained only for gamma-rays with energies below 0.5 MeV, for which the interaction points in the liquid are localized.  

\begin{center}
\begin{figure}
\includegraphics*[width=0.5\textwidth]{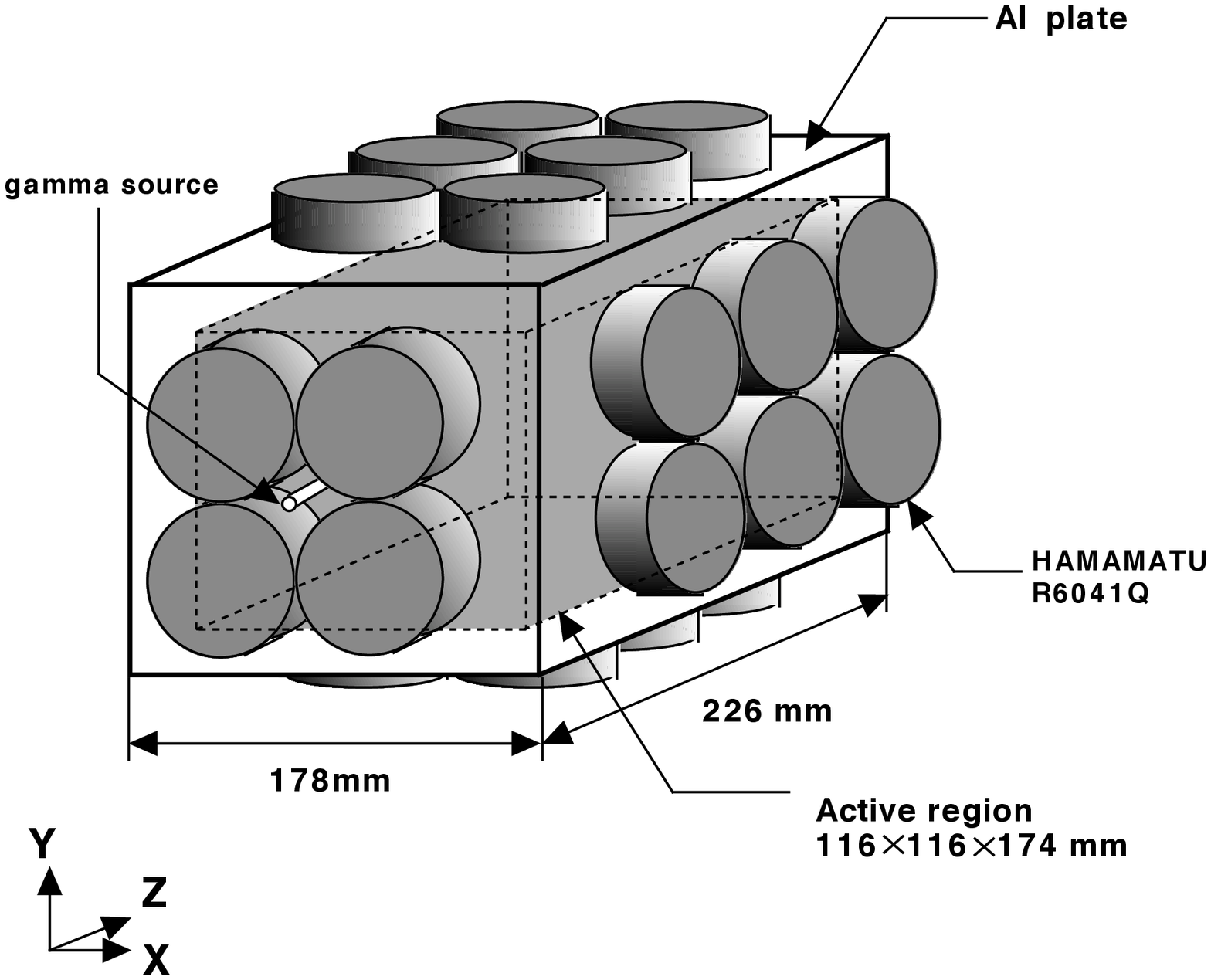}
\caption{Schematic of the LXe scintillation prototype with 32 2" PMTs (Hamamatsu R6041Q) developed for the MEG experiment \citep{Mihara:2002}.}
\label{fig:Fig45}
\end{figure}
\end{center}

\begin{center}
\begin{figure}
\includegraphics*[width=0.4\textwidth]{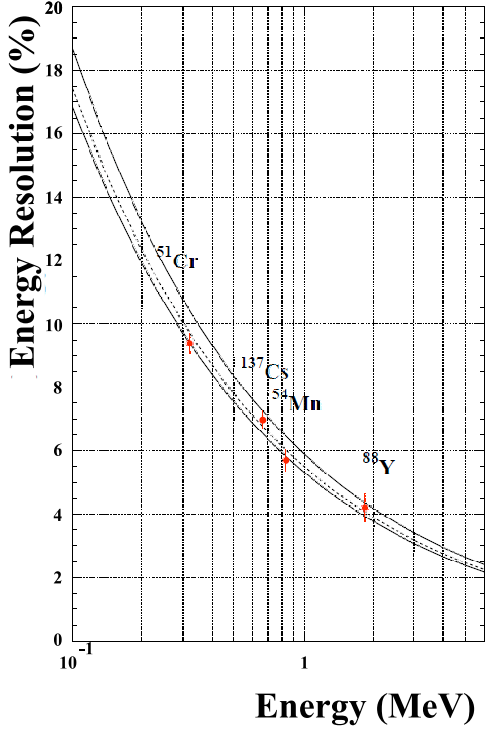}
\caption{Energy resolution ($1\sigma$) as a function of gamma-ray energy, measured with the MEG LXe scintillation prototype. Solid curves are from Monte Carlo simulations \citep{Mihara:2002}. The data follows $1/\sqrt{N_{phe}}$ as expected.}
\label{fig:Fig46}
\end{figure}
\end{center}

In 1998, a LXe scintillation prototype, shown in Figure \ref{fig:Fig45}, was built to study the detector concept later proposed for the MEG experiment \citep{PSI:1999, Baldini:2002}, discussed in \ref{sec:meg}.  The prototype used 32$\times$2'' PMTs (Hamamatsu R6041Q), with a quantum  efficiency (QE) of about 6\% \citep{Terasawa:1998}, to view a 2.3 liter volume of LXe. 
Figure \ref{fig:Fig46} shows the energy resolution as a function of gamma-ray energy between 0.3 and 2 MeV, measured with this prototype \citep{Mihara:2002}. 

In \cite{Doke:2001}, it was suggested that an energy resolution better than 1\% ($1 \sigma$) could be achieved for 52.8 MeV gamma-rays of interest to the MEG experiment, by using a large sensitive volume of LXe surrounded by PMTs. The distribution of the light pattern on the PMTs would also allow a precise determination of  the gamma-ray interaction points in the LXe volume.  These conclusions led to the final MEG detector's design. The expected resolution of 0.8\% ($1 \sigma$) at 52.8 MeV was indeed directly measured in 2005 with the larger MEG prototype (see \ref{sec:meg}). 

\subsection{Sum signal mode}
\label{sec:sum_signal}
We now consider a LXe detector in which one simultaneously observes both ionization and scintillation signals produced by a particle interaction in the liquid, and uses their sum to infer a better energy resolution than allowed by each mode separately. 
As described in \ref{sec:fundamental}, the ionization signal and scintillation signals are complementary and anti-correlated, therefore, the fluctuations of ionization and scintillation are also anti-correlated.
As a result, the fluctuation of the sum signal is smaller than that of the individual signals. We have already discussed the first observations of this anti-correlation in LAr irradiated with relativistic heavy ions \citep{Crawford:1987, Masuda:1989}, later proven also in LXe irradiated with  relativistic electrons \citep{Doke:2002}. Here we discuss more recent measurements exploiting the sum signal mode in LXe, irradiated with  high energy electrons and gamma-rays.  A clear anti-correlation and a substantial improvement of the energy resolution was observed by two independent groups. Conti et al \citep{Conti:2003} used a single PMT coupled with an optical window to a LXe gridded ionization chamber irradiated by Bi$^{207}$. A resolution of 3\% (1$\sigma$) for 570 keV gamma-rays was achieved at a field of 4 kV/cm. 

\begin{center}
\begin{figure}
\includegraphics*[width=0.45\textwidth]{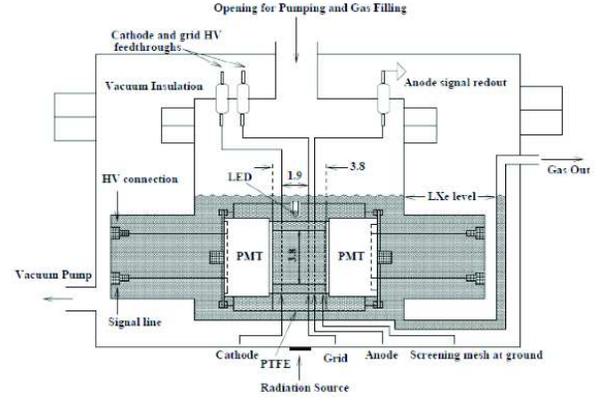}
\caption{Schematic of a ``sum signal mode" detector \citep{Aprile:NIM2007}.}
\label{fig:Fig25}
\end{figure}
\end{center}

Better results were obtained by Aprile et al. \citep{Aprile:NIM2007}, using a gridded ionization chamber with PMTs immersed in the liquid. 
Figure \ref{fig:Fig25} shows a schematic view of this detector's inner structure. The ionization chamber consists of three optically transparent mesh electrodes that serve as the cathode, Frish grid and anode, enclosed by a PTFE (Polytetrafluoroethylene) tube. PTFE is used for its high reflectivity to VUV light \citep{Yamashita:2004}. Two Hamamatsu R9288 PMTs, mounted on both sides of the PTFE tube, detect the LXe scintillation light. One additional shielding mesh is placed between the PMT and the anode mesh to avoid induction between the PMT and the anode-charge signals. The 1.9 cm drift region between the Frish grid and the cathode defines the liquid xenon sensitive volume. The distance between the grid and the anode is 3 mm. The structure is mounted in a stainless steel vessel and surrounded by a vacuum cryostat for thermal insulation.

\begin{center}
\begin{figure}
\includegraphics*[width=0.4\textwidth]{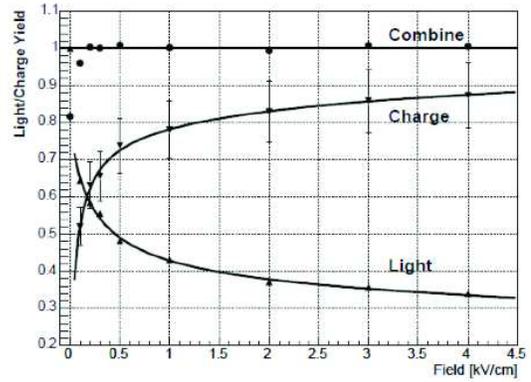}
\caption{Light and charge as a function of drift field for  $^{137}$Cs  662 keV $\gamma$-rays \citep{Aprile:NIM2007}.}
\label{fig:Fig26}
\end{figure}
\end{center}

\begin{center}
\begin{figure}
\includegraphics*[width=0.45\textwidth]{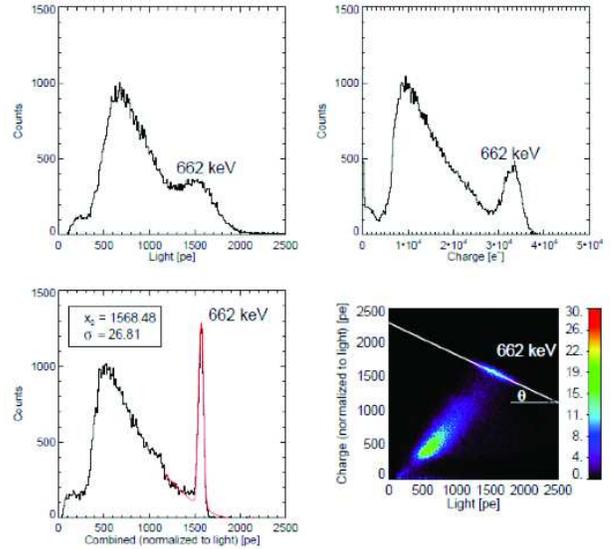}
\caption{Energy spectra of Cs$^{137}$ 662 keV $\gamma$-rays measured with 1 kV/cm field in LXe, separately from charge and light (top panels) and from combined charge and light (lower panel, left); charge-light anti-correlation and its angle (lower panel, right) \citep{Aprile:NIM2007}.}
\label{fig:Fig27}
\end{figure}
\end{center}

The sum of the light and charge yields should be constant with applied electric field if the light yield measured in this apparatus is normalized so that it represents the number of photons emitted from the liquid xenon, and the charge yield is obtained by the number of electrons collected, as illustrated in Figure \ref{fig:Fig26}. The variation of the light and charge yields with field is also shown in the figure. The energy spectra of $^{137}$Cs $\gamma$-rays measured with a 1 kV/cm drift field  are shown in Figure \ref{fig:Fig27}. The top two spectra were obtained separately from scintillation and ionization signals, with an energy resolution of 10.3\% (1$\sigma$) and 4.8\% (1$\sigma$), respectively.  The strong charge-light anti-correlation is indicated in the bottom-right plot of Figure \ref{fig:Fig27}. 
The straight line indicates the charge-light correlation angle. The charge-light combined spectrum (bottom-left) revealed a greatly improved energy resolution with a value of 1.7\% (1$\sigma$). 

\subsection{Time Projection Chamber Mode}
\label{ss:TPC:mode}

The concept of a liquid time projection chamber (TPC) was first proposed by Rubbia in 1977 \citep{Rubbia:1977} for a large-scale liquid argon detector dedicated to proton decay, solar neutrino and other rare phenomena in  particle physics. This type of detector allows three-dimensional event imaging, energy measurement and particle identification. The theory of electron imaging of ionizing events and the design of the electrode system of a liquid TPC was developed in \citep{Gatti:1979}. This type of non-destructive electrode system was successfully implemented in a LXeTPC, proposed for the first time by Aprile in 1989 for spectroscopy and Compton imaging of MeV gamma-rays from astrophysical sources \citep{EAprile:89, EAprile:95}. We use this example to describe  the basic principle of a TPC with three-dimensional position sensitivity. 

\begin{center}
\begin{figure}
\includegraphics*[width=0.53\textwidth]{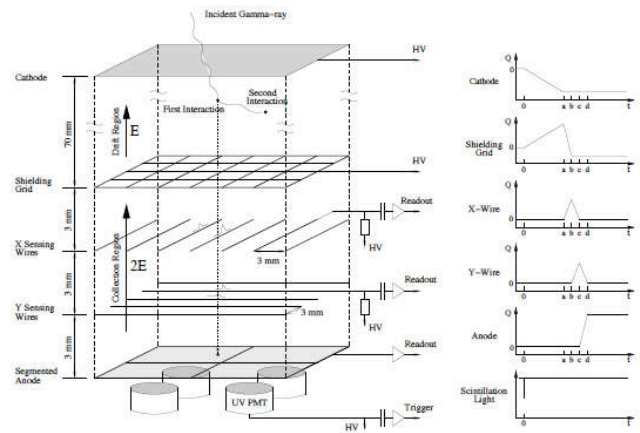}
\caption{Schematic of the LXeTPC electrodes structure (not to scale) with corresponding pulse shapes \citep{EAprile:98.nim}.}
\label{fig:Fig29}
\end{figure}
\end{center}

\begin{center}
\begin{figure}
\includegraphics*[width=0.5\textwidth]{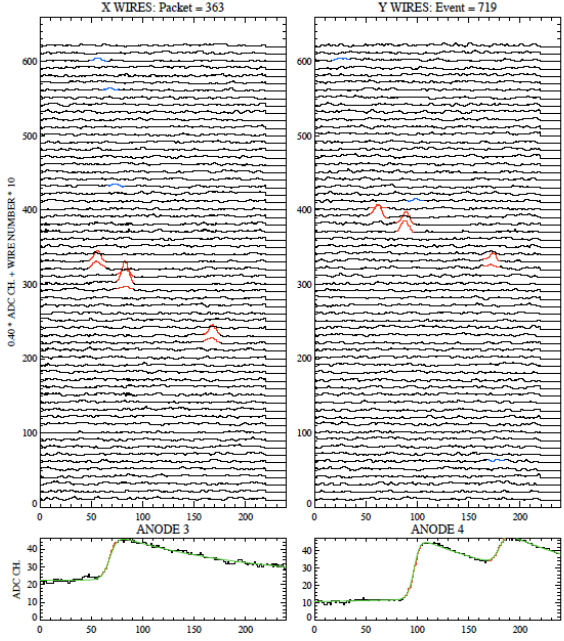}
\caption{Display of a 1.8 MeV $\gamma$-ray event with multiple Compton interactions, recorded with the LXeTPC \citep{EAprile:98.nim}.}
\label{fig:Fig30}
\end{figure}
\end{center}

Figure \ref{fig:Fig29} shows a schematic of the TPC electrodes arrangement and of the expected pulse shapes. The LXe sensitive volume, a box of $\sim$19$\times$19$\times$7 cm$^3$, is defined by a cathode, a series of field shaping rings, a wire-electrode system, and four independent anodes made of wire meshes.  Four VUV sensitive photomultipliers (PMTs), coupled to the TPC vessel with quartz windows, are used to detect the scintillation light providing  the initial event time, \textit{t$_0$}.  
Ionization electrons produced by a gamma-ray interaction in the sensitive volume, drift under an applied electric field of
$\sim$1 kV/cm, inducing a signal on two orthogonal planes of parallel wires, with a 3 mm pitch, before collection on the anodes. The signals from the wires in each plane and from the anodes are amplified and digitized at a sampling rate of 5MHz, to record the pulse shape. The location of the hit wire(s) in the two planes
provide the $x$ and $y$ coordinates in the TPC reference frame, while
the time, measured starting from \textit{t$_0$}, and the known drift velocity, give the interaction depth
($z$ coordinate). The energy of the event is measured from the total charge collected by the anode(s). From the measured spatial coordinates and energy deposited in each interaction point, Compton kinematics allow reconstruction of the incoming gamma-ray direction, within the ambiguity due to the unknown direction of the Compton scattered electron. 
An example of the event imaging capability of this TPC is shown in Figure \ref{fig:Fig30}. The event  shows a 1.8 MeV $\gamma$-ray with multiple Compton interactions, fully contained in the sensitive volume. The digitized pulses (0.2 $\mu$s/sample) on the wires and anodes are plotted as a function of the drift time. 
A new TPC of similar design as the one described above but with PMTs immersed in the liquid was built by the Waseda group \citep{Takizawa:2002}. It uses a wire-less structure for $x$ and $y$ read out and the first VUV PMTs operating at LXe temperature which were developed within the LXeGRIT R\&D program in collaboration with Hamamatsu.

\subsection{Two-Phase Time Projection Chamber Mode} 
\label{sec:two-phase}
Excess electrons liberated in LXe by ionizing radiation can be efficiently extracted from the liquid into the gas \citep{Bolozdynya:1995} for effective amplification. The process by which amplification occurs in the gas phase is called electroluminescence or proportional scintillation. Proportional scintillation of LXe was first observed by Dolgoshein \citep{Dolgoshein:1967}. The detector that uses this secondary scintillation emission is known as a two-phase detector and it is typically operated as a TPC. 
\begin{center}
\begin{figure}
\includegraphics*[width=0.45\textwidth]{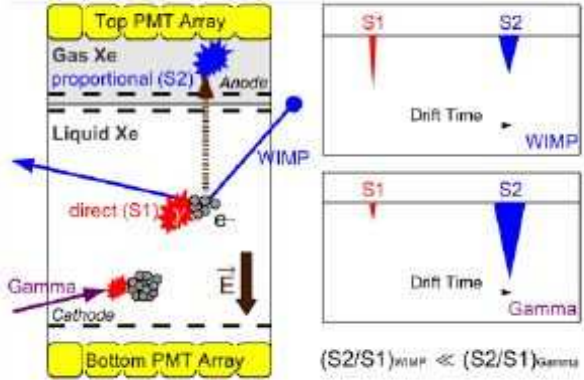}
\caption{Schematic of a two-phase xenon TPC.}
\label{fig:FigXenon_principle}
\end{figure}
\end{center}

\begin{center}
\begin{figure}
\includegraphics*[width=0.5\textwidth]{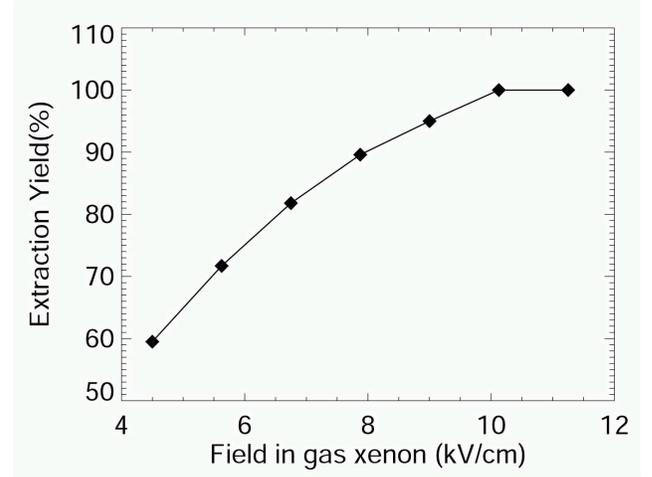}
\caption{Electron-extraction yield in LXe as a function of the electric field in the gas phase \citep{Aprile:2004}.} 
\label{fig:Fig32}
\end{figure}
\end{center}

Figure \ref{fig:FigXenon_principle} presents the principle of operation of a typical two-phase TPC, developed for the detection of dark matter weakly interacting massive particles (WIMPs), discussed in \ref{sec:DM}. Both the scintillation and the ionization, typically called the S1 and S2 signals, produced by radiation in the sensitive LXe volume are detected. Like in the previous LXeTPC, ionization electrons drift under an applied electric field but once they reach the liquid surface, they are extracted into the gas phase  where they emit proportional scintillation light. Figure \ref{fig:Fig32} shows the extraction yield of electrons from LXe as a function of the electric field in the gas phase \citep{Aprile:2004}. It is 100\% for a field greater than 10 kV/cm. 

The number $N_{\gamma}$ of  proportional scintillation photons is given by the expression
\begin{equation}
N_{\gamma}=\alpha N_e(E/p-\beta)pd.
\label{eq:prop}
\end{equation}
Here $p$ is the gas pressure, $E$ is the extraction field and $d$ is the distance traveled by the extracted electrons in gas, \(N_e\) is the number of electrons extracted from the liquid to the gas phase,  \(\alpha\) is the amplification factor, and \(\beta\) is the threshold of the reduced field for proportional light production \citep{Bolozdynya:1995}. 
Typically, both the direct scintillation and the proportional scintillation are detected with VUV sensitive PMTs, located in the gas and in the liquid. Given the different index of refraction of liquid and gaseous Xe, the direct light suffers total internal reflection at the interface, and thus is most effectively detected by the PMTs in the liquid. The PMTs in the gas detect mostly the proportional scintillation light and provide  x-y event localization, with a resolution  ranging from millimeters to centimeters, depending on the PMTs size and granularity of the array. 
The third coordinate, along the drift direction, is inferred from the time difference between the direct  and the proportional scintillation signals and the known electron drift velocity in the liquid, with a resolution better than $\sim$1mm. 

The three dimensional (3D) position sensitivity of the TPC allows to correct the position dependence of both the direct and proportional scintillation signals, which in turn results in improved energy resolution.

\begin{figure}
\begin{center}
\includegraphics[width=0.55\textwidth]{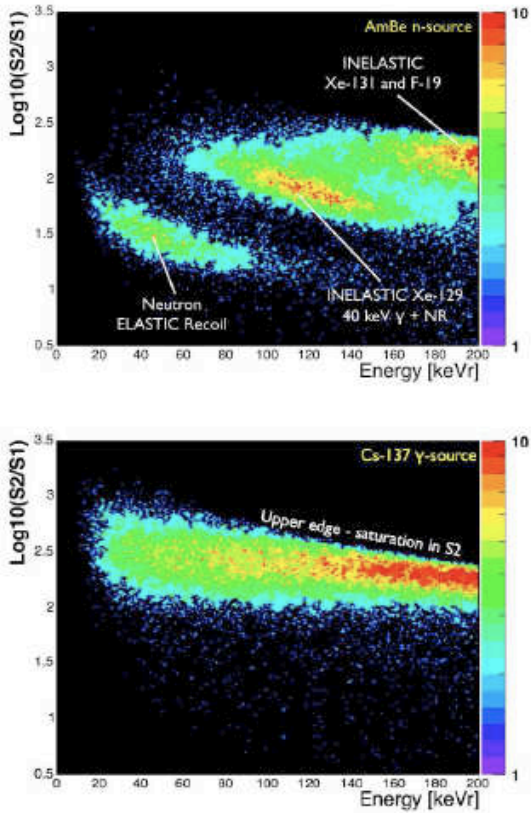}
\caption{\label{2Dplots} A two-phase xenon detector's response to AmBe neutrons (top) and $^{137}$Cs gamma rays (bottom), at 2 kV/cm drift field \cite{Aprile:2006}.}
\end{center}
\end{figure}

The key criterion of these two phase XeTPCs in the search for rare processes is their ability to discriminate nuclear from electron recoils. This discrimination relies on the different ratio of ionization and scintillation for nuclear and electron recoils in LXe, expected from the different ionization density and different geometry of the track structure for these two types of particles. This was first verified using a small two-phase prototype \cite{Aprile:2006} developed as part of the R\&D work for the XENON experiment (see \ref{sec:XENON}). Figure \ref{2Dplots} shows the response of the two-phase prototype to neutrons and low energy Compton scattered gamma rays. The logarithm of the ratio S2/S1 is plotted as a function of nuclear recoil energy. The elastic nuclear recoil band is clearly separated from the electron recoil band, providing the basis for discrimination against background gammas and betas in the two-phase Xe detectors for dark matter searches.

\section{Applications of Liquid Xenon Detectors}
\label{sec:applications}
Several liquid xenon (LXe) detectors based on the operating principles discussed in the previous chapters have been developed to address some of the most intriguing and fundamental questions in physics today, from the search for dark matter in the  form of weakly interacting massive particles (WIMPs), to the search for neutrinoless double beta decay of $^{136}$Xe to determine the neutrino mass. Additionally, a search for the rare $\mu \rightarrow$ e$\gamma$ decay is currently underway with the largest ever built LXe detector, probing new physics beyond the standard model of particle physics. Finally, several R\&D projects are aiming to establish LXe detectors for positron emission tomography in medicine, with direct benefit to society. In this section we review these applications of LXe detectors, covering especially the developments of  the past ten years. 

\subsection{Applications to Gamma-Ray Astrophysics}
\subsubsection{LXe Compton Telescope for 0.3 to 20 MeV $\gamma$-rays}
\label{sec:compton_tele}
Gamma-ray astrophysics in the medium energy range is concerned with nuclear transition lines and the 511 keV positron annihilation line, in addition to high energy continuum emission from accreting black holes and cosmic ray interactions with the interstellar medium. Gamma-ray lines make this band of the electromagnetic spectrum particularly powerful for the study of the lifecycle of matter: formation and evolution of stars result in generation and ejection of heavy nuclei during the hydrostatic burning stages of stars, or in their explosive end stages as supernovae. The chemically enriched interstellar medium, in turn, becomes the birth place for a new generation of stars, providing the building blocks of life in the universe. Gamma-ray lines, through spectroscopy and imaging, provide the deepest look into the explosion mechanism of both gravitational core collapse supernovae and thermonuclear white dwarf supernovae \citep{RDiehl:2006:NPA:g-ray:spec}.

The various astrophysical sources impose different requirements on the energy resolution of an instrument: for gamma-ray line spectroscopy of white dwarf supernovae, an energy resolution of $\sim$3\% FWHM at $\sim$847 keV is necessary. For other sources such as diffuse $^{26}$Al or $^{60}$Fe emission, gamma-ray lines are below the resolution of even germanium spectrometers \citep{RDiehl:2008:NAR:26Al,WWang:2007:AA:60Fe}, and hence the accessible astrophysical information lies predominantly in the observed intensity distribution, which must be determined with a resolution of order 1$^o$. 

The low source fluxes combined with high background from interactions of cosmic rays or trapped protons with structural and detector materials require a large field-of-view (FOV), a large effective area, low intrinsic background, and excellent background discrimination. Given the dominance of the Compton cross-section at these energies, a compact Compton telescope, ideally with time-of-flight (TOF) capability, is the most promising concept. The idea of a LXeTPC as a Compton telescope for MeV $\gamma$--ray astrophysics was first proposed in \citep{EAprile:89:SPIE}. A series of experiments with LXe detectors followed, aimed at establishing the charge and light yields, with specific emphasis on the
use of LXe in a position sensitive detector \citep{Aprile:1991_307, Aprile:1991_300, Aprile:1991, Aprile:1990}. The development of the first LXeTPC prototype, described briefly in subsection \ref{ss:TPC:mode} led to the realization of a balloon-borne instrument, the liquid xenon gamma-ray imaging telescope (LXeGRIT) \citep{ACurioni:07:NIM:LXeGRIT:eff, EAprile:98.nim, Aprile:2000}. Figure ref{fig:compton} shows the schematics of this experiment as a prototype of a compton  telescope.

\begin{center}
\begin{figure}
\includegraphics*[width=0.4\textwidth]{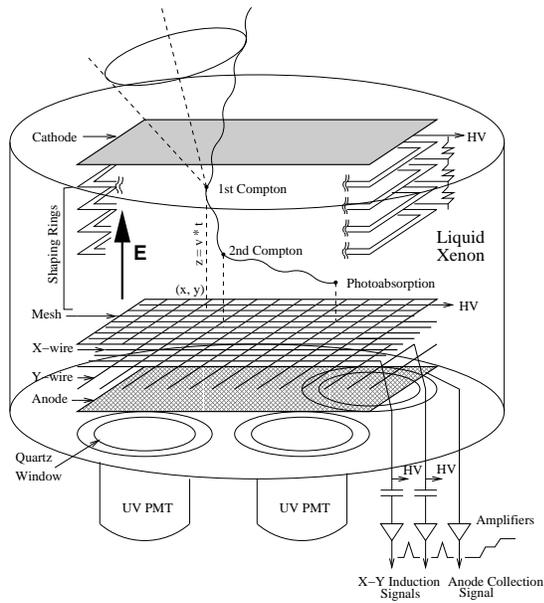}
\caption{Schematics of a compton telescope, here the LXeGRIT detector \citep{EAprile:03:newastr}.}
\label{fig:compton}
\end{figure}
\end{center}

\begin{center}
\begin{figure}
\includegraphics*[angle=180,clip,width=1\columnwidth,trim=20 60 60 20]{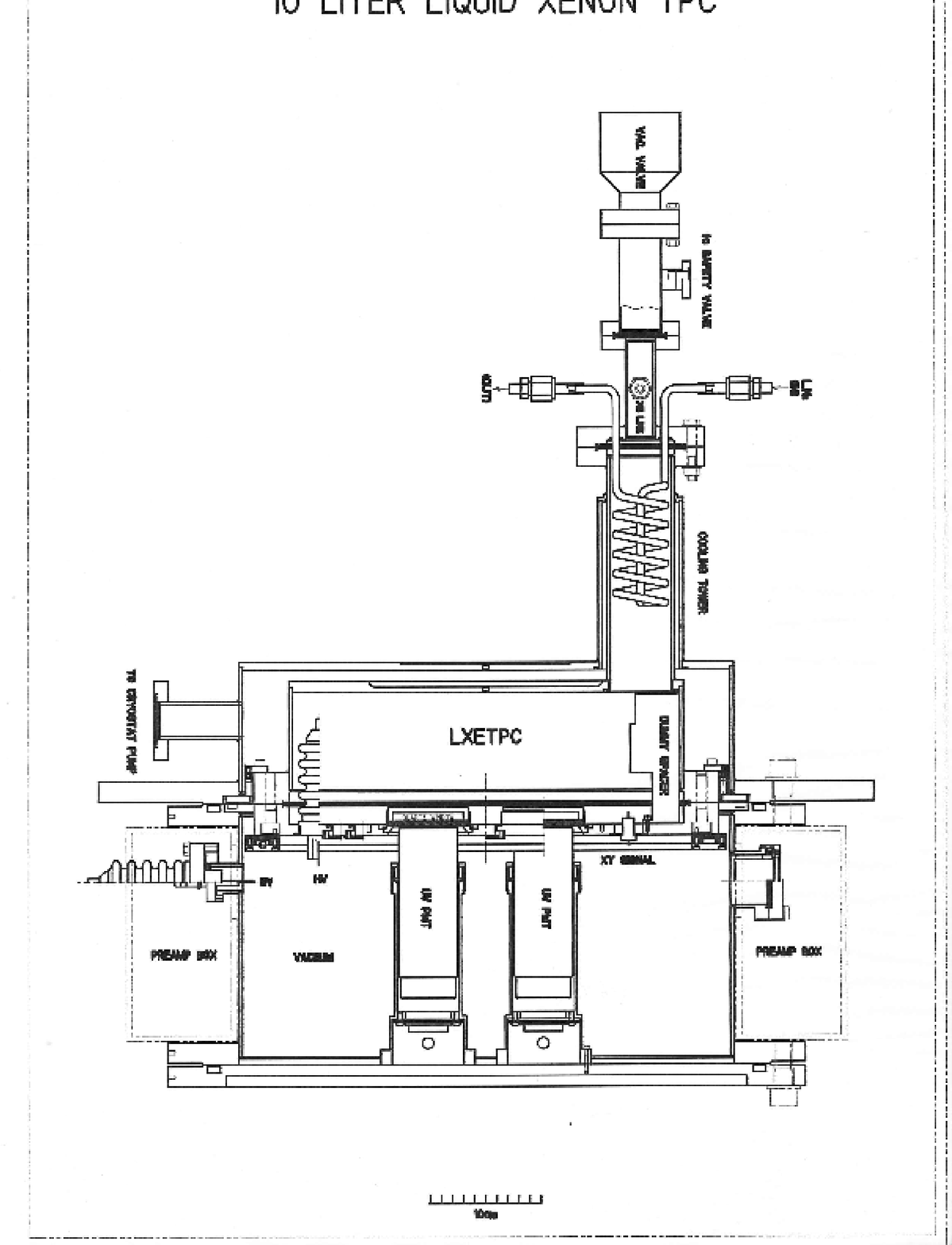}
\caption{Schematic of the LXeGRIT TPC \citep{EAprile:03:newastr}.}
\label{fig:TPC}
\end{figure}
\end{center}

\begin{center}
\begin{figure}
\includegraphics*[width=0.4\textwidth]{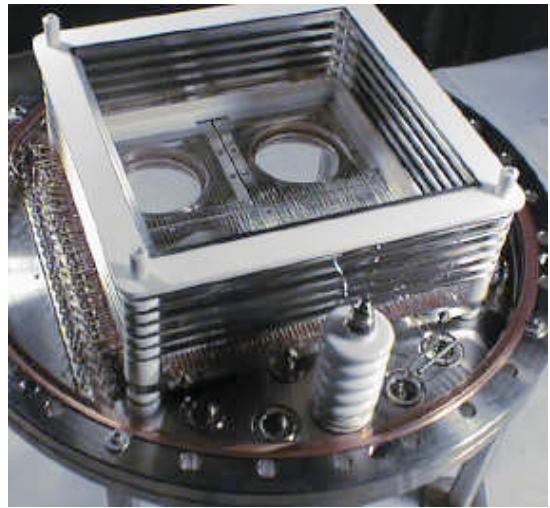}
\caption{The LXeGRIT TPC structure (cathode removed) \citep{EAprile:03:newastr}.}
\label{fig:TPC_images}
\end{figure}
\end{center}

A mechanical drawing of the TPC is shown in Figure \ref{fig:TPC}. The TPC electrodes are mounted on a 42 cm diameter flange (see Figure \ref{fig:TPC_images}) which encloses a 10 liter cylindrical vessel. Xe gas, purified to $<1$ ppb O$_2$ equivalent, by a combination of a hot metal getter and a cooled molecular sieve filter, is liquified into the vessel by a controlled flow of liquid nitrogen (LN$_2$) through the copper coil of a condenser on the top. Thermal insulation of the cold vessel is provided by a vacuum cryostat. To maintain the insulation and the cooling during a balloon flight, a sorption pump embedded in the custom-built LN$_2$ dewar carried on board was used. The liquid temperature of $\sim$-95$^o$C, corresponding to a pressure of 1.5 atm, was reliably maintained for periods of many days. As the launch date and time are not known a priori, the detector must be filled and operational well in advance (a
couple of days) of a launch opportunity. This requires a good purity of the gas, a low outgassing rate for the detector, and a reliable cooling system. We note that this first LXeTPC did not use continuous gas circulation through purifiers, yet long electron lifetime was achieved and the charge stability with time was verified both in laboratory testing and after the recovery of the payload. 

LXeGRIT was tested in balloon flights in 1999 and 2000 \citep{EAprile:03:spie:flight, ACurioni:03:spie:bgd, EAprile:03:newastr, ACurioni:07:NIM:LXeGRIT:eff}, measuring for the first time the internal and atmospheric background spectrum at balloon altitude, confirming the
comparatively low intrinsic background of a LXe instrument. Table \ref{t:lxegrit} shows the characteristics of the payload in the 1999 flight configuration.
\begin{center}
\begin{table}[htbp]
\caption{LXeGRIT Payload Characteristics.}
\label{t:lxegrit}

\vspace{\baselineskip}
\centering \sloppy 
\abovecaptionskip 5pt
\tabcolsep 1.8pt
\begin{tabular}{|ll|}\hline
\tabcolsep 1.8pt
\raggedright
Energy Range & $0.2$ -- $20$ MeV\\
Energy Resolution $(FWHM)$  & $8.8\%\times(1\mathrm{MeV}/E)^{1/2}$ \\
Position Resolution $(1\sigma)$ & 1 mm (3 dimensions)\\
Angular Resolution $(1\sigma)$ & 3$^o$ at $\sim$1.8 MeV \\
Field of View (FWHM) & 1~sr \\
Effective Area (Imaging) & 16~cm$^2$ @ 1MeV \\
LXeTPC Active Volume & 20 cm $\times$ 20 cm $\times$ 7 cm \\
Active Back Shield & 2730 cm$^2$, 10 cm thick NaI(Tl) \\
Active Side Shield & 4750 cm$^2$, 10 cm thick NaI(Tl)  \\
Active Top Shield & 1600 cm$^2$, 1.2 cm thick plastic \\
LN$_2$ Dewar & 100 liter \\
Instrument Mass,\enspace Power & 1100 kg,\enspace 450 W \\
Telemetry,\enspace Onboard Storage & $\sim$2 $\times$ 500 kbps,\enspace 
                                     $\sim$2 $\times$ 9 GB \\
\hline
\end{tabular}
\end{table}
\end{center}

\begin{figure} 
\centering
\includegraphics[width=0.48\textwidth,clip]{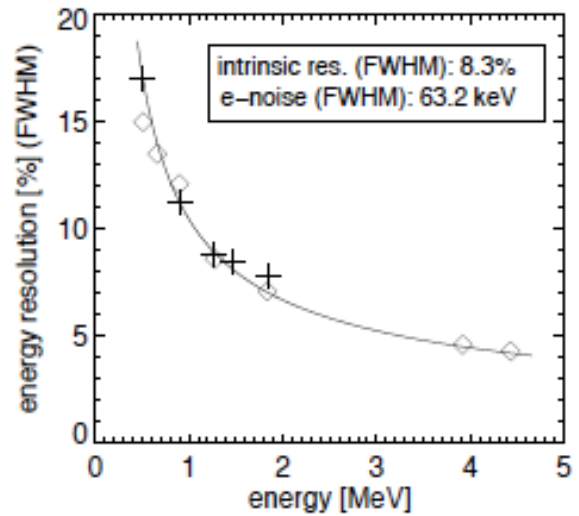}
\caption{\label{f:TPC_resolution} The energy resolution of LXeGRIT as a function of energy, as measured during two different balloon flights (crosses and diamonds).}
\end{figure}

\begin{figure} 
\centering
\includegraphics[width=0.48\textwidth,clip]{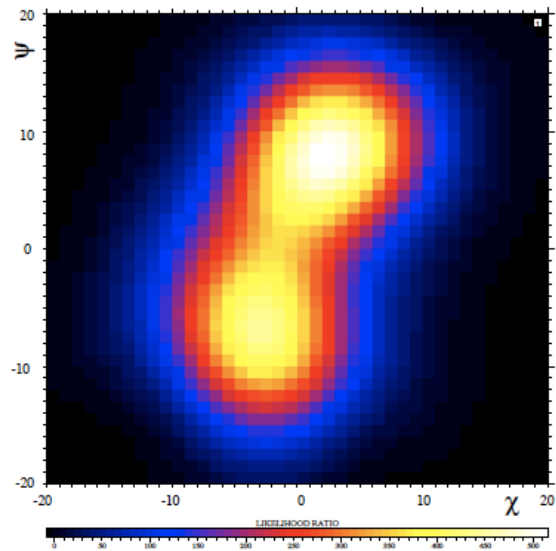}
\caption{\label{f:LXeGRIT_imaging}
LXeGRIT image (reconstructed angles in degrees) of two calibration sources, $^{60}$Co and $^{22}$Na, using a maximum likelihood imaging technique. Reprinted with permission from \citep{EAprile:2008:NIMA:imaging}.}
\end{figure}

A Compton telescope works by reconstructing statistically the arrival direction of multiple source gamma-rays from measurement of the Compton scatter angle and the direction of the scattered gamma-ray. This requires identification of the locations of the first Compton scatter location, and the second interaction point, including their sequence. For determination of the Compton scatter angle $\varphi$, the energy deposit in the first interaction point $E_1$ and measurement of the total gamma-ray energy are required, according to:
\begin{equation}
\cos\varphi= 1 - \frac{E_0}{E_\mathrm{tot}-E_1}+\frac{E_0}{E_\mathrm{tot}}
\end{equation}

where $E_0$ is the electron rest energy. The angular resolution is determined not only by the uncertainty in the locations of the first and second interaction points but also by the energy resolution of the detector. For best efficiency, the energy threshold should be as low as 50 keV, placing stringent requirements on the noise performance of the charge readout system. LXeGRIT achieved a position resolution of 0.85 mm ($1 \sigma$) in X/Y and 0.25 mm in Z, an energy resolution of 4.3\% ($1 \sigma$) at 1 MeV \citep{ACurioni:07:NIM:LXeGRIT:eff} (see  Figure \ref{f:TPC_resolution}), and an energy threshold for imaging of 150 keV (below 100 keV for spectroscopy). 
LXeGRIT demonstrated successful reconstruction of the interaction sequence for three or four interactions in the sensitive volume \citep{UOberlack:00:spie00:CSR} and imaging performance with 3$^o$ resolution ($1 \sigma$) at $\sim$1.8 MeV \citep{EAprile:2008:NIMA:imaging}. Figure \ref{f:LXeGRIT_imaging} shows an image of two calibration sources, imaged using a maximum likelihood technique. The angular resolution is dominated by the selection of scatter angles (large scatter angles were preferred due to the limited drift gap of $\sim$7 cm)  and the energy resolution of the instrument, whereas the position resolution for individual interactions plays a role mainly for scatter angles below 40$^o$.

LXeGRIT was truly a prototype, in that it successfully demonstrated the principle while also indicating paths for future improvements. 
Lacking powerful event rejection, the digital readout system turned out to be a bottleneck for the performance of the instrument \citep{EAprile:98.nim}. The light yield, based on the readout of 4 PMT's mounted outside the cryostat, was too limited to provide effective trigger cuts in energy, and could therefore not effectively discriminate against charged particles at the trigger level.

Various improvements over LXeGRIT have been suggested and tested in smaller setups, from enhanced light collection with PMTs in the liquid, to enhanced energy resolution based  on light-charge anti-correlation as discussed in section \ref{sec:fundamental}. Moreover, the fast scintillation light  should enable TOF in a compact double scatter Compton telescope with 10-15 cm separation between detector planes. This would not only result in powerful discrimination against background at the trigger level, by distinguishing between forward and backward scattered events, but also in greatly enhanced efficiency.   

\subsubsection{LXe Ionization Calorimeter for 10 to 100 GeV $\gamma$-rays}
\begin{center}
\begin{figure}
\includegraphics*[width=0.5\textwidth]{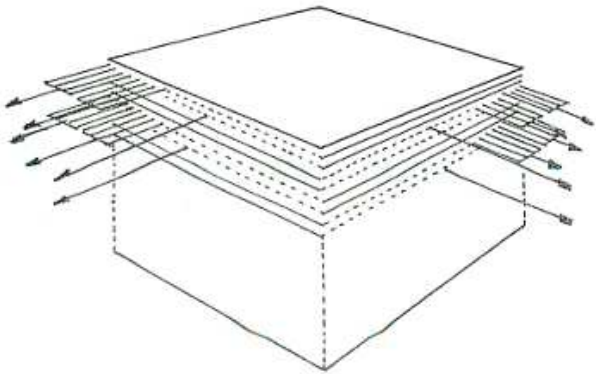}
\caption{Schematic of the wire electrode system for the LXe ionization calorimeter for GeV gamma rays \citep{Doke:1989,Doke:1992b}.}
\label{fig:Fig34}
\end{figure}
\end{center}

A design study of a LXe homogeneous calorimeter for astrophysical gamma ray lines in the 10 -100 GeV energy range was carried out in the early 1990 by the Waseda group \citep{Doke:1989,Doke:1992b}.  The goals were a geometric factor of $\sim$1 m$^2$sr,  requiring a LXe volume of $\sim$0.6 m$^3$ (1.7 tons) and an energy resolution better than 1\% ($\sigma$) for energies in the 3 -70 GeV range.
By comparison,  the energy resolution of the crystal calorimeter on board the recently launched Fermi observatory is about 10\% ($\sigma$) at 80 GeV \cite{Atwood:2009}.  A schematic drawing of the LXe calorimeter  is shown in Figure \ref{fig:Fig34} including a cross-section view of the wire electrodes.  Monte Carlo simulations have shown that an energy resolution $<0.5$\% ($\sigma$), a position resolution of 2.4 mm, and an angular resolution of 0.3-1.6 degrees, could be achieved for gamma-rays at normal incidence, in the 3-70 GeV energy range.  
With its large effective area and superior energy resolution, such a detector would be ideal to search for gamma ray lines in the range 10-100 GeV which may result from the annihilation of WIMPs in the Galactic halo (see next section for methods of direct detection of these). Fast timing is necessary for rejection of charged particle background, and scintillation light detection by fast PMTs or LAAPDs embedded in the calorimeter is needed for a reliable trigger of the gamma events.

While there are currently no plans for the realization of such a LXe telescope for a space mission, we believe that the advancement of efficient photodetectors operating in the liquid, and of efficient cryocoolers able to maintain the cryogenic liquid for extended periods of time, are two enabling technologies for this application. On the other hand,  the intrinsic long dead time limitation of this type of  calorimeter can be resolved by a LXe scintillating calorimeter, with energy and spatial resolution still superior to most other detector approaches. This has recently been in fact achieved by the MEG experiment discussed in section \ref{sec:meg}.

\subsection{Applications to Dark Matter Direct Detection}
\label{sec:DM}

A major experimental effort worldwide aims at the direct detection of Galactic dark matter, with a variety of detectors and methods \cite{Jungman:1996}. The motion of stars and gas in galaxies, including our own, indicates that the stars are immersed in a dominant component of non-luminous matter, 5 -10 times the total mass of the stars, and responsible for most of the gravitational binding. Other evidence indicates that this ``dark matter'' is non-baryonic and cold. The leading theoretical suggestion, based on supersymmetry theory, is that it consists of new particles (WIMPs) in the mass range 10 -1000 GeV. Such particles could be detected through collisions with targets of normal matter, producing nuclear recoils, with energies $<100$ keV. Xenon is a particularly suitable dark matter target because of its high atomic number (coherent weak interaction rates being proportional to A$^2$) and because it contains odd and even isotopes allowing a study of both spin-dependent and spin-independent interactions.

The principal experimental obstacle is that the  predicted interaction cross sections are typically in the range 10$^{-6}$-10$^{-10}$ pb (1pb = 10$^{-36}$ cm$^2$) giving event rates $\sim$1$\times$10$^{-4}$/kg/day in a Xe target, compared with ambient background rates typically 10$^6$ times higher from radioactivity in detector materials, shielding structures and the surrounding environment, as well as from cosmic ray muons. The latter can be reduced by operating detectors deep underground, while the former can be reduced by using high-purity detector and shielding materials, veto techniques, and detectors with discriminating power against gamma, electron (beta decay) and alpha background. 
To achieve the large mass and the low energy threshold which is required for Xe as a WIMP target, the favorite approach is a TPC operated in two-phase (liquid/gas), originally proposed by Barabash and Bolozdynya in 1989 \citep{Barabash:1989} and discussed in subsection \ref{sec:two-phase}. Simultaneous detection of both ionization and scintillation signals provides greater information than detection of either ionization or scintillation alone. Both signals are typically detected with photomultipliers placed around the target region. The direct scintillation signal is referred to as S1 and, the ionization signal turned into a proportional scintillation signal is referred to as S2, as seen from Figure \ref{fig:FigXenon_principle}. The ratio S2/S1 is found to differ significantly for nuclear and electron recoils (the latter from gammas or beta decay) and a 
mixture of signal and background separates into two populations of events with an overlap of only 0.1-1\%, dependent on the geometry and electric fields. The event energy can be obtained, after calibration with gamma sources, from the amplitude of the S1 signal, while the event position can be reconstructed in the transverse \textit{x}-\textit{y} plane from the distribution of the S2 signal in the PMT array, and in the vertical direction from the time difference between the prompt S1 signal and the ionization drift time. Thus the two-phase XeTPC approach  provides full 3-D event localization with a position resolution varying from millimeter to several centimeters, determined largely by the granularity of the PMTs array. The energy threshold  is determined by the collection efficiency of the direct light.

\begin{center}
\begin{figure}
\includegraphics*[width=0.4\textwidth]{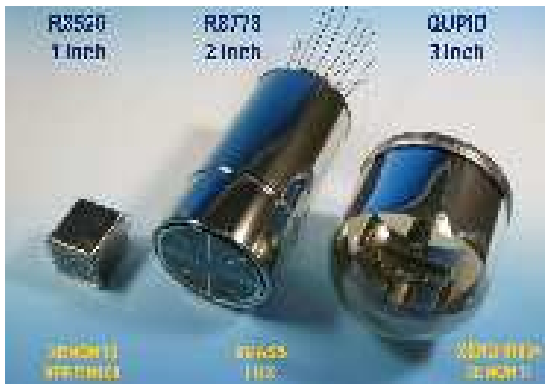}
\caption{A photo of the different Hamamatsu PMTs developed for the XENON and XMASS dark matter experiments.}
\label{fig:xenon_pmt}
\end{figure}
\end{center}

Four collaborations are currently using or developing LXe detectors for dark matter: the XENON, ZEPLIN and LUX collaborations have chosen the two-phase TPC approach, while the XMASS collaboration has chosen the single-phase detector approach, using only the scintillation signal. The XENON detectors have so far adopted  a  geometry  with the drift length similar to the detector diameter, and with PMT arrays in the gas and in the liquid, to achieve a very low energy threshold for nuclear recoils. A moderate field of $\sim$1 kV/cm is used and by adopting small size metal channel PMTs with 1'' square geometry, good photocathode coverage for maximum S1 signal detection and high spatial resolution with S2 signal detection  is emphasized. For larger scale devices, radioactivity and cost considerations have led to a design based on a larger size (3'' spherical shape) novel type PMT, the QUartz Photon Intensifying Detector (QUPID), developed by UCLA and Hamamatsu \citep{Arisaka:2008}. Figure \ref{fig:xenon_pmt}  shows a photo of the QUPID device, in comparison with the other special PMTs developed by Hamamatsu for the XENON and XMASS experiments.  

The ZEPLIN detectors have adopted  a single array of  large size PMTs. ZEPLIN II used  3'' diameter PMTs placed  in the gas phase, resulting in comparatively low light yield due to the loss of light by total internal reflection at the liquid/gas interface. ZEPLIN III focused on high light yield, viewing the sensitive volume from below with an array of 2'' diameter  PMTs, and on high field operation, resulting in a shallow geometry, with a short drift gap and hence a more limited mass of LXe. 

The LUX detector, currently in development, follows the XENON design with larger size PMTs both in liquid and gas. A taller geometry is used to allow a large cut in the drift direction, required to suppress backgrounds from the PMTs radioactivity.  Finally, XMASS uses a single  large volume of LXe surrounded on all sides by PMTs,  to efficiently detect the scintillation light for both energy and position measurement. The original 2" diameter PMT developed by Hamamatsu for XMASS (R8778) has been modified into an "hexagonal" geometry to enable better photocathode coverage of a near spherical volume of LXe. 

The sensitivity to dark matter is dependent  on target mass, running time, achieved background levels, and energy threshold.  So far, no signal has been seen in a few months running of any of the detectors.  These searches have chosen to make faster and more cost-effective progress by upgrading or constructing new detectors to achieve improvements of an order of magnitude at each step. Thus the detectors reviewed below are often in the form of a series of detectors, progressively improving the sensitivity limits, using larger target masses and/or improvements in background. 

The measure of the sensitivity of a given experiment is normally given by the 90\% confidence upper limit on the hypothetical WIMP-nucleon dark matter cross section.  The determination of this signal upper limit, in the presence of residual background, is more difficult than in the case of a measured positive signal. There is no unique definition of ``90\% confidence'' when one wishes to look both for a non-zero signal and at the same time find an upper limit \citep{Feldman:1998}. There is also a difference of definition between a ``one-sided'' or ``two-sided'' limit \citep{Feldman:1998} so that the quoted value from a given experiment can vary by a factor 2 depending on the method and definition used, and in particular on the method used to take account of residual backgrounds.  

\subsubsection{The XENON Detectors}
\label{sec:XENON}
The XENON Dark Matter Search is a phased program  aiming at progressively improved sensitivity by the series of two-phase TPCs, XENON10, XENON100, and XENON1000, the numbers referring to the order of magnitude of fiducial target mass in kg. 

\begin{center}
\begin{figure}
\includegraphics*[width=0.37\textwidth]{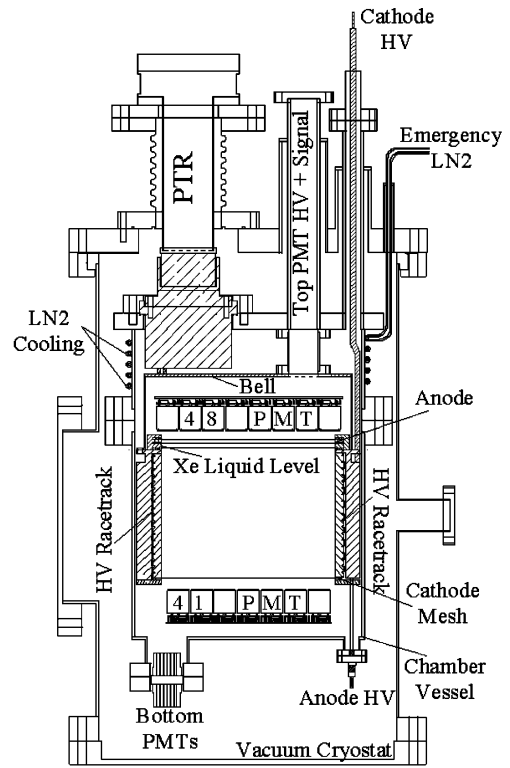}
\caption{Schematic of the XENON10 detector \cite{Aprile:2009_instrument}.}
\label{fig:xenon10_sch}
\end{figure}
\end{center}

The XENON10 experiment \citep{Angle:2008} contained a total of 15 kg Xe within a PTFE cylinder of 20 cm inner diameter and 15 cm height. PTFE is used for its high reflectivity in the VUV regime \citep{Yamashita:2004}. Figure \ref{fig:xenon10_sch} shows the detector's schematic. The TPC was equipped with four meshes: the cathode and one mesh below the liquid surface set up the drift field, together with field shaping rings in the cylinder. One mesh just above the liquid surface is kept at positive high voltage, providing a field of $\sim$10 kV/cm for fully efficient electron extraction. A fourth mesh on top serves to close the field lines. The active volume was observed by top and bottom arrays totaling 88 (Hamamatsu R8520-06-Al) PMTs, selected for low U/Th radioactivity. The majority of the S1 signal is seen by the lower array (located in the liquid), due to total internal reflection from the liquid surface, while the S2 signal (from the gas) is seen mainly by the top array. The light yield in XENON10 was 2.2 photoelectrons/keV allowing 5 keV nuclear recoil energy threshold. To eliminate background events at the edges of the detectors, the volume was fiducialised (by radial, top, and bottom software cuts) to a mass of 5.4 kg. The maximum drift length in the liquid was 15 cm, and the drift field 0.7 kV/cm.  

\begin{center}
\begin{figure}
\includegraphics*[width=0.45\textwidth]{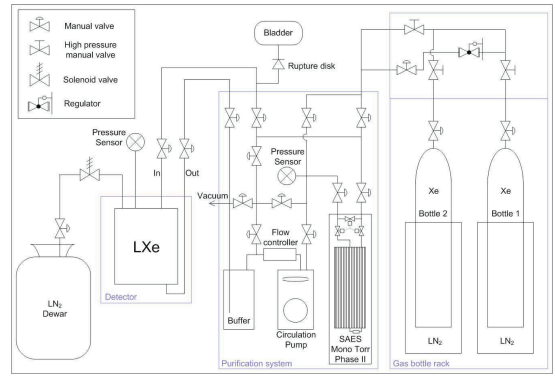}
\caption{The XENON10 detector purification system \cite{Aprile:2009_instrument}.}
\label{fig:xenon10_purif}
\end{figure}
\end{center}

XENON10 was cooled by a 65 W pulse tube refrigerator (PTR) \citep{Haruyama:2004} coupled to the liquid volume via a copper head in the top of the detector, as shown in Figure \ref{fig:xenon10_sch}. Xenon was purified in the gas phase by continuous circulation through a commercial hot getter. An electron lifetime in excess of 2 ms was achieved with this system. Figure \ref{fig:xenon10_purif} shows a schematic of the purification system. 
The experiment was located in the Gran 
Sasso National Laboratory at a depth of 3800 meters water equivalent (m.w.e.). The XENON10 cryostat, made of stainless steel, was enclosed in a structure made of 20 cm thick lead and 20 cm thick polyethylene, to shield it from gamma-rays and neutron background.   
After calibration with gamma and neutron sources, the experiment ran for two months, obtaining in the predefined signal region a background of 6 events as expected from the gamma background, and 4 spurious events which could be rejected as possible signal events. The corresponding dark matter cross section limits were 5$\times$10$^{-8}$ pb (1-sided) and 1$\times$10$^{-7}$ pb (2-sided) at a WIMP mass of 30 GeV, and about a factor 2 higher at 100 GeV. The XENON10 experiment, together with the ZEPLIN II experiment described below, have proven the concept of reliable XeTPC operation for dark matter searches.

\begin{center}
\begin{figure}
\includegraphics*[width=0.5\textwidth]{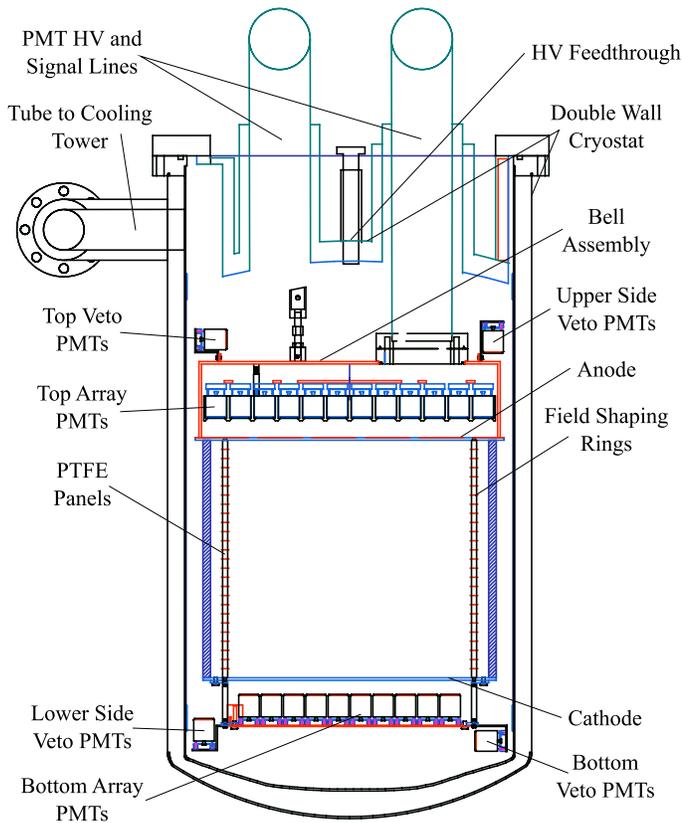}
\caption{Schematic of the XENON100 detector \citep{IDM08:XENON100}.}
\label{fig:xe100_photo}
\end{figure}
\end{center}

\begin{figure}[htbp]
	\begin{center}
		\begin{tabular}{cc}
			\includegraphics[width=0.265\textwidth]{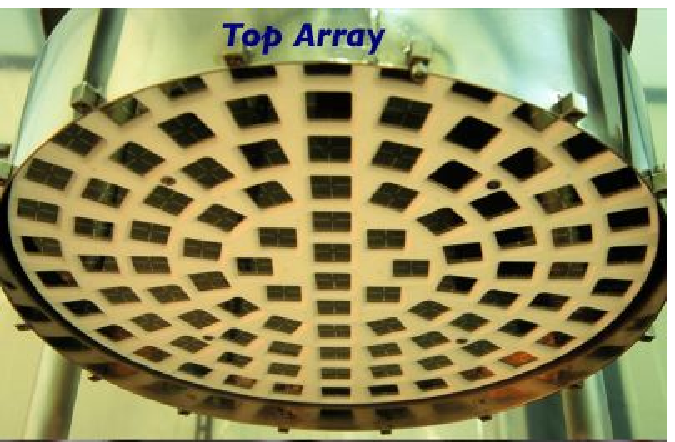} &
			\includegraphics[width=0.215\textwidth]{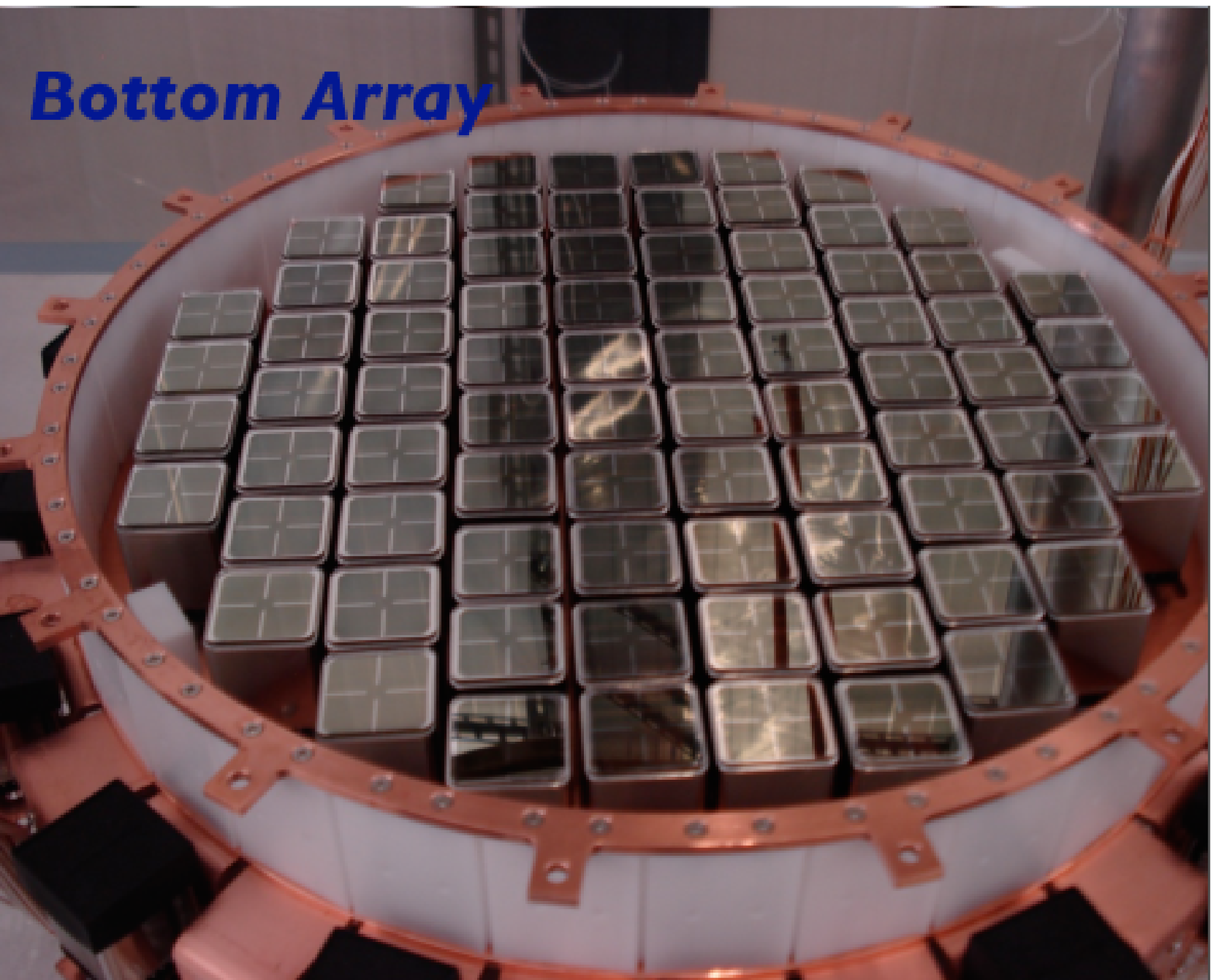}
		\end{tabular}
		\caption{The top and bottom PMT arrays of the XENON100 detector.}
		\label{fig:xe100_pmt}
	\end{center}
\end{figure}

The XENON100 detector  has replaced XENON10 and is now operating underground at the Gran Sasso laboratory. The new TPC incorporates the successful design features of XENON10, but aims at improving the background rate by a factor 100 by the use of (a) lower background version of the same type of PMT (Hamamatsu R8520-06-Al-mod), (b) reduction of the mass and radioactivity of the cryostat, including selection of low-activity stainless steel, (c) improved screening of other construction materials, and (d) a new cryogenic system designed to remove the refrigerator system from the vicinity of the experiment. In addition, an active LXe veto was added around the target volume. The active veto and target volume are separated by a PTFE cylinder. The target is viewed from the top and the bottom by a total of 178$\times$1'' low background PMTs.  The Xe veto region is independently viewed by 64 PMTs, from the XENON10 detector. The XENON100 TPC has a drift length of 30 cm and requires a total of $\sim$170 kg of LXe of which $\sim$70 kg are in the target. The LXe is contained in a double wall stainless steel vessel and cooling is provided by a 170 W PTR, the same type as used on the MEG detector discussed later in this chapter \citep{Haruyama:2004}. The PTR is used both to liquefy the gas and to maintain the liquid temperature during operation. Figure \ref{fig:xe100_photo} shows a schematic of the XENON100 TPC and its overall assembly in the vacuum cryostat. Figure \ref{fig:xe100_pmt} shows the assembled top (a) and bottom (b) PMT arrays used in XENON100. 

\begin{center}
\begin{figure}
\includegraphics*[width=0.45\textwidth]{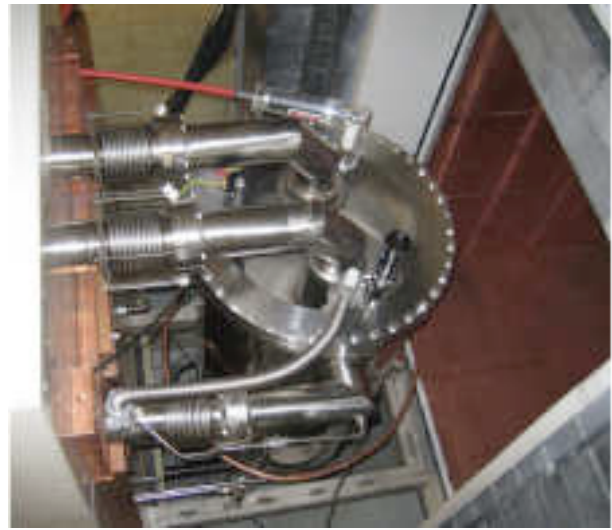}
\caption{The XENON100 detector installed in its shield \cite{IDM08:XENON100}.}
\label{fig:xe100_shield}
\end{figure}
\end{center}

\begin{center}
\begin{figure}
\includegraphics*[width=0.4\textwidth]{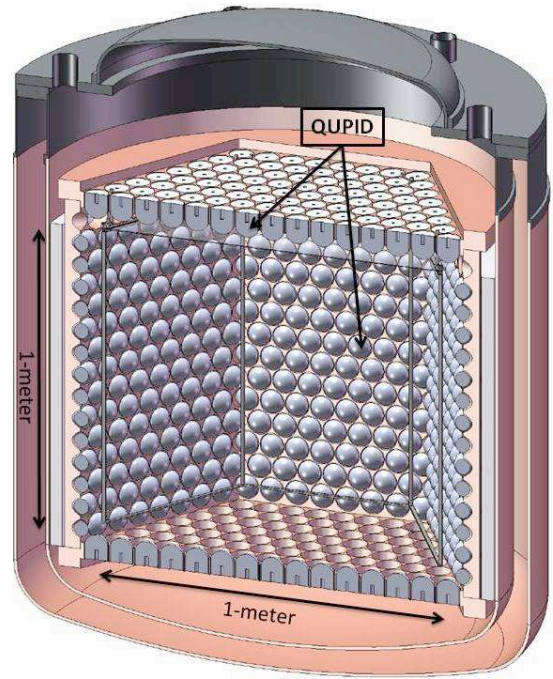}
\caption{Schematic  of the XENON1T detector.}
\label{fig:xenon1t}
\end{figure}
\end{center}

The detector is enclosed in a passive shield of polyethylene and lead to reduce the neutron and gamma background. Figure \ref{fig:xe100_shield} shows the assembled XENON100 mounted inside the open shield cavity. From the  improvements in design and materials selection,  the background objective is to reach $<10^{-3}$ gammas/kg/d/keV, and $<1$ neutron/100 kg/y, in 50 kg fiducial mass. High energy neutrons from muon spallation in the rock are calculated to be $<0.3$/100kg/y. The intrinsic rate of electron recoils from radioactive $^{85}$Kr in Xe will be reduced to 10$^{-4}$ events/kg/d/keV, corresponding to a Kr/Xe level of 5 parts per trillion (ppt), using a dedicated cryogenic distillation column. After full commissioning and calibration with neutron and gamma sources, this detector will become operational for science runs in late 2009, aiming at a spin-independent sensitivity of 2$\times$10$^{-9}$ pb for 100 GeV WIMPs. This large gain over XENON10 is achievable partly through the increase in fiducial mass and partly through the reduction in backgrounds, while maintaining the same low energy threshold. An upgrade of XENON100 is envisaged by doubling the target fiducial mass to 100 kg, and further reductions in background. With a full two years of run time, this could reach sensitivity levels in the range of 10$^{-9}$ -10$^{-10}$ pb. 

If a dark matter signal is seen at this cross section level, there will be a clear need to obtain a sufficient number of events to make a spectrum, from which the WIMP mass could be estimated. Studies of a larger 1 ton detector are already in progress, and the XENON100 phase provides the testing of many of the key technologies and performance characteristics relevant to the realization of 1-10 ton Xe detectors. To achieve the necessary further background reduction, an increased amount of Xe self-shielding  and a full $4\pi$ coverage with QUPID photodetectors \citep{Arisaka:2008} is planned (Figure \ref{fig:xenon1t}). 

Recent Monte Carlo studies indicate that for an adequate amount of self shielding, a xenon TPC with an inner fiducial volume of 1 ton would require a total mass of 2.4 tons. In addition to the intended capability of measuring a spectrum of events at 10$^{-10}$ pb, this detector would have an ultimate sensitivity of 10$^{-11}$ pb. Studies have also been made of a multi-ton, two target system (Xe and Ar) for dark matter, double beta decay, and pp solar neutrinos \citep{Arisaka:2008}. 

\subsubsection{The ZEPLIN Detectors}
\begin{figure}
\includegraphics[width=0.3\textwidth]{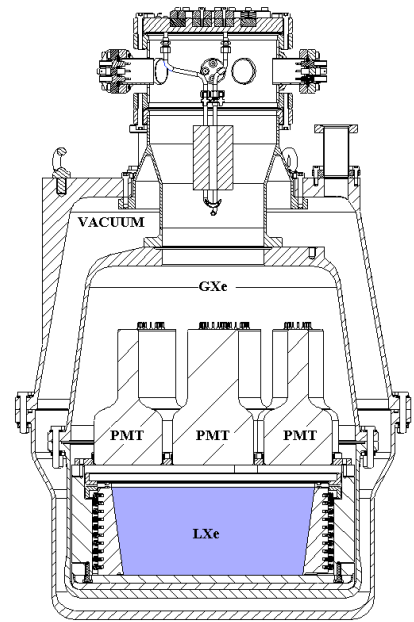}
\caption{Schematic of the ZEPLIN II detector \citep{Alner:2007}.}
\label{fig:zep_ii}
\end{figure}

The ZEPLIN collaboration has been operating LXe detectors for dark matter at the UK Boulby Mine underground site, already prior to the XENON collaboration. After ZEPLIN I, a 3 kg single phase detector that reached a sensitivity 10$^{-6}$ pb in 90 days of operation, the collaboration focused on constructing the first two-phase Xe TPC. ZEPLIN II used a 35 kg LXe target  viewed by seven 3''  low background PMTs (ETL D742) located in the gas above the liquid (Figure \ref{fig:zep_ii}). The electron drift distance was 14 cm, similar to that of XENON10.  Shielding was 20 cm lead outside of a 30 cm thickness liquid scintillator active veto, which provided both passive neutron shielding and active gamma shielding to remove residual gammas penetrating the lead shield. The target region was defined by a tapered PTFE annulus serving as a light reflector and support for electric field-shaping rings. Grids above and below the liquid provided a drift field of 1kV/cm in the liquid, and a third grid in the gas provided an 8 kV/cm field to extract electrons from the liquid with 90\% efficiency. The light yield was limited to 1.5 photoelectrons/keV, giving an energy threshold of 5 keV for  electron recoils. The whole assembly was located in the 2800 m.w.e. Boulby salt mine. The radon-related wall events had not been anticipated at that time, and to remove these necessitated a quite severe software cut, reducing the mass to an inner 7 kg fiducial region, and hence reducing the effective target mass and sensitivity by a factor 4. Nevertheless, a 31 day run produced a (90\%, 2-sided) lower limit of zero and upper limit of 10 signal events in a 50\% efficiency event region between 5 and 15 keV, which translates into a dark matter cross section 90\% upper limit (2-sided) of 7$\times$10$^{-7}$ pb at 60 GeV WIMP mass, or a 1-sided upper limit of 4$\times$10$^{-7}$pb \citep{Alner:2007}.

\begin{center}
\begin{figure}
\includegraphics*[width=0.5\textwidth]{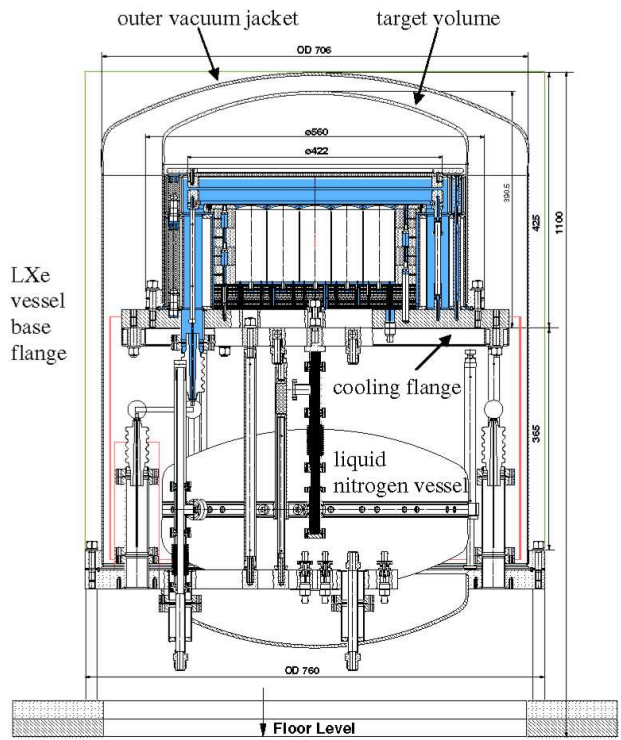}
\caption{Schematic of the ZEPLIN III detector \citep{Lebedenko:2008}.}
\label{fig:ziii_detector}
\end{figure}
\end{center}

Another detector based on the same two-phase TPC principle, but aiming at a lower energy threshold and higher field operation, was developed by the same collaboration in parallel with ZEPLIN II. The completed ZEPLIN III detector (See Figure \ref{fig:ziii_detector}) was deployed underground in 2008. It uses a thin layer of liquid Xe (thickness 3.5 cm, diameter 38 cm) viewed by an array of 31$\times$2''  PMTs (ETL 730) immersed in the liquid, to achieve a light collection $\sim$3 photoelectron/keV.
In addition, a higher drift field (several kV/cm) can be used with the shallower target, producing a larger separation between the gamma and nuclear recoil populations (an overlap of only 0.1\%, compared with 1\% for ZEPLIN II). The fiducial mass, defined by the outer radius for which position reconstruction was possible, was 7 kg. The detector was set up in the same shielding system as that for ZEPLIN II. A three-month run produced 7 events close to the edge of an otherwise empty 50\% signal region between 2 and 12 keV. These events are compatible with being background, and using a likelihood technique, a 90\% 2-sided WIMP cross-section limit of 7$\times$10$^{-8}$ pb was achieved \citep{Lebedenko:2008}.

ZEPLIN III is being upgraded by the substitution of lower activity PMTs, and by the addition of an active Gd-loaded hydrocarbon veto. The anticipated reduced background is expected to enable the background-free signal region to be increased from 50\% to 70\%, with a possible reduction in electron recoil energy threshold to 1.5 keV. A 1-2 year run should improve the ZEPLIN III WIMP cross-section sensitivity to 10$^{-8}$ pb. 

\subsubsection{The LUX Detector}
\begin{center}
\begin{figure}
\includegraphics*[width=0.3\textwidth]{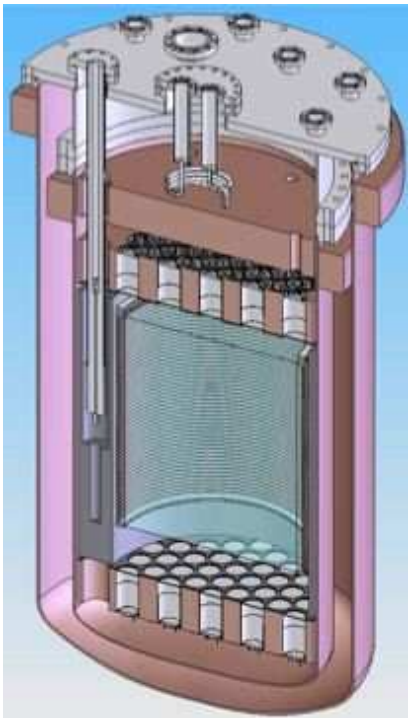}
\caption{Schematic of the LUX detector \cite{IDM08:LUX}.}
\label{fig:FigLUX}
\end{figure}
\end{center}
The LUX experiment \cite{IDM08:LUX} aims at  deploying a 100 kg fiducial mass, two-phase XeTPC  in a water shield to be constructed in the 4800 m.w.e. Sandford Underground Laboratory of the Homestake Mine, in South Dakota. Homestake has been selected as the site for a Deep Underground Science and Engineering Laboratory (DUSEL) in the USA, and is currently under study.  The LUX detector adopts the same principle as the other two-phase detectors, but makes greater use of self-shielding to reduce the large  background contributed by the PMTs in the fiducial region. 200 kg out of the 300 kg total mass is used for self-shielding. The TPC volume is viewed by 120 Hamamatsu R8778 PMTs, located above and below the target region (Figure \ref{fig:FigLUX}). This 2'' diameter PMT was originally developed for the XMASS 100 kg prototype, discussed next. It features high quantum efficiency ($>30$\%) and low radioactivity, compared to other VUV PMTs such as those used in ZEPLIN III, and can operate at LXe temperature. 

The goal of this experiment is to reach a WIMP cross section sensitivity better than $\sim$7$\times$10$^{-9}$ pb in 1 year of operation \citep{IDM08:XENON100}. A 50 kg prototype with four PMTs has been built and is currently being tested above ground. 

\subsubsection{The XMASS Detectors}

\begin{center}
\begin{figure}
\includegraphics*[width=0.3\textwidth]{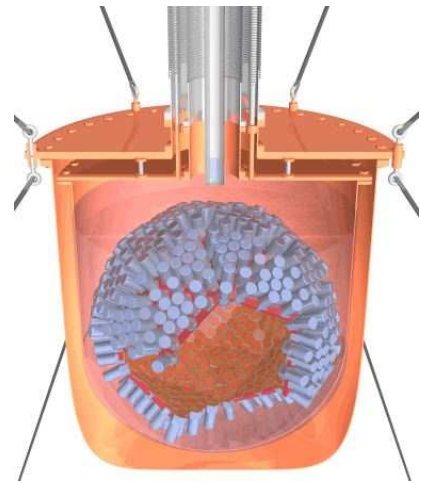}
\caption{Schematic of the XMASS detector \citep{IDM08:XMASS}.}
\label{fig:xmass800}
\end{figure}
\end{center}
\begin{center}
\begin{figure}
\includegraphics*[width=0.3\textwidth]{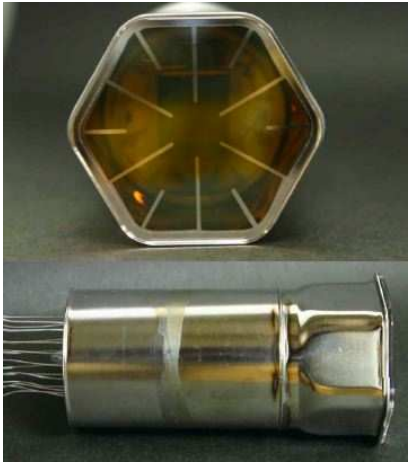}
\caption{The hexagonal PMT of the XMASS detector.}
\label{fig:xmass800_pmt}
\end{figure}
\end{center}

The XMASS experiment, proposed in 2000 aims at neutrinoless double beta decay, pp and $^{7}$Be solar neutrinos detection, in addition to dark matter direct detection, with a 10 ton LXe detector based solely on scintillation detection. The strategy is to use the excellent self-shielding capability of the liquid to achieve 
an effectively background-free inner volume. In the first phase of the XMASS program, a 100 kg prototype for dark matter detection was built and deployed at the Kamioka underground laboratory, in Japan. 

The detector, in the form of a cube of 30 cm side, was contained in a low activity copper vessel, further shielded for gamma and neutron background reduction. To detect the scintillation of LXe, 54$\times$2'' low-background Hamamatsu R8778 PMTs were coupled to the copper vessel with MgF$_2$ windows. The PMT photocathode coverage was 17\%.
A picture of the PMT developed for XMASS is shown in Figure \ref{fig:xenon_pmt}, together with the other PMTs developed  for XENON.  Position reconstruction with a vertex resolution of $\sim$6.5 cm for 10 keV events was demonstrated and used to reduce the gamma background from the PMTs activity. To reduce the $^{85}$Kr contamination in Xe, a cryogenic distillation system was developed and successfully used to achieve a Kr/Xe level below 3 ppt \cite{Abe:2008}.

The second phase, currently under development \cite{IDM08:XMASS}, consists of a 857 kg LXe  detector with the main physics goal of detecting dark matter (Figure \ref{fig:xmass800}). 
The large liquid volume is viewed by 642 hexagonal PMTs (Figure \ref{fig:xmass800_pmt}), arranged in an approximately spherical shape with 67\% photocathode coverage of the inner surface of the detector. Simulations predict a light yield of 8 photoelectrons/keV. If the new PMTs achieve their specification of a factor 10 reduction in U/Th activity compared with their predecessors, the expected reduction of background events with radius should leave a central mass of 100 kg free of background, for at least a 1 year running period. The whole experiment would be installed in the new underground area in the Kamioka mine, and surrounded by a water shield to remove external gamma and neutron background. The projected dark matter cross section limit would reach 10$^{-9}$ pb for a running time of 5 years.  
The advantage of a single phase detector like XMASS, compared with the previously-discussed two-phase detectors, is clearly the simpler design, but this comes at the expense of a large amount of non-fiducial liquid Xe. Without the diagnostic value of the ionization signal, the whole experiment relies on absolute suppression of both external and internal backgrounds. 

\subsection{Applications to Particle Physics}
\subsubsection{Prototypes of LXe Ionization and Scintillation Calorimeters}
\begin{center}
\begin{figure}
\includegraphics*[width=0.5\textwidth]{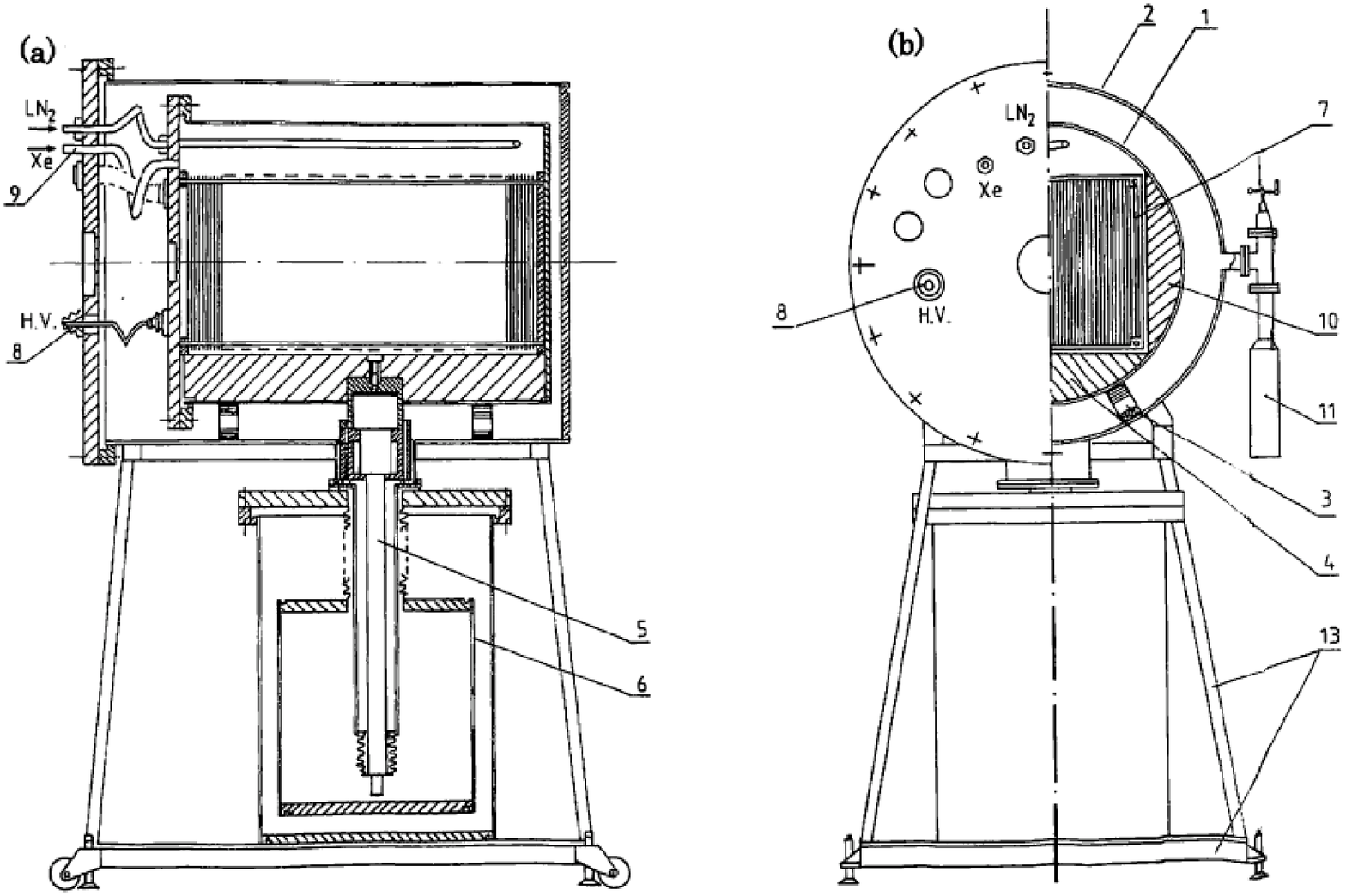}
\caption{(a) and (b), Schematic of the Russian LXe ionization calorimeter: inner (1) and outer (2) cylindrical vessel, (3) multi-plate supports, (4) and (10) aluminum segments, (5) copper bar, (6) cryostat with LN$_2$, (7) electrode system, (8) HV connector, (9) Xe gas supply tube, (11) zeolite trap \citep{Baranov:1990}.}
\label{fig:Fig35}
\end{figure}
\end{center}

\begin{figure}[htbp]
	\begin{center}
		\begin{tabular}{cc}
			\includegraphics[height=0.231\textwidth]{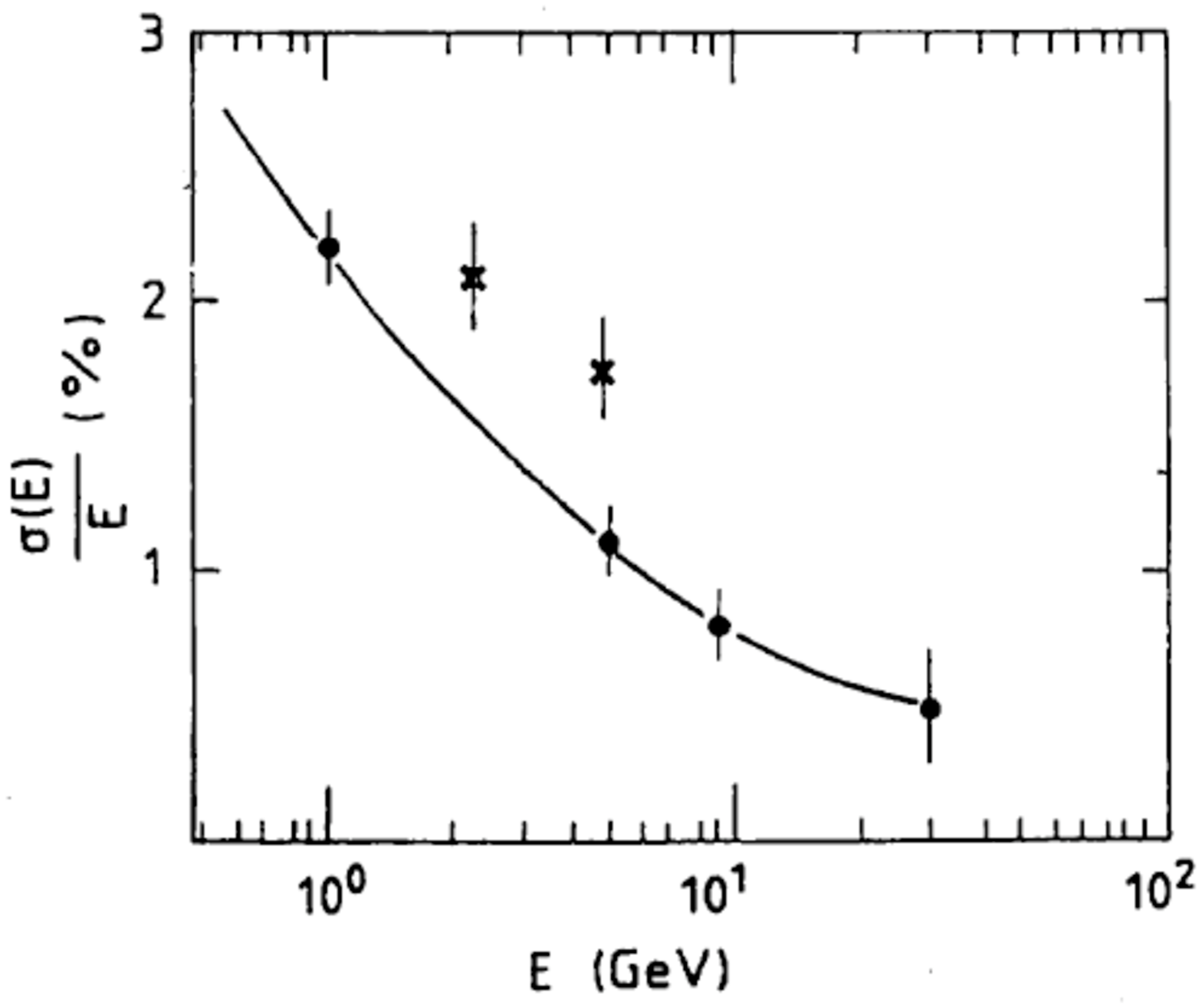} &
			\includegraphics[height=0.239\textwidth]{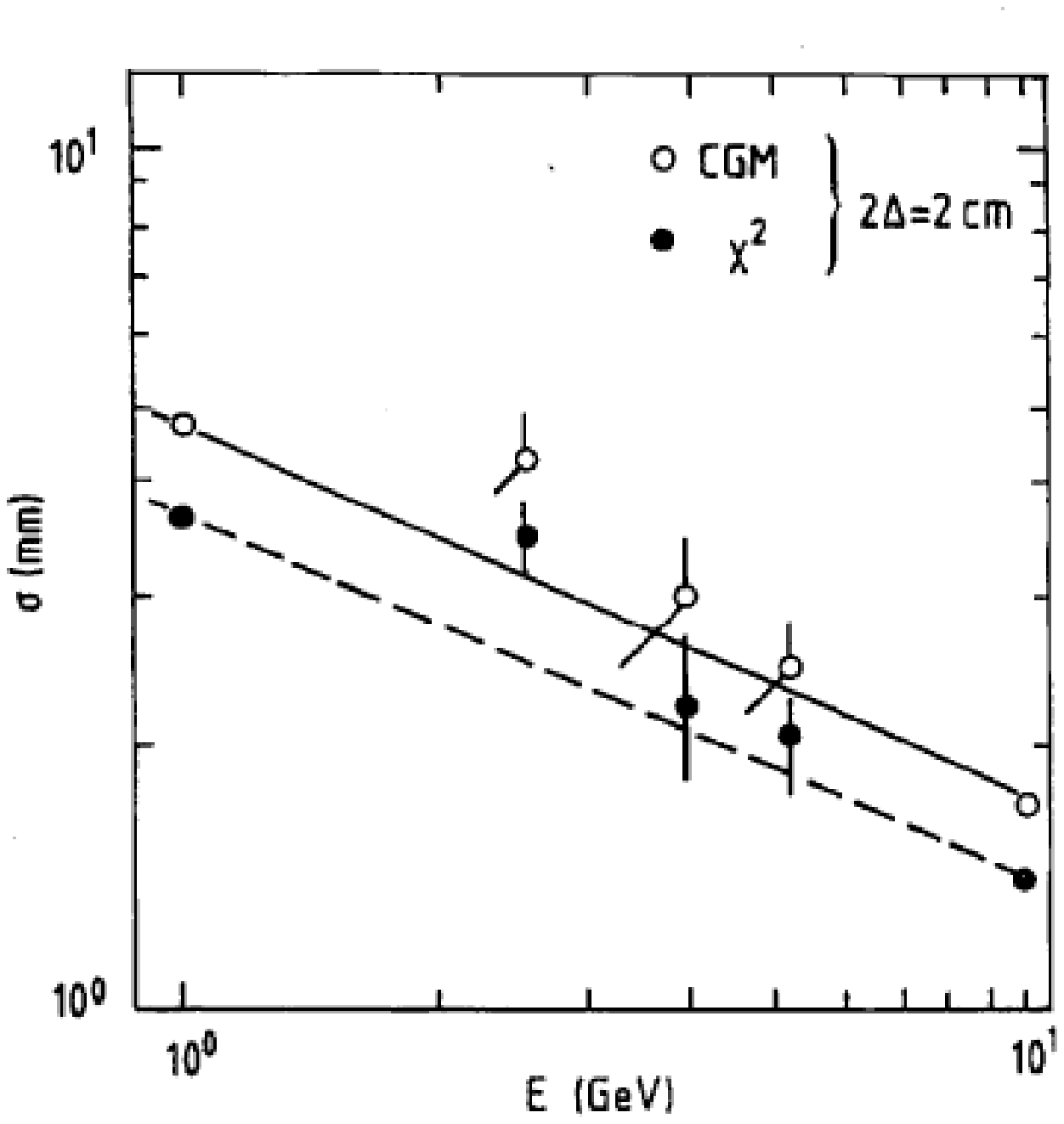}
		\end{tabular}
		\caption{(left) Energy resolution dependence on electron energy. Crosses are the experimental points and circles are points from Monte Carlo simulations. (right) Position resolution dependence on electron energy based on two different reconstruction algorithms. Full and open circles are data for two different position reconstruction algorithms, and the dashed and solid lines the corresponding predictions from simulation \citep{Baranov:1990}.}
		\label{fig:Fig36}
	\end{center}
\end{figure}

Due to its high cost, LXe has not found widespread use in calorimeters for high energy physics experiments, despite the superior energy resolution and compact scale of a LXe ionization calorimeter, as discussed in \ref{sec:ion_chamber}. Two prototypes have been built to date. The first  was built by a Russian group \citep{Baranov:1990} and is shown in Figure \ref{fig:Fig35}. It is 25 cm in diameter and 50 cm in length. Two different electrode configurations were tested, one a multi-plate type for energy measurement, and the other consisting of strip electrodes for position measurement.  Figure \ref{fig:Fig36} compares measured results with a Monte Carlo simulation. The measured resolutions in the range 1- 6 GeV were $\sigma_E$/\textit{E}  = 3.4/\textit{E}$^{0.5}$\% for energy and $\sigma_x$/\textit{E}  = 4.6/\textit{E}$^{0.5}$ mm for position, with \textit{E} in GeV. 

\begin{center}
\begin{figure}
\includegraphics*[width=0.5\textwidth]{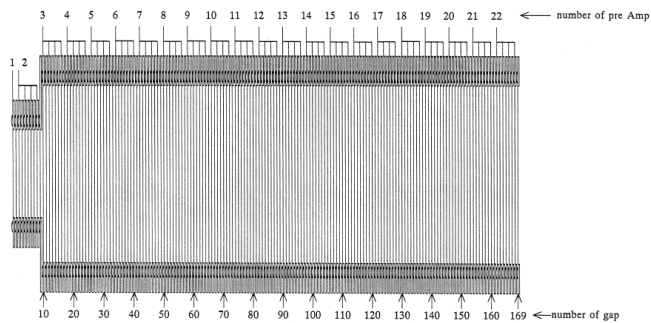}
\caption{Schematic of the Waseda multi-plate LXe ionization calorimeter \citep{Okada:2000}.}
\label{fig:Fig37}
\end{figure}
\end{center}

The second homogeneous liquid Xe calorimeter prototype was built by the Waseda group \citep{Okada:2000} and tested in a high energy electron beam at CERN. Figure \ref{fig:Fig37} shows the electrode system, consisting of 169 cells and their readout arrangement. An energy resolution of 0.9 \% ($1 \sigma$) was measured with 70 GeV electrons, close to the simulated value of 0.7 \%.  

In 1988, a LXe scintillation calorimeter with fast response, excellent energy and position resolutions, and high radiation resistance was proposed \citep{Chen:1988} for an experiment at the Superconducting Super Collider (SSC), under discussion at that time. The design \citep{Radermacher:1992}, included a long barrel homogeneous calorimeter and two end-cap sampling calorimeters. 

\begin{center}
\begin{figure}
\includegraphics*[width=0.5\textwidth]{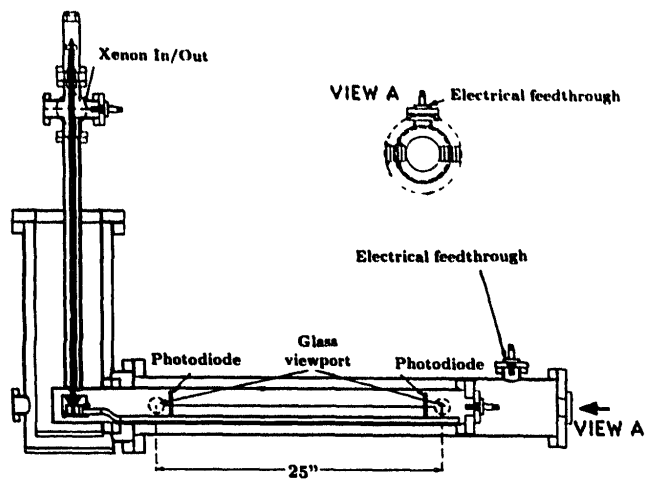}
\caption{Schematic of a full length liquid xenon/krypton cell for uniformity tests with mirrors and diodes \citep{Chen:1993}.}
\label{fig:Fig40b}
\end{figure}
\end{center}

A full length (66 cm) LXe prototype cell with a volume of 4 liters was built to investigate the light collection uniformity in the homogeneous section of the calorimeter (see Figure \ref{fig:Fig40b}). The cell walls, made of MgF$_2$-coated Al, were used as UV reflectors with 85-90\% reflectivity. Three large area, UV-sensitive, silicon photodiodes with mesh 
type surface electrode, detected the LXe scintillation \cite{Kashiwagi:1993}. 
 
\begin{center}
\begin{figure}
\includegraphics*[width=0.5\textwidth]{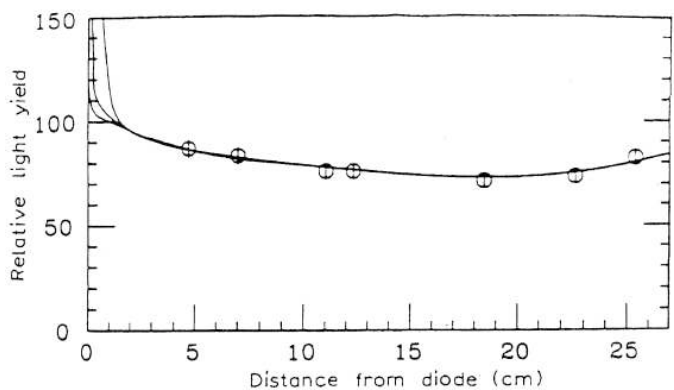}
\caption{Relative light yield as a function of the distance from the Si photodiode, measured with cosmic ray muons. Monte Carlo curves are shown for different light collimators \cite{Chen:1993}.}
\label{fig:Fig41}
\end{figure}
\end{center}

\begin{center}
\begin{figure}
\includegraphics*[width=0.5\textwidth]{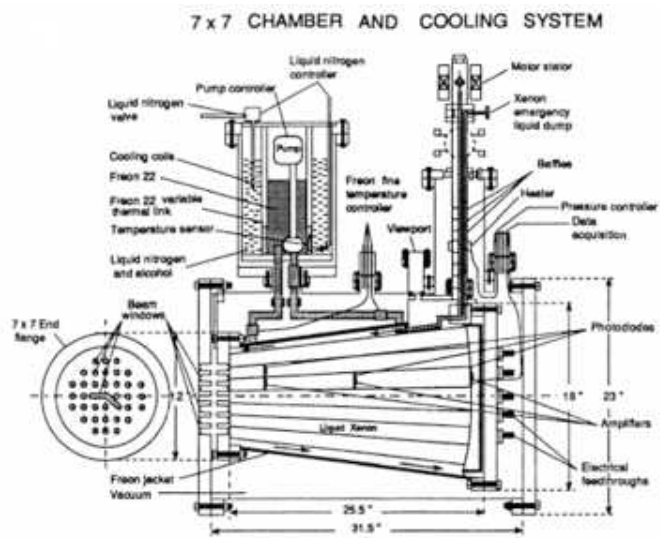}
\caption{Schematic of the MIT-ITEP-Dubna LXe scintillation calorimeter prototype \citep{Akimov:1996}.}
\label{fig:Fig40_mod}
\end{figure}
\end{center}
To achieve sufficient longitudinal uniformity, the cell had a tapered shape for a focusing effect, in addition to using tiny light collimators  in the form of grids coated with Al, placed in front of the photodiodes. The relative light collection uniformity was measured with such LXe cell, using cosmic ray muons and alpha particles. The results are shown in Figure \ref{fig:Fig41} \citep{Chen:1993}. Good uniformity was obtained, at the expense, however of a drastic reduction in number of photons collected. As a result, the energy resolution, estimated from Monte Carlo simulations \citep{Chen:1993}, was about  0.5\% ($1 \sigma$) for energies above 50 GeV, almost the same as that of a LAr calorimeter \citep{Doke:1985}.  

At the same time, the MIT-ITEP-Dubna collaboration, constructed a full-scale prototype of a LXe/LKr scintillation calorimeter, consisting of 45 unit cells, schematically shown in Figure \ref{fig:Fig40_mod} \citep{Akimov:1996}. 

\begin{center}
\begin{figure}
\includegraphics*[width=0.4\textwidth]{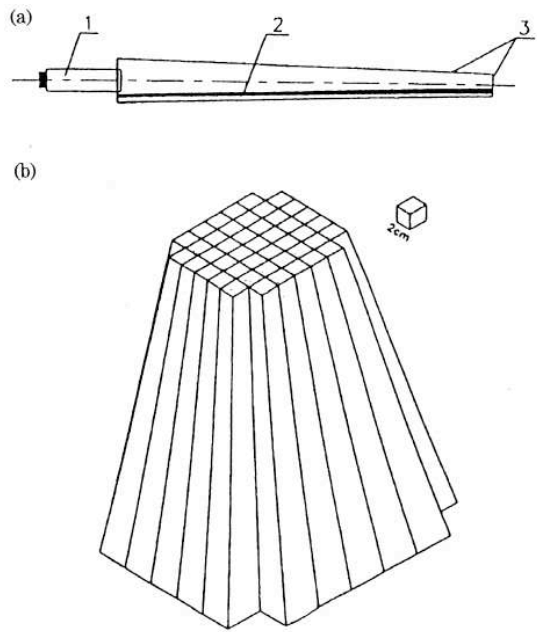}
\caption{(a) Unit cell of the LXe scintillation calorimeter:(1) PMT, (2) fragment of p-terphenyle strip, (3) Mylar  reflector. (b) Assembly of the full calorimeter reflector structure \cite{Akimov:1995}.}
\label{fig:Fig42}
\end{figure}
\end{center}

The walls of each unit cell were partly coated with strips of wavelength shifter (p-therphenyl) to convert the VUV light to visible light \citep{Akimov:1995, Akimov:1996} and the silicon photodiodes were replaced with PMTs. Figure \ref{fig:Fig42} shows a schematic of the unit cell and of the full calorimeter reflector structure \citep {Akimov:1995}.

\begin{center}
\begin{figure}
\includegraphics*[width=0.5\textwidth]{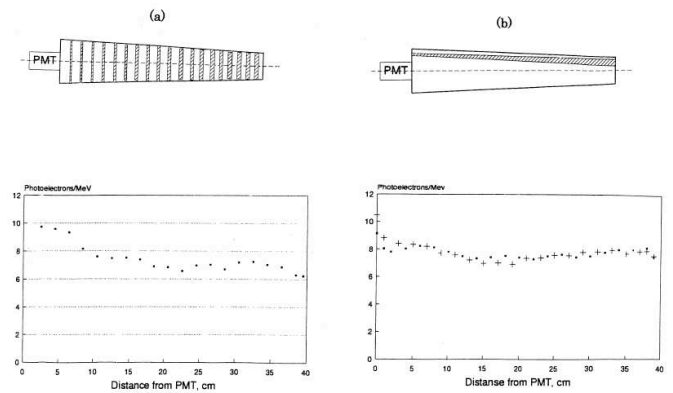}
\caption{Efficiency of light collection measured with the movable alpha-source for the unit cell shown in Figure~\ref{fig:Fig42}, for two different shapes of wave length shifter coating \cite{Akimov:1996}}
\label{fig:Fig43}
\end{figure}
\end{center}

 Figure \ref{fig:Fig43} show the variation of light collection efficiency versus distance from the PMT window for two type of cells, using different wavelength shifters \citep{Akimov:1995, Akimov:1996}. The maximum number of 10 photoelectrons/ MeV close to the PMT corresponds to only 0.02\% collection of the photons produced. 
The best energy resolution was 7.5\% ($1 \sigma$) for 348 MeV electrons, with an energy dependence $\Delta E/E  = 5/\sqrt{E}$\% where \textit{E} (GeV) is the incident electron energy. This energy resolution was not as good as that typical of the liquid xenon ionization calorimeter discussed above. In subsequent years,  much improvement was achieved as a result of new developments in photodetectors for the Xe VUV light, which could be operated directly in the liquid.  In the following we discuss two experiments dedicated to two rare processes expected from new physics beyond the Standard Model (SM) of particle physics.

\subsubsection{The RAPID Detector  for $\pi \rightarrow\mu\nu\gamma$ Decay}
\label{sec:rapid}
\begin{center}
\begin{figure}
\includegraphics*[width=0.45\textwidth]{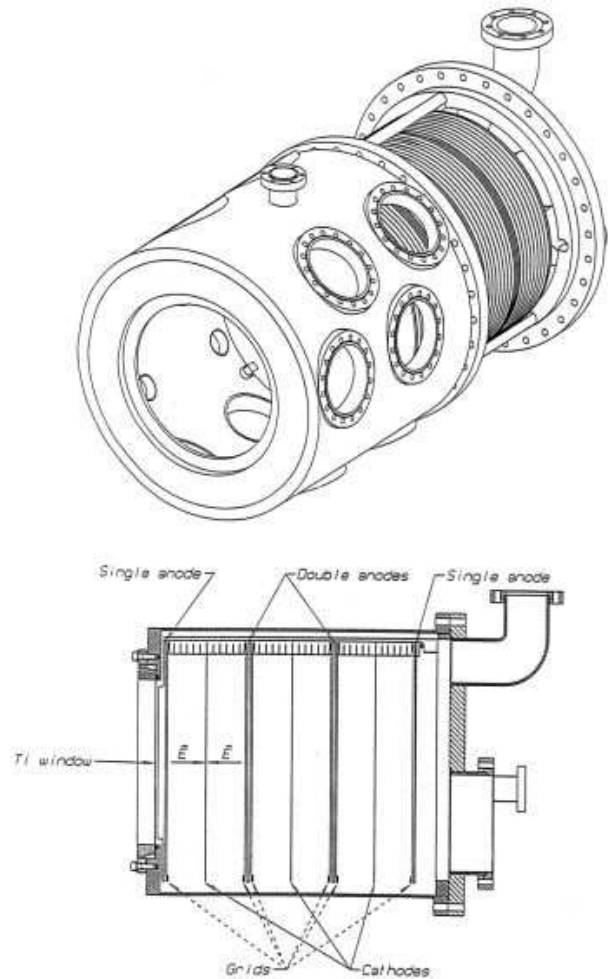}
\caption{(\textit{top}) Schematic of the RAPID LXeTPC vessel with Ti window for incident $\gamma$-rays and quartz windows for PMTs. 
(\textit{bottom}) Cross section of the electrodes for charge read out\citep{Bressi:1998}.}
\label{fig:rapid}
\end{figure}
\end{center}

In the early 1990 a LXeTPC based on ionization and scintillation light was developed by a Padova University group for detection of gamma-rays from the rare 
$\pi \rightarrow\mu\nu\gamma$ decay \citep{Brown:1964, Bressi:1998}.
The gamma-rays can be emitted either from the charged pion and muon (this term is called Internal Bremsstrahlung IB) or directly from the decay vertex due to the composite nature of the pion (Structure Dependent term SD). When squaring the decay matrix to get the decay probability, an Interference Term (INT) appears, which is proportional to SD$\times$IB.
The IB term is a pure QED term, well described and calculable within the frame of this theory. The SD term takes into account the fact that the pion is not an elementary particle but is constituted by quarks.
In the pion to muon radiative decay, however, both the SD term and the INT term are heavily suppressed compared to the IB term. With a very good approximation, the gamma-rays emitted from this decay can be thought of as a pure QED process.
QED calculations show that for a pion decay at rest the gamma spectrum is strongly peaked at zero,  falling approximately as 1/\textit{E}$_{\gamma}$ up to the maximum energy of 30 MeV. The goal of the RAPID experiment was to measure the energy of both the muon and the gamma
produced in the decay, in order to obtain not only the branching ratio but also the
full two-dimensional distribution. For the detection of the gamma-ray, LXe was recognized as an optimum medium and an ionization TPC with a 38 liter sensitive volume, self-triggered by the scintillation light was adopted 
to achieve low energy threshold, high detection efficiency and good energy resolution. The TPC, shown in Figure \ref{fig:rapid}, is described in detail in \citep{Carugno:1996}. It consists of a 64 liter cylindrical vessel, with a gamma-ray entrance window made of one 1 mm-thick titanium and 12 quartz windows for the detection of the scintillation light with VUV-sensitive PMTs. The electrode structure, for the detection of the ionization signal, consists of 6 drift regions, each 6.3 cm long with a shielding grid at 3 mm from the anode. With an applied electric field of 0.5 kV/cm, the energy threshold achieved was 230 keV. 
This was the first large volume TPC successfully implemented for a physics experiment, and the first to address the demanding requirement of high purity of the liquid for charge drift and a long-term stability of the charge yield. The Padova group successfully implemented cleaning and purification procedures, including continuous  recirculation of Xe gas  through hot metal getters by means of a clean circulation pump. An electron lifetime in excess of milliseconds was achieved and the TPC response was stable for a period of 2 months. The energy resolution obtained with this detector was about 7\% ($1 \sigma$) at 1 MeV: 

\begin{equation}
\frac{\sigma(E_{\gamma})}{E_{\gamma}} = \frac{7\%}{\sqrt{E_{\gamma}}}\oplus\frac{0.14 MeV}{E_{\gamma}}
\label{Eqn9}
\end{equation}
where \textit{E}$_{\gamma}$ is in MeV and $\oplus$ indicates a quadratic sum.
Despite the high detection threshold and  the large low-energy gamma-ray background, the branching ratio for gamma-rays of energy larger than 1 MeV was measured to be 2.0$\times$10$^{-4}\pm$12\% (stat)$\pm$4\% (syst), in good agreement with the theoretical QED expectation \citep{Bressi:1998}. 

\subsubsection{The MEG Detector for $\mu \rightarrow$ e$\gamma$ Decay}
\label{sec:meg}
Lepton flavor violation is predicted in supersymmetric grand unified theories (SUSY-GUTs) at a measurable level. The rare $\mu \rightarrow$ e$\gamma$ decay is the most promising lepton flavor violating process, with a predicted branching ratio above 10$^{-14}$ \citep{Orito:1997}. The first (and present) experimental upper limit to this branching ratio was reported in 1999 at a level of 1.2$\times$10$^{-11}$ \citep{Brooks:1999}. In the same year, an experiment capable to improve on this limit by at least two orders of magnitude was proposed \citep{PSI:1999}, based on the original suggestion by Doke to use a position sensitive LXe scintillating calorimeter, with excellent energy and position resolution, later to be known as the MEG detector \citep{Baldini:2002}. A $\mu \rightarrow$ e$\gamma$ decay event is characterized by the clear 2-body final state where the decay positron and the gamma ray are emitted in opposite directions with energies equal to half the muon mass (\textit{E} = 52.8 MeV). While positrons of this energy are abundant from the standard Michel decays of muons, gamma rays with such high energies are very rare. Therefore the key requirement for the MEG detector is high energy resolution for gamma rays, since the accidental background rate decreases, at least, with the square of the energy resolution \citep{PSI:1999,Baldini:2002}. 

\begin{center}
\begin{figure}
\includegraphics*[width=0.5\textwidth]{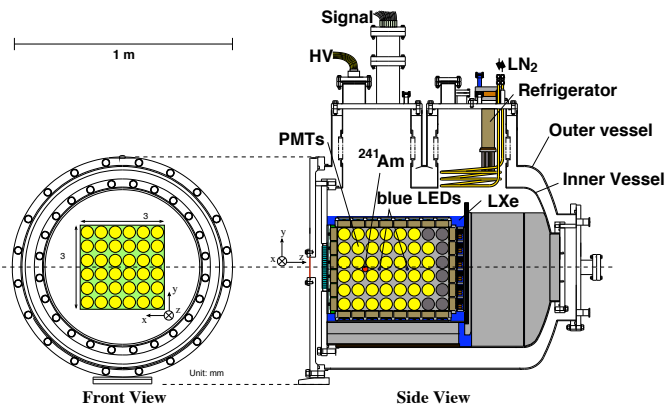}
\caption{Schematic of the large MEG prototype \cite{Mihara:2004a}.}
\label{fig:Fig47}
\end{figure}
\end{center}

\begin{center}
\begin{figure}
\includegraphics*[width=0.5\textwidth]{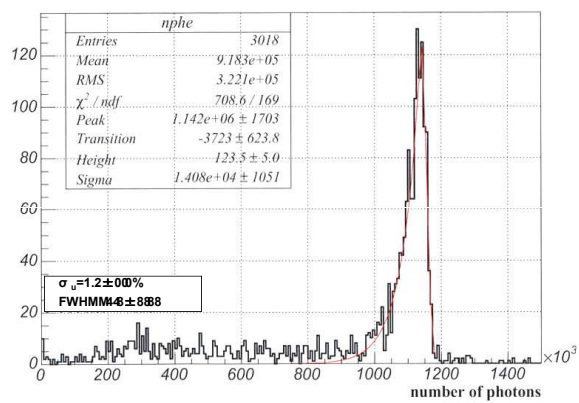}
\caption{Energy spectrum for 55 MeV gamma-rays measured with the large MEG prototype \citep{Baldini:2005a}.}
\label{fig:Fig48}
\end{figure}
\end{center}

The excellent properties of LXe as scintillator and the developments of new PMTs with high quantum efficiency, fast timing and their possible operation in LXe, motivated the choice of the detector for the MEG experiment. 
With the Xe scintillation light detected by a large number of PMTs immersed in the liquid volume, the detector's energy resolution is proportional to 1/$\sqrt{N_{pe}}$, where $N_{pe}$ is the number of photoelectrons, and was estimated to be better than 1\% ($1 \sigma$) for 52.8 MeV gamma-rays \citep{Doke:1999}. In addition, the gamma-ray interaction points can also be determined from the light distribution on individual PMTs \citep{Orito:1997}. 
Following the R\&D with a small prototype \citep{Mihara:2002} and with further improvement of the metal channel type PMT,  a large LXe prototype (68.6 l) with 228$\times$2'' PMTs was constructed and tested with gamma rays of 10 to 83 MeV \citep{Mihara:2004a}. The prototype and associated cooling and purification apparatus, shown schematically in Figure \ref{fig:Fig47}, were tested at the Paul Scherrer Institute in Switzerland, where the final MEG experiment is located. In this prototype, a pulse tube refrigerator (PTR) was used for the first time to condense Xe gas and maintain the liquid temperature stable over long periods of time. The PTR, optimized for liquid xenon, had a cooling power of 189 W at 165 K \citep{Haruyama:2004}.  

This prototype was large enough to test both the reliability and stability of the cryogenics system based on the PTR and the efficiency of the purification system required for the calorimeter's performance. The study of the impact  of impurities on Xe scintillation light transmission was particularly important and valuable to later LXe detectors, such as those developed for dark matter searches discussed earlier. While the transmission of the scintillation light is also affected by Rayleigh scattering, this process does not cause any loss of light and hence does not  affect the energy resolution. Two types of purification systems were developed for studies with this prototype. 
One is a gaseous purification system where liquid xenon is evaporated, purified by a heated metal getter, and then liquefied back in the cryostat \citep{Mihara:2004b}. This method is effective in removing all types of impurities but not efficient, since the purification speed is limited by the cooling power. The
 other method was developed specifically to remove water, which was identified as the major contributor to light absorption, at a much higher rate (100 $l$/hr) by circulating the LXe with a cryogenic centrifugal pump. Water is efficiently removed by a filter containing molecular sieves. Using this method, the total impurity concentration was reduced in 5 hours from 250 ppb to 40 ppb in the total 100 liters LXe volume of the MEG prototype \citep{Mihara:2006}.

Novel calibration and monitoring techniques have also been developed with this prototype study. In particular, a lattice of radioactive point-sources was developed by permanently suspending in the active volume (100 micron diameter) gold-plated tungsten wires carrying a small amount of $^{241}$Am \citep{Baldini:2006}. This method was successfully implemented and used to determine the relative quantum efficiencies of all the PMTs and monitor the stability of the detector during the experiment. In the final MEG detector, the PMT configuration is similar to that tested in the prototype, hence the methods of reconstructing the gamma-ray energy and interaction points were proven with the prototype. The reconstructed energy distribution for 55 MeV gamma rays from $\pi^o$ decay is shown in Figure \ref{fig:Fig48} \citep{Baldini:2005a, Ootani:2005, Sawada:2007}. The spectrum is asymmetric with a low-energy tail, caused mainly by gamma-ray interactions with materials in front of the sensitive region and by the leakage of shower components. The estimated energy resolution at the upper edge is 1.2 $\pm$ 0.1\% ($1 \sigma$) \citep{Ootani:2005, Sawada:2007} as shown in Figure \ref{fig:Fig48}, which is roughly equal to the value obtained by extrapolation of the small prototype data. The events remaining above 55 MeV are known to be due to mis-tagging of $\pi^o$ events. The position and time resolutions are evaluated to be 5 mm and $\sim$120 ps (FWHM) respectively \cite{Baldini:2006}. 

\begin{center}
\begin{figure}
\includegraphics*[width=0.35\textwidth]{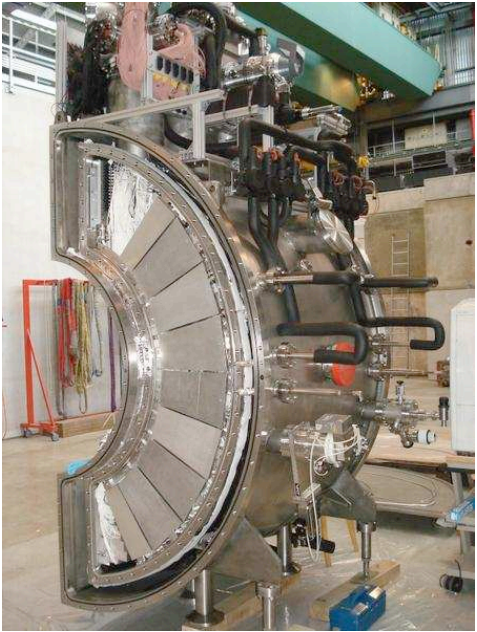}
\caption{Image of the full scale MEG LXe scintillation calorimeter.}
\label{fig:Fig50}
\end{figure}
\end{center}

The MEG cryostat, consisting of cold and warm vessels, is designed to reduce passive materials at the gamma-ray entrance. The same PTR tested on the prototype will be used for cooling, with an emergency cooling system based on liquid nitrogen flowing through pipes installed both inside the cold vessel and on its outer warm vessel. Gaseous and liquid purification systems 
are located on site for removing impurities possibly retained after liquefaction. Table \ref{Table10} shows the basic parameters of the LXe scintillation calorimeter for the MEG experiment. 
The construction of the calorimeter was completed in Fall 2007. Calibration started the same year and physics data taking continues to date. MEG, with a total LXe volume of  almost 2.7 tons and about 850 PMTs is currently the largest LXe detector in operation worldwide.

\begin{center}
\begin{table}[b!]
\caption{Basic parameters of the liquid xenon scintillation calorimeter for the MEG experiment.}
\label{Table10}
\begin{tabular}{|ll|}
\hline
\T Inner radius of the vessel	& 634 cm \\
Outer radius of the vessel	& 1104 cm \\
Acceptance angle & $\pm$60$^\circ$ \\
Volume of liquid xenon & 0.9 m$^3$ \\
Number of photomultipliers & 846 \\
Total mass of PMTs with their holders & 450 kg \\
Weight of the vessel	& 2.7 tons \\
Total weight of the calorimeter & 6 tons \\
\hline
\end{tabular}
\end{table}
\end{center}

\subsubsection{The EXO Detector for  0$\nu\beta\beta$ Decay of $^{136}$Xe}
A key topic in particle physics is the nature of neutrinos, their relationship to the charged leptons, and their mass. Oscillation experiments have confirmed mixing between the three neutrino types, from which mass differences can be estimated. The absolute mass of the neutrino remains unknown but is now expected to be in the region 0.01-0.1 eV \citep{Avignone:2005}. If neutrinos are Majorana in nature, i.e. they are their own antiparticle, the absolute mass scale can be determined by observations of nuclei which undergo double beta decay (2$\nu\beta\beta$), a decay with the emission of two electrons and two neutrinos. As Majorana particles, they can also decay into two electrons without emitting neutrinos, giving a signal at a single well-defined energy. One such nucleus is $^{136}$Xe, an 8\% constituent of natural Xe. Enrichment by centrifuging allows to increase this fraction to at least 80\%. Neutrinoless double beta decay of $^{136}$Xe will occur at an energy of 2479 keV, at a rate which depends on the square of the neutrino mass scale. Hence observations of this mode of decay could fix the neutrino mass scale for the first time. The sensitivity of such experiments is quoted as the reciprocal of the decay rate, or as the lifetime of the observed decay in years. For the expected neutrino mass range of 0.01- 0.1 eV, the lifetime would be in the region of 10$^{27}$-10$^{28}$ years, which in the case of $^{136}$Xe requires 1-10 tons of that isotope for observation of a few events in 1-2 years. 

\begin{center}
\begin{figure}
\includegraphics*[width=0.5\textwidth]{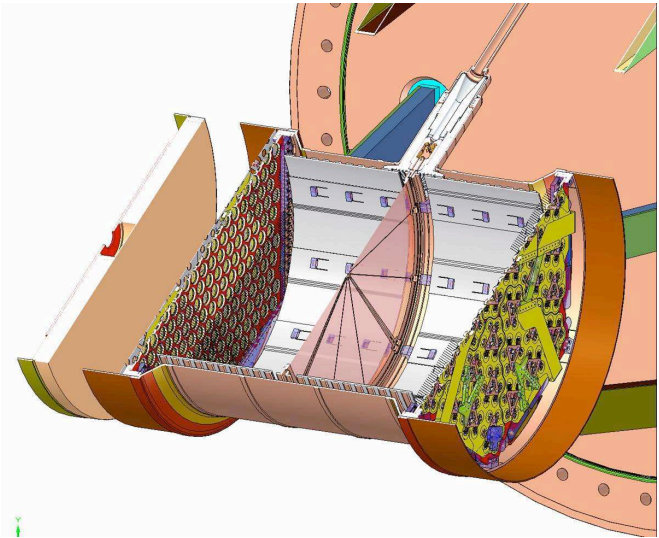}
\caption{Schematic of the EXO-200 TPC \cite{Gratta:2009}. The length of the TPC as well as the diameter of the TPC planes are 40 cm.}
\label{fig:exo-200_tpc}
\end{figure}
\end{center}

\begin{center}
\begin{figure}
\includegraphics*[width=0.5\textwidth]{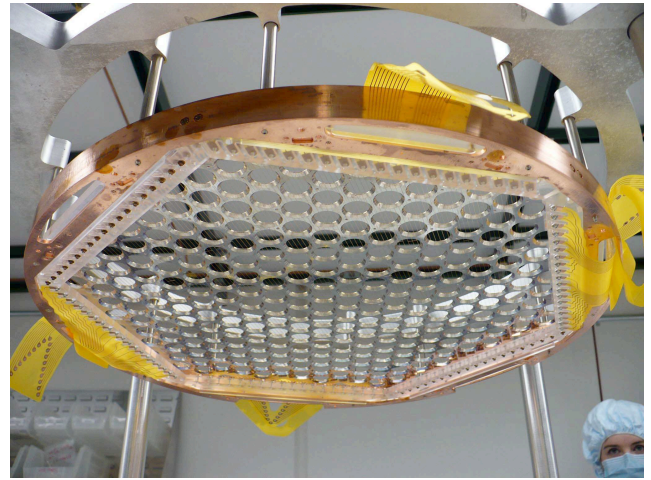}
\caption{Image of one EXO-200 anode plane. Induction and charge collection grids cross at 120$^o$. The LAAPD support platter is visible behind the grids. \cite{Gratta:2009}.}
\label{fig:exo-200_anode}
\end{figure}
\end{center}

The experimental effort using LXe to search for the 0$\nu\beta\beta$ decay of $^{136}$Xe was pioneered more than 20 years ago by Barabanov et al. \citep{Barabanov:1986}. A small detector filled with $^{136}$Xe enriched LXe was used by Bernabei et al. \citep{Bernabei:2002}. The EXO experiment plans to reach the multi-ton level in two stages, the first of which is EXO-200, a 0.2 ton prototype.  Although the search is for a signal at a single specific energy, it is necessary to have excellent energy resolution to separate this signal from the tail of the $2 \nu\beta\beta$ decay spectrum, for which the 0$\nu\beta\beta$ line is at the end point. As discussed in \ref{sec:two-phase}, the simultaneous detection of ionization and scintillation promises the best energy resolution for LXe detector, hence the EXO-200 experiment is baselined on  a liquid xenon time projection chamber (LXeTPC)  with detection of both charge and  light signals. To minimize instrumental background, the cryostat is made from a double wall of selected low activity copper, inside a 25 cm thick lead enclosure for gamma shielding, and located in the underground site of the Waste Isolation Pilot Plant (WIPP), at a depth of 2000 m.w.e. to minimize muon induced background. All components used in the EXO-200 have been qualified for low radioactivity with a variety of techniques \citep{EXO-NIM}. The cold vessel of the cryostat is cooled with heat exchangers fed by three industrial refrigerators, together capable of 4.5 kW at 170 K. The system offers double redundancy without the use of liquid nitrogen, discouraged at the underground site for safety reasons.  Xe is recirculated and purified in the gas phase by hot metal getters, with a custom-built pump, a Xe heater and condenser, cooled by a fourth refrigerator that is also used by a radon trap. A control system maintains the pressure difference across the thin TPC vessel to less than 0.3 bar. The TPC vessel (see Figure \ref{fig:exo-200_tpc}) is shaped to minimize dead volumes and hence to make efficient use of the enriched Xe which will be ultimately used, after testing with natural Xe.  The drift region, designed to support up to 75 kV with a central cathode plane, is defined by field shaping rings. About 150 kg of Xe are in a fiducial region, where the field is sufficiently uniform and the background sufficiently low for the rare event search. Each end of the cylindrical vessel flares out radially to make space for two wire grids for XY position reconstruction (See Figure \ref{fig:exo-200_anode}). An array of about 250 large area avalanche photodiodes (LAAPDs \citep{Neilson:2009}), mounted on a platter behind the grids, is used to read out the Xe light.  Based on Monte Carlo simulations, the overall light detection efficiency is between 15\% and 20\%, depending on event position in the volume. The readout electronics, for the charge induction and collection wires and for the light sensors, is located outside the lead shielding. The EXO-200 TPC is in the final stages of assembly in a clean room at Stanford, while the support and cryogenic systems are being commissioned underground. It is expected that an engineering run, with natural Xe, will take place in 2009 and running with enriched $^{136}$Xe will follow. For an estimated energy resolution of $\sim$1.6\% ($1\sigma$) at the 2.5 MeV endpoint energy, and with the estimated overall background, the sensitivity of this prototype to the 0$\nu\beta\beta$ lifetime is projected as 6.4$\times$10$^{25}$y (90\% confidence) after two years of data taking. This corresponds to an effective Majorana mass of the neutrino of 0.13 (0.19) eV depending on the nuclear matrix models. EXO-200 should also be able to gather information on the $\nu\nu\beta\beta$ decay, which has not yet been detected.
The required increase in sensitivity to the level of 10$^{27}$ -10$^{28}$ years, to match the expected Majorana mass level, can in principle be achieved with the same technique by increasing the detector mass to the 10 ton level. However, production of enriched xenon requires about an order of magnitude more xenon to start with. Given that the world production is only about 30 tons of xenon per year, it is clear that the procurement of such large amounts of xenon enriched in Xe-136 poses a major challenge for the realization of these experiments.

A more positive signal identification can be obtained by the detection of the Ba$^{++}$ ion to which the $^{136}$Xe decays, in coincidence with the electron ionization signal. This was first suggested in \cite{Miyajima:1974}. Individual atoms of $^{136}$Ba may be detected using fluorescent spectrometry with laser excitation. Different methods for extraction of ions from liquid xenon have been considered by the EXO collaboration, including the use of a cryo-tip or an optical fiber tip, both of which have stringent technical requirements \citep{KSullivan:2007}. EXO-200 will not test the Ba tagging method.

An alternative approach has been recently proposed by Arisaka \citep{Arisaka:2008} with the XAX ($^{136}$Xe-Ar-$^{129/131}$Xe) detector concept, which combines dark matter detection with 0$\nu \beta\beta$ decay, and includes a concentric arrangement of $<5$ tons enriched $^{136}$Xe surrounded by 5 tons depleted Xe ($^{129/131}$Xe) which provides simultaneously a dark matter target and shielding for the 0$\nu\beta\beta$ decay target.  The $^{129/131}$Xe also provides a pp solar neutrino detector, free from the $2 \nu\beta\beta$ decay background in natural Xe. A separate Ar detector is proposed as a comparison target for dark matter signals. Background studies for this scheme indicate that the 0$\nu\beta\beta$ decay sensitivity would be competitive with the EXO scheme without the need for the complex Ba tagging.

\subsection{Applications to Medical Imaging}
The excellent properties of liquid xenon for gamma-ray detection make it also very attractive for single photon emission computed tomography (SPECT) and positron emission tomography (PET),  the two primary modalities in use for functional medical imaging. The large majority of SPECT and PET systems in use today are  based on crystal scintillators as detection media. The first application of liquid xenon for SPECT \citep{Zaklad:1972}, used a multi-wire proportional counter with a position resolution of 4 mm FWHM. However, the counter suffered significant instabilities precluding its use as a practical gamma camera. In 1983, Egorov et al. constructed a gamma-camera using electroluminescence emission in high pressure xenon gas. An intrinsic spatial resolution of 3.5 mm (FWHM) and an energy resolution of 15\% (FWHM) for 122 keV gamma-rays were obtained \citep{Egorov:1983}.   
     
Most of the subsequent R\&D on liquid xenon detectors for medical imaging has focused on PET systems, with two approaches: 1) detectors which measure the energy and location of gamma-rays from the ionization signal, with the scintillation signal used only as an event trigger, and 2) detectors which use only the scintillation signal to measure the energy and location of gamma-rays. We discuss recent developments in both approaches. While it will be clear that using all the features of LXe (good energy resolution from combined ionization and light, high spatial resolution and Compton reconstruction) in a high rate, high sensitivity PET system is potentially very important for medical imaging, there are several practical issues such as cost, safety, complexity compared to conventional systems, plus the added requirement to produce fast images, complicated by the Compton reconstruction, which will require additional studies before this technology is accepted for such applications. 

\subsubsection{LXe Compton PET}
\label{sec:ComptonPET}
The principle of a PET system as diagnostic tool in medicine is to infer the location of a tumor from the measurement of the 3-D location of a radiotracer previously injected into the patient and labeled with positron emitting nuclides, such as $^{11}$C, $^{13}$N, $^{15}$O or $^{18}$F. The emitted positrons slow down in the surrounding patient's tissue annihilating with atomic electrons to give two back-to-back 511 keV annihilation photons. The annihilation occurs within a few millimeters from the  positron source. By detecting the two photons in coincidence and the coordinates of their interaction points in a detector, it is possible to define a line along which the positron emitting source is located in the patient. A set of such intersecting lines allows 3-D reconstruction of the source. Thus the main requirements for a PET detector are: (a) high photon detection efficiency ($\sim$80\% for each 511 keV gamma), (b) position resolution of a few millimeters, (c) time resolution to reduce the rate of false coincidences, (d) good energy resolution ($<100$ keV FWHM) to discriminate photons scattered in the patient and, (e) a high count rate capability ($\sim$10$^5$ -10$^6$ s$^{-1}$ per cm$^2$ of detecting surface). 

A LXe based PET detector should be able to improve on the energy and position resolution currently achieved with commercial whole-body PET scanners based on crystal scintillators. The scintillation of LXe is also faster than that from most crystal scintillators, allowing time of flight measurement, as will be discussed later. In a typical high Z, high density detector medium, the 511 keV gamma rays interact predominantly via Compton scattering, i.e. the energy is deposited in one or more interaction points before the gamma ray is photo-absorbed. The interaction points are too close to be resolved and the average location is taken approximately as the first interaction point, which is then the only point on the original gamma ray trajectory. 

If the detector is a LXeTPC such as the one discussed in \ref{sec:compton_tele}, the combined measurement of spatial coordinates and energy loss in each interaction point, allows one to use Compton kinematics to determine the sequence of interactions (compare figure \ref{fig:compton}). With the first interaction point known, the lower Z and density of LXe compared to typical crystal scintillators used in PET, is no longer a disadvantage.

\begin{center}
\begin{figure}
\includegraphics*[width=0.45\textwidth]{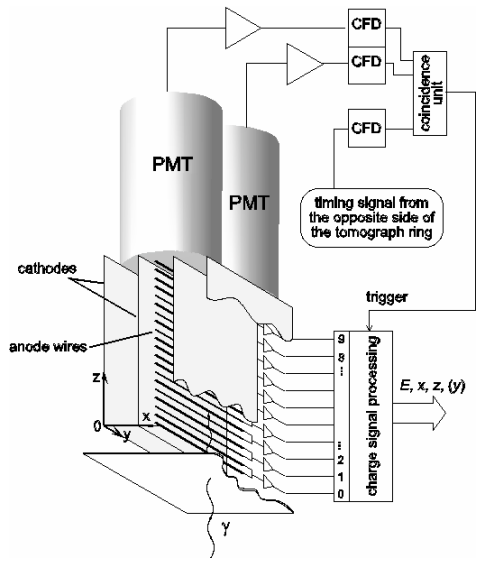}
\caption{Schematic of the Coimbra PET prototype \cite{Chepel:2001}.}
\label{fig:pet_coimbra}
\end{figure}
\end{center}

The first idea of using a LXeTPC for PET was proposed in 1993 by Chepel with subsequent R\&D documented in \citep{Chepel:1995, Chepel:1994b, Chepel:1997, Chepel:1999, Chepel:2001, Lopes:1995, Crespo:1998, Crespo:2000}. The setup used for the tests of this first Compton PET is schematically shown in Figure \ref{fig:pet_coimbra}. 
The detector was a LXe multi-wire detector consisting of six ionization cells, each formed by two parallel cathode plates with a multi-wire anode in the middle. Negative voltage is applied to the cathode, with the anode wires (50 $\mu$m diameter, 2.5 mm spacing) at ground. The 20 anode wires were connected in pairs to reduce the number of readout channels. The signal from each charge channel was fed to a low-noise charge-sensitive preamplifier followed by a linear amplifier, connected to a leading edge discriminator and a peak-sensing ADC. A typical noise level of 800 electrons FWHM was achieved for the charge readout. 

\begin{center}
\begin{figure}
\includegraphics*[width=0.5\textwidth]{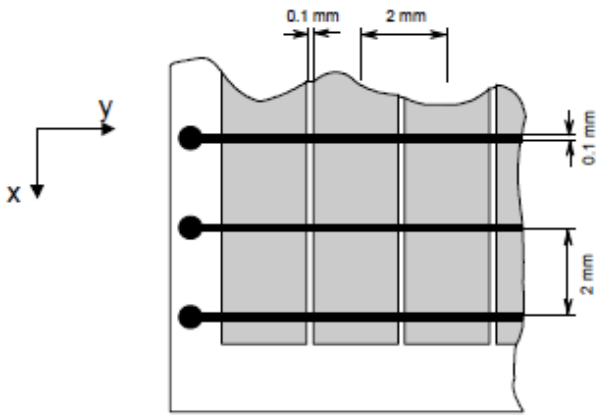}
\caption{The mini-strip plate of the Coimbra PET prototype \citep{Solovov:2002a}.}
\label{fig:strip_plate}
\end{figure}
\end{center}

The scintillation light produced in the active LXe filling the cell was detected by two Hamamastu R1668 PMTs, coupled to the cell through quartz windows. For each PMT, the fast signal from the last dynode was fed into a constant fraction discriminator (CFD) for triggering and the anode signal was amplified and shaped for measuring the light amplitude. The X-coordinate of the interaction point is determined with a precision better than 1 mm ($1 \sigma$) by measuring the electron drift time, triggered by the light signal. The identification of the pair of wires on which the charge is collected gives the depth of the interaction (Z-coordinate), with a resolution estimated at 2-5 mm ($1 \sigma$). For measurement of the Y-coordinate, at first, the ratio of the amplitudes of the two PMTs was used but the position resolution was only $\sim$10 mm ($1 \sigma$). To improve this resolution by about a factor of ten, the use of a mini-strip plate (Figure \ref{fig:strip_plate}) was considered \citep{Solovov:2002a}, and several tests for this electrode system were conducted. However, to our knowledge this position resolution has not yet been achieved for 511 keV gamma rays, although a multi-strip ionization chamber was tested with 122 keV gamma rays, obtaining a position resolution of $<2$ mm ($1 \sigma$) \citep{Solovov:2002b, Solovov:2003}. 

Recently two new groups have undertaken further R\&D towards this challenge, the Nantes-Subatech group in France \citep{Thers:2003} and the Columbia group \citep{Giboni:2007}. While single point position resolution is an important parameter of a LXe detector for PET, the double point resolution, i.e. the minimum distance required between two sources to be still separated in the image. As discussed by Giboni et al. \citep{Giboni:2007}, this double point resolution depends not only on position resolution, but also on statistics of events from the sources, as well as on the background from unrelated events. In an image, these factors determine the contrast besides the background from wrong Compton sequencing. There is a very strong background from genuine positron annihilation events with at least one gamma ray Compton scattering even before leaving the patient. Most of these scatters are very forward, with a small angle to the original direction, and with only a small change in gamma ray energy.

\begin{center}
\begin{figure}
\includegraphics*[width=0.4\textwidth]{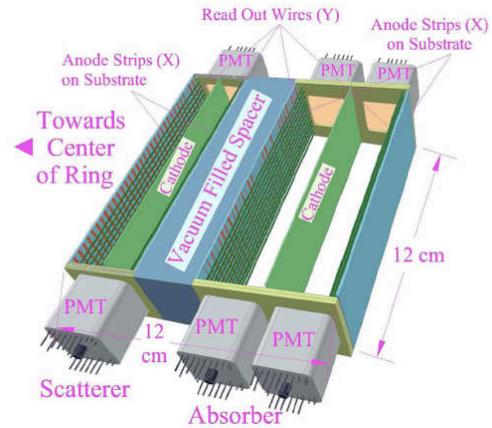}
\caption{Schematic of  the Columbia LXe Compton PET unit cell \citep{Giboni:2007}.}
\label{fig:pet_1}
\end{figure}
\end{center}

\begin{center}
\begin{figure}
\includegraphics*[width=0.4\textwidth]{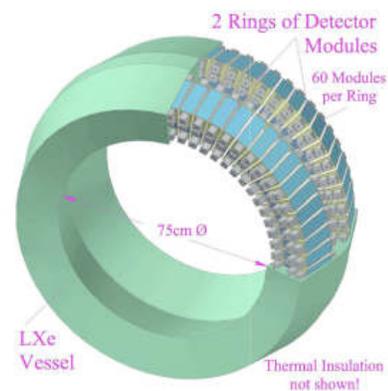}
\caption{Schematic of the Columbia full scale LXe Compton PET \citep{Giboni:2007}.}
\label{fig:pet_2}
\end{figure}
\end{center}

To identify and reject this background as far as possible, energy resolution is by far the most important factor. In \citep{Giboni:2007}, the LXe TPC proposed as Compton PET (see Figure \ref{fig:pet_1} and Figure \ref {fig:pet_2}) relies on the improved energy resolution that results from the sum of the  anti-correlated ionization and scintillation signals demonstrated in \citep{Aprile:2007}. The estimated FWHM energy resolution of 4\% must be compared with 15\% and 25\% for present GSO and BGO crystal scintillators.

Efficient light detection with PMTs immersed in the LXe helps not only the energy resolution but also provides very good coincidence timing. Giboni et al. \citep{Giboni:2005} measured a timing resolution of about 270 ps for 511 keV gamma rays in LXe. This might not yet be sufficient for a pure TOF-PET, but is sufficient for TOF-assisted PET. Compton kinematics not only determines the sequence of interaction points, but also constrains the direction of the incoming photon to the surface of a cone. The opening angle of the cone is the Compton scattering angle in the first interaction point. If the incoming directions are thus determined for both annihilation gamma rays, even if one of the gamma rays scatters within the body, one can still determine the original line on which the annihilation gamma rays were emitted. The number of events with one of the gamma rays scattering once in the body amounts to about twice the number of non-scattered events. Therefore, with Compton reconstruction, the number of acceptable events is nearly a factor 3 larger than the number of good events in a standard PET detector \citep{Giboni:2007}. This increased detection efficiency results in a reduction of the radiation dose for the patient.

\begin{center}
\begin{figure}
\includegraphics*[width=0.55\textwidth]{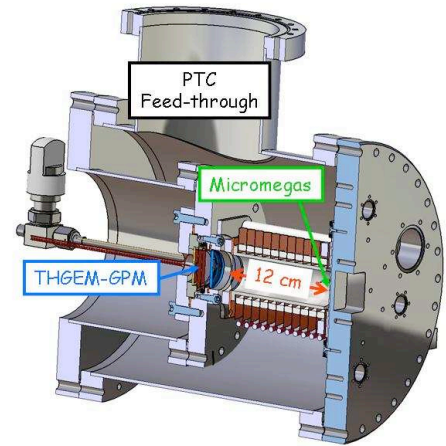}
\caption{Schematic of the Nantes LXe Compton PET prototype \citep{Grignon:2007}.}
\label{fig:Fig1Dom}
\end{figure}
\end{center}

The approach used by a Nantes-Subatech group uses  a micro-structure pattern readout for the charge and a gaseous photomultiplier (GPM) \citep{Breskin:2009} for the light signal. The aim is to achieve an energy resolution similar to that of Aprile et al. \cite{Aprile:2005}, and a $\sigma_x \sim \sigma_y \sim \sigma_z \leq$ 0.1 mm position resolution, in order to reconstruct the incident angle of the 511 keV gamma-rays with an angular resolution less than 2$^o$ ($1 \sigma$).  
Moreover, another medical imaging technique has been proposed by the same group \citep{Grignon:2007}, aimed at reducing the examination time and the radiation dose for the patient. This technique is based on precise location in organs of a 3-photon emitter (Scandium-44). The method would provide high resolution  imaging by combining  the information from a PET with the direction cone defined by a third gamma ray.
The first LXe Compton telescope prototype  has been built and is under testing. It features a 3$\times$3$\times$12 cm$^3$ active volume (Figure \ref{fig:Fig1Dom}). The LXe cryogenics are controlled by a dedicated Pulse Tube Cryocooler \citep{Haruyama:2004}, the same machine used on the XENON100 and MEG experiments.  
First results are very promising, demonstrating for the first time the complete collection of the ionization electrons by using a so-called Micromegas detector, fully immersed in LXe \citep{Giomataris:1996}. The scintillation signal has been so far detected by a standard PMT, to be replaced by a GPM \citep{Breskin:2009}.  

\subsubsection{LXe TOF PET}
\begin{center}
\begin{figure}
\includegraphics*[width=0.4\textwidth]{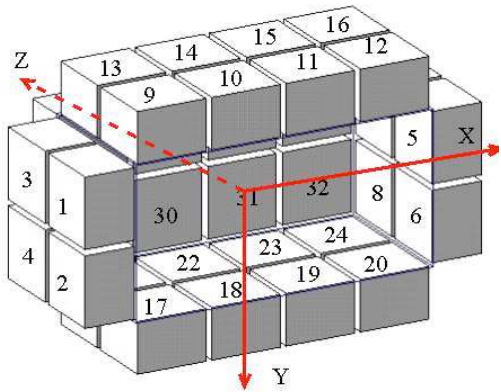}
\caption{Arrangement of 32 PMTs in Waseda TOF PET prototype \cite{Nishikido:2004}.}
\label{fig:Fig53}
\end{figure}
\end{center}

\begin{center}
\begin{figure}
\includegraphics*[width=0.45\textwidth]{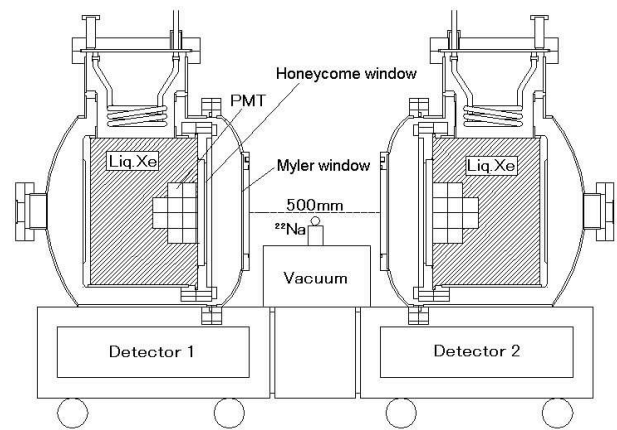}
\caption{Schematic of the Waseda TOF PET prototype \citep{Nishikido:2004}.}
\label{fig:Fig54}
\end{figure}
\end{center}

\begin{center}
\begin{figure}
\includegraphics*[angle=270,width=1\columnwidth]{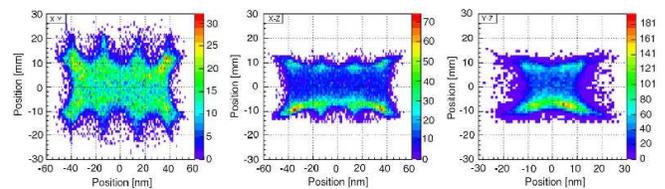}
\caption{Spatial distributions of 511 keV interaction points reconstructed with the Waseda prototype \cite{Doke:2006}.}
\label{fig:Fig56}
\end{figure}
\end{center}

The possibility of a TOF PET, based on the excellent properties of LXe as scintillator, was first suggested by Lavoie in 1976 \citep{Lavoie:1976}, and the study of this type of PET was carried out in 1997 by the Waseda group. A prototype, consisting of two LXe chambers, was constructed in 2000. In its initial implementation, each chamber has a LXe sensitive volume of 120$\times$60$\times$60 mm$^3$ viewed by only 32$\times$1'' PMTs, with one side left uncovered as an entrance window for gamma rays (Figure \ref{fig:Fig53}).  
A schematic of the whole assembly is shown in Figure \ref{fig:Fig54}. The distance between the entrance planes is 70 cm. Testing began in 2001 \citep{Nishikido:2004}, and soon thereafter, improved PMTs became available and were used in the prototype \citep{Nishikido:2005a}. 
Testing of the prototype with the new PMTs was carried out with a $^{22}$Na source placed in the center of the setup. The output signals from the 32 PMTs of each chamber were recorded by charge-sensitive analog-to-digital converters (QADCs), and the TOF information was recorded with time-to-digital converters (TDCs). The position of a gamma ray interaction was determined by the time of flight of the gamma ray to the PMTs. The position distributions of interaction points in X-Y, X-Z and Y-Z planes inside the sensitive volume, reconstructed with this method, are shown in Figure \ref{fig:Fig56}. 

\begin{center}
\begin{figure}
\includegraphics*[width=0.4\textwidth]{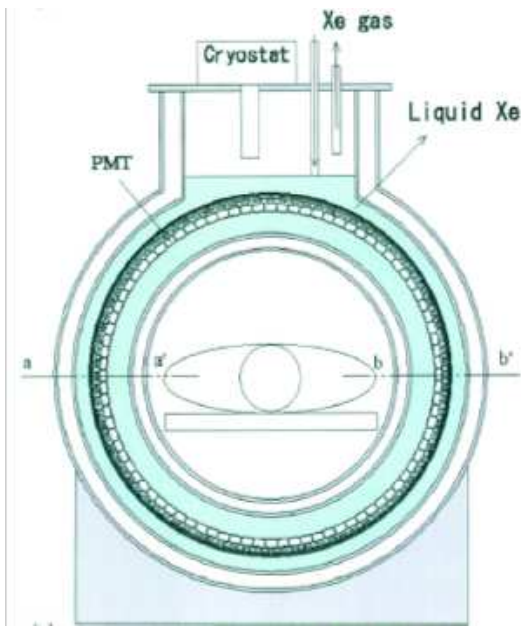}
\caption{Schematic of the Waseda full scale LXe TOF-PET \citep{Doke:2006}.}
\label{fig:Fig59}
\end{figure}
\end{center}

The energy resolution obtained from the sum of 32 PMTs was 26.0\% (FWHM) \citep{Doke:2006}. The resolution of the time difference spectrum between the two detectors was 514 ps (FWHM) for annihilation gamma rays. For a central volume of 5$\times$5$\times$5 mm$^3$ an even better time resolution of 253 ps was obtained. It is expected that the time resolution improves inversely proportional to $\sqrt{N_{pe}}$, i.e. an increase in the quantum efficiency or the light collection efficiency will improve the timing. This is within reach, based on available PMTs with significantly better quantum efficiency, compared to the PMTs used in the Waseda prototype. With a time resolution of 300 ps,  one will be able to reduce the number of background events. A full scale TOF PET, with an axial sensitive length of 24 cm, has been proposed by the Waseda group (see Figure \ref{fig:Fig59}) \citep{Doke:2006, Nishikido:2005b}. 

Another proposal for a LXe TOF PET was recently made by the Grenoble group \citep{Gallin-Martel:2006, Gallin-Martel:2008}, applying a light division method to measure the position in the axial direction. However, the reported position resolution was only about 10 mm (FWHM) in the central region, not competitive with what is achieved in commercially available crystal PET. Simulation results show much better performance, but unfortunately the large discrepancy is not understood or explained yet. 

Clearly, more studies are required to establish LXe Compton/TOF detectors as an alternative to classical crystal PET systems. LXe based detectors require associated systems (from cryogenics to gas handling and purification) which are typically perceived as complex and costly, contributing to some reluctance in their use, especially as diagnostic tools in medicine. In the last few years, however, many significant technological advances have resulted in large LXe detector systems operating for many months with excellent and stable performance. In particular, the LXe detectors deployed underground for dark matter searches (see \ref{sec:DM}) have proven that long term stability, reliability, and also safety issues  can be confidently solved.  


\section{Summary}

We have reviewed the current state-of-the-art in liquid xenon detectors and associated principles of operation and performance characteristics. We could not expand on many topics to the extent we would have liked, yet we have attempted  to gather the most relevant information from past and recent experimental work on liquid xenon detectors developed for a variety of applications. It is our wish that this article may be a useful resource especially for the new generation of investigators who are turning with enthusiasm to this material for the detection and measurement of different particles in different fields. Recent years have seen an increased interest in liquid xenon detectors and an aggressive development of related technologies, driven by rare event searches from dark matter to neutrinoless double beta decay. The rate at which the performance of liquid xenon detectors has improved in the last ten years is very encouraging. Yet, much work remains to be done in many areas, including a better understanding of the response of LXe to low energy particles and of the factors affecting the energy resolution. As the scale of some of these experiments grows, so does the challenge to push the performance limit in order to improve the detection sensitivity. Additional challenges for multi-ton scale LXe detectors will be associated with the availability of xenon, its cost and purification. The next ten years will bring to fruition many of the ongoing experiments and others now in the planning phase. We are confident that these will make far-reaching contributions to some of the most important questions in physics today. It is also encouraging to see a rapid development of liquid xenon detectors for improved imaging in medicine, a practical application benefiting society as a whole.

\begin{acknowledgments}
We would like to thank the many colleagues who share our passion for this topic and who have contributed to the advancement of liquid xenon detectors. Special thanks go to the many students and young researchers who are working on the experiments covered in this article, committed to their success. One of us (E. Aprile) would like to thank her students Bin Choi and Guillaume Plante for their assistance in preparing the article. 
\end{acknowledgments}

\bibliographystyle{apsrmp}


\end{document}